\documentclass[journal,comsoc]{IEEEtran}
\usepackage[T1]{fontenc}

\ifCLASSINFOpdf
\else
\fi
\usepackage{amsmath}
\usepackage[cmintegrals]{newtxmath}

\usepackage{bm}
\usepackage{amsmath}
\usepackage{algorithm}
\usepackage[noend]{algpseudocode}
\usepackage{graphicx}
\usepackage{epsfig,subfigure}
\usepackage{balance}
\usepackage{mathrsfs}

\usepackage{rotating}
\usepackage{xcolor}

\begin{document}
%
\title{Optimizing Bloom Filter: Challenges, Solutions, and Comparisons}
%
%
%

\author{Lailong~Luo,
        Deke~Guo,
        Richard~T.B.~Ma,
        Ori~Rottenstreich,
        and~Xueshan~Luo 
\thanks{This work is partially supported by National Natural Science Foundation of China under Grant No.61772544, and the Hunan Provincial Natural Science Fund for Distinguished Young Scholars under Grant No.2016JJ1002. \emph{Corresponding author: Deke Guo.}}  
\thanks{Lailong Luo,  Deke Guo, and Xueshan Luo are with the Science and Technology Laboratory on Information Systems Engineering, National University of Defense Technology, Changsha, Hunan, 410073, China. E-mail: \{luolailong09, dekeguo, xsluo\}@nudt.edu.cn.}
\thanks{Deke Guo is also with the College of Intelligence and Computing, Tianjin University, Tianjin, 300350, P. R. China.}
\thanks{Richard T. B. Ma is with the School of Computing, National University of Singapore, Singapore. E-mail: tbma@comp.nus.edu.sg.}
\thanks{Ori Rottenstreich is with the Technion and ORBS Research, Israel. E-mail: or@cs.technion.ac.il.}
}

\maketitle

\begin{abstract}
Bloom filter (BF) has been widely used to support membership query, i.e., to judge whether a given element $x$ is a member of a given set $S$ or not. Recent years have seen a flourish design explosion of BF due to its characteristic of space-efficiency and the functionality of constant-time membership query. The existing reviews or surveys mainly focus on the applications of BF, but fall short in covering the current trends, thereby lacking intrinsic understanding of their design philosophy. To this end, this survey provides an overview of BF and its variants, with an emphasis on the optimization techniques. Basically, we survey the existing variants from two dimensions, i.e., performance and generalization. To improve the performance, dozens of variants devote themselves to reducing the false positives and implementation costs. Besides, tens of variants generalize the BF framework in more scenarios by diversifying the input sets and enriching the output functionalities. To summarize the existing efforts, we conduct an in-depth study of the existing literature on BF optimization, covering more than 60 variants. We unearth the design philosophy of these variants and elaborate how the employed optimization techniques improve BF. Furthermore, comprehensive analysis and qualitative comparison are conducted from the perspectives of BF components. Lastly, we highlight the future trends of designing BFs. This is, to the best of our knowledge, the first survey that accomplishes such goals. 
\end{abstract}


\IEEEpeerreviewmaketitle

\section{Introduction}
%
%
%
%
\IEEEPARstart{B}{loom} filter \cite{BF} is a space-efficient probabilistic data structure for representing a set of elements with supporting membership queries with an acceptable false positive rate. Hitherto, the applications of BF and its variants are manyfold. In the field of networking, BF has been employed to enable routing and forwarding \cite{NDN_1} \cite{Multiclass_BF} \cite{Multicast_1} \cite{Multicast_2} \cite{Routing_1} \cite{Multicast_ICL} \cite{routing_comnet}, web caching \cite{Paradox_BF} \cite{WebCache_1}, network monitoring \cite{Monitoring_1}, security enhancement \cite{Security_1} \cite{Concatenated_BF}, content delivering \cite{ContentDlivery_1}, etc. In the area of databases, BF is a proper option to support query and search \cite{BooleanQueries_1} \cite{PublishSubscribe_1}, privacy preservation \cite{Spatial_BF}, key-value store \cite{BloomStore} \cite{KBF}, content synchronization \cite{ICBF} \cite{IBLT} \cite{SetRecon} \cite{Set_sync_tkde} \cite{set_sync_CBF}, duplicate detection \cite{Duplication_1} and so on. Beyond the general fields of networking and databases, BFs have been recently used to resolve biometric issues \cite{Biometric_1} \cite{Biometric_2}, and even navigation tasks in the mobile computing scenarios \cite{Navigation_1}. Other detailed applications can be found in several surveys \cite{ExistingSurvey_1} \cite{ExistingSurvey_2} \cite{ExistingSurvey_3}. 

The major motivation for conducting this survey is two-fold. First of all, tens of BF variants have been proposed in recent years. Existing surveys, however, are somehow out-of-date and do not cover these new proposals. The latest survey \cite{ExistingSurvey_1}, which covers about 25 variants, was published five years ago. Secondly, the existing surveys \cite{ExistingSurvey_1} \cite{ExistingSurvey_2} \cite{ExistingSurvey_3} mainly focus on the applications of BF, but lack essential understandings of the optimization techniques of the proposed variants. Therefore, they do not provide operational advice to the users in reality. As shown in Table \ref{table:comparison}, compared with the existing surveys, our survey covers more BF variants (more than 60) and introduces general applications rather than focus on a single scenario. In particular, our survey concentrates on the optimization techniques employed in these variants. 

\begin{table}[tbp]
\centering
\begin{footnotesize}
\caption{Comparison with existing surveys.} \label{table:comparison} \vspace{-0.1in}
\begin{tabular}{c|c|c|c}\hline
First author &  \# variants & Scenarios & Optimizations   \\ \hline
Broder \cite{ExistingSurvey_2}  & $\leq 5$ & Networks & Not mentioned  \\  \hline
Geravand \cite{ExistingSurvey_3}   &  $\leq 15$& Network security &  Not mentioned \\  \hline
Tarkoma  \cite{ExistingSurvey_1}   & $\leq 25$ & Distributed system  &  Not mentioned   \\  \hline
This survey  & $\geq 60$ & General & Detailed    \\  \hline
\end{tabular} \vspace{-0.1in}
\end{footnotesize}
\end{table}

To this end, we go back to the design philosophy of BF and thoroughly analyze the existing optimization techniques to improve BF. In this manner, we expect to provide a guidance to potential users whenever BFs are within their considerations. Specifically, we survey the existing variants from two dimensions, i.e., performance and generalization. To improve the performance, dozens of variants devote themselves to reducing the false positives and easing the implementation. Besides, tens of variants generalize the BF framework in more scenarios by diversifying the input sets and enriching the output functionalities. In our survey, more than 60 up-to-date designs are reviewed and qualitatively yet systematically analyzed.  As far as we know, this is the first survey which systematically summarizes the optimization techniques of BFs. 

\begin{figure}
  \centering
  \includegraphics[width=3.2in]{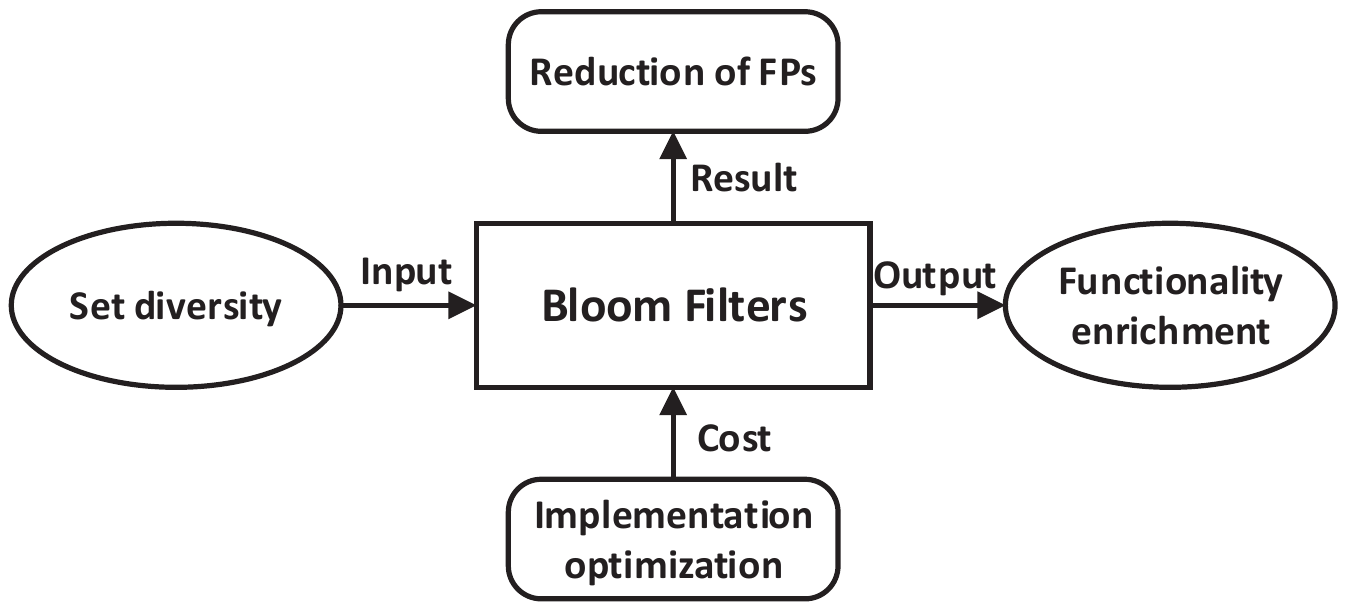}\\ \vspace{-0.05in}
  \caption{The top view of the optimization angles for Bloom filter.}\label{fig:logic} \vspace{-0.15in}
\end{figure}

Despite its space-efficiency, BF still faces some challenges related to false positives, implementation, elasticity, and functionality to some extent. To ease the potential challenges and further improve the performance of BF, many interesting techniques have been proposed and diverse variants have been designed. In this survey, we present the prior arts of improving BF from four angles, i.e., techniques to reduce the false positives, optimizations in a real implementation, dedicated designs for diverse datasets, and proposals to enable more functionalities. The intrinsic logic of the four angles is shown in Fig. \ref{fig:logic}. Basically, from the perspective of performance, the BF can be improved by reducing the false positives and optimizing the implementation. Besides, the BF framework can also be generalized from both the input and output aspects by representing more types of sets and enabling more functionalities beyond set membership queries. 

Following the above directions, we organize the rest of this survey as follows. Section \ref{sec:BF} details the design philosophy of BF and analyzes the challenging issues while Section \ref{sec:Applications} briefly introduces the applications of BFs. Thereafter, we detail the variants designed for reducing false positives (FPs) in Section \ref{sec:ReducingFPP}, implementation optimizations in Section \ref{sec:Implementation},  generalizing the set diversity in Section \ref{sec:SetDiversity}, and functionality enrichment in Section \ref{sec:Functionality}. Following that, Section \ref{sec:Analysis} analyzes and compares all mentioned variants from a quality perspective. We summarize this survey and enumerate several open issues about the BF framework in Section \ref{sec:Summary} and then conclude this survey in Section \ref{sec:Conclusion}.

\section{Bloom Filters}\label{sec:BF}
In this section, we detail the basic theory of Bloom filter in terms of its framework, characteristics,  and challenges. 

\subsection{Framework of Bloom filter}
Bloom filter (BF) is a space-efficient probabilistic data structure that enables constant-time membership queries \cite{BF}. Let $S\mathrm{=}\{x_{1},x_{2},...,x_{n}\}$ be a set of $n$ elements such that $S\mathrm{\subseteq} U$, where $U$ is a universal set.  BF represents such $n$ elements using a bit vector of length $m$. All of the $m$ bits in the vector are initialized to 0. Specifically, to insert an element $x$, a group of $k$ independent hash functions, $\{h_{1},h_{2},...,h_{k}\}$, are employed to randomly map $x$ into $k$ positions $\{h_{1}(x),h_{2}(x),...,h_{k}(x)\}$ (where $h_i(x)\mathrm{\in} [0,m\mathrm{-}1]$) in the bit vector. Then the bits in these $k$ vector positions are all set to 1. To query whether an arbitrary element is a member of set $S$, BF maps the element into its bit vector with the $k$ hash functions and thereafter checks whether all the $k$ bits are 1s. If any bit at the $k$ hashed positions of the element is 0, the BF concludes that this element does not belong to the set; otherwise, the BF indicates that the queried element belongs to the set $S$. 

\begin{figure}
  \centering
  \includegraphics[width=3.2in]{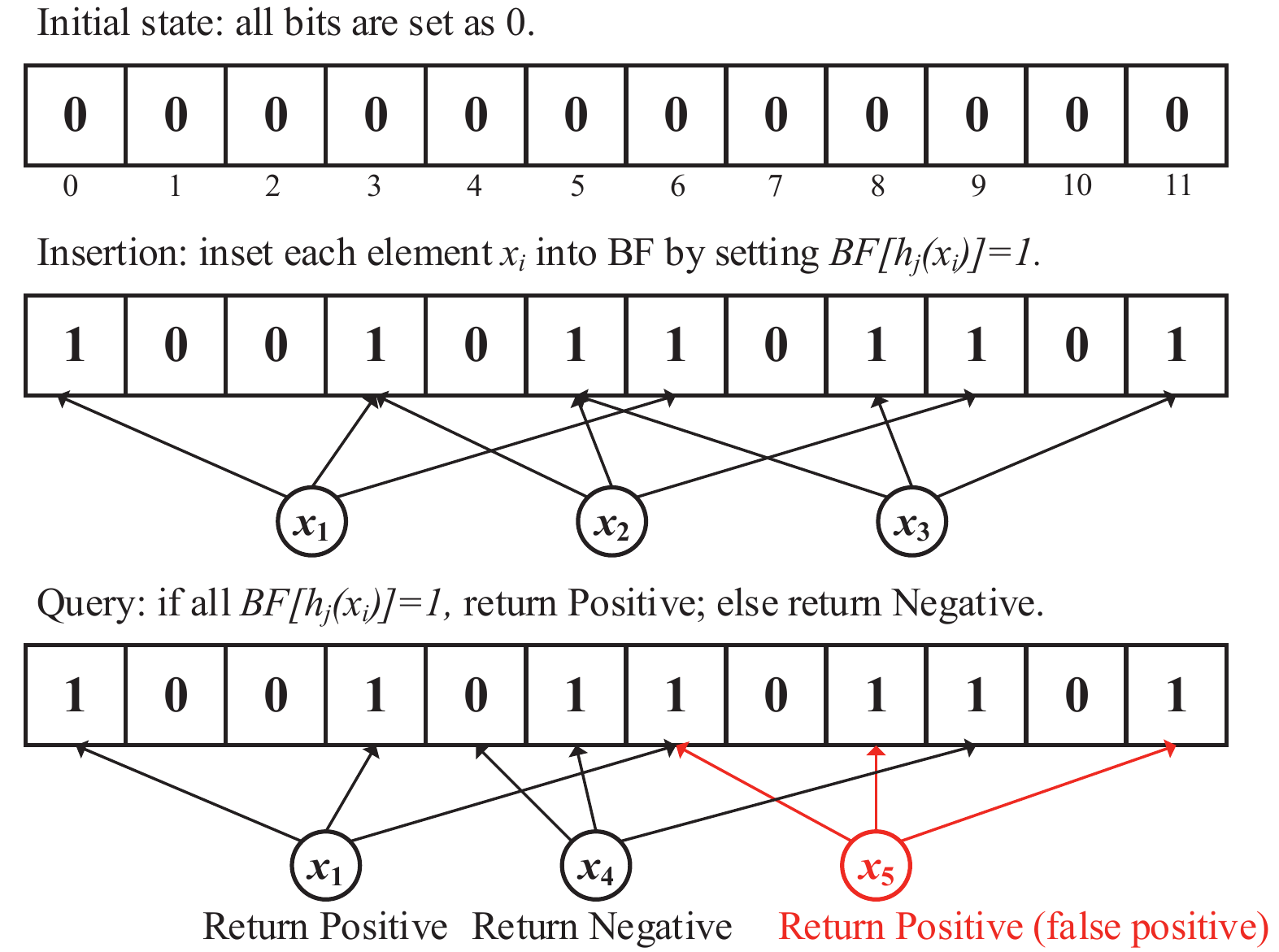}\\ \vspace{-0.05in}
  \caption{An illustrative example of BF with $m\mathrm{=}12$ and $k\mathrm{=}3$ to represent set $S\mathrm{=}\{x_1,x_2,x_3 \}$. Note that the query of $x_5$ results in a false positive error.}\label{fig:BF} \vspace{-0.15in}
\end{figure}

Fig. \ref{fig:BF} presents an example of BF with $m\mathrm{=}12$ bits and $k\mathrm{=}3$ hash functions to represent the set $S\mathrm{=}\{x_1,x_2,x_3 \}$. To insert these elements, the 3 corresponding bits for each element are set to 1. When querying, the BF checks the 3 corresponding bits for the queried element. For $x_1$, the bits at position 0, 3 and 6 are all 1, thus BF returns ``Positive'' for the query. The membership query of $x_4$ returns ``Negative'' since the bit at position 4 is 0. Note that, due to the unavoidable hash conflicts (generating same hash value for diverse input elements), the membership query based on BF may incur false positive errors, i.e., wrongly indicating that a non-member element is a member of $S$. In Fig. \ref{fig:BF}, the query of $x_5$ returns positive because the bits at position 6, 8, and 11 are all 1, though $x_5\notin S$. Although BF incurs false positives, for many applications the space savings and constant locating time outweigh this drawback when the probability of false positive is small.

To realize acceptable false positive rate $f_r$, the parameters of BF, i.e., the BF length $m$, the number of employed hash functions $k$, and the number of elements in the set $n$, call for careful design. Theoretically, the value of false positive rate can be calculated as:
\begin{equation}\label{equ:fpr}
  f_r = \left[1-\left(1- \frac{1}{m}\right)^{nk}\right]^k \approx \left(1-e^{-\frac{kn}{m}}\right)^k,
\end{equation}
where $(1-1/m)^{nk}$ is approximated by $e^{-kn/m}$. Consequently, to realize the minimum value of $f_r$, $e^{-kn/m}$ should be minimized. With this insight, the optimal value of $k$ is derived as:
\begin{equation}\label{equ:opt_k}
  k_{opt} = \frac {m} {n} \ln 2 \approx \frac{9m}{13n}.
\end{equation}
Followed by $k_{opt}$, the resulted false positive rate equals $0.5^k \approx 0.6185^{m/n}$. That is, with the given $k$, to maintain a fixed false positive rate, the value of $m$ should be increased linearly with the value of $n$.

However, as reported in \cite{BF_fpr1} and \cite{BF_fpr2}, the false positive rate in real deployments is higher than the value given by Equ. \ref{equ:fpr}. Theoretically, it has been proven in \cite{BF_fpr3} that Equ. \ref{equ:fpr} offered a lower bound of the false positive rate. After formulating the BF framework as a typical problem of balls and urns, the authors in \cite{BF_fpr3} present a more accurate false positive rate of BF as:
\begin{equation}\label{equ:real_fpr}
 f_r = \frac {1} {m^{k(n+1)}} \sum_{i=1}^{m} i^ki! \binom m i  \left\lbrace\begin{array}{c} kn\\i\end{array}\right\rbrace,
\end{equation} 
where $\left\lbrace\begin{array}{c} kn\\i\end{array}\right\rbrace = \frac 1 {i!} \sum_{j=0}^i (-1)^j \binom i j j^{kn}$ is called a \emph{Stirling number of the second kind} \cite{BF_fpr4}.

Naturally, in the framework of BF, deleting an element is not permitted. The reason is that resetting the corresponding 1s to 0s directly may lead to false negative results for other elements. Therefore, Counting BF (CBF) \cite{Counting_BF} extends the BF by replacing each bit as a counter with multiple bits. When inserting an element, the corresponding $k$ bits will be increased by 1. In contrast, the deletion of an element will be supported via decreasing the corresponding counters by 1. In this way, the deletion of an element will not affect the existence of other elements. It has been proved that 4 bits for a counter are enough to achieve eligible overflow probability (less than $1.37\mathrm{\times} 10^{-15}\mathrm{\times} m$, where $m$ is the number of counters in CBF). CBF also supports constant-time membership query. To answer a membership query, the CBF checks the $k$ corresponding counters. If all of them are non-zero, CBF judges that the queried element is a member; otherwise, negative. 

\subsection{Intrinsic characteristics of Bloom filter} \label{subsec:characteristics}
As a probabilistic data structure, BF supports fast membership query with potential false positive errors. For better understanding, we highlight the intrinsic characteristics of BF as follows. 

\textbf{Space-efficient}. BF programmes each element in a given set with a $m$-bit vector, irrespective of the number of bits in a bit representation of an element. With each element as input, the $k$ independent hash functions will select $k$ bits in the bit vector and set the chosen bits as 1s. The caused space overhead, i.e., the value of $m$, is only proportional to the number of elements $n$, and will not be affected by the length of the elements. For example, given the bpe (bits per element) as 5, i.e., $m/n\mathrm{=}5$, the optimized $k$ can be derived out as 3 or 4 according to Equ. \ref{equ:opt_k}. Then the bit vector represents the $n$ elements, without concerning the size of each element.



\textbf{Constant-time query}. By employing BF, querying the membership of elements in $U$ can be simplified as binary checking of the corresponding $k$ bits. If all the $k$ bits are 1s, BF believes the queried element belongs to the set $S$, otherwise not. Thus the time-complexity of querying an element is $O(k)$, which is much faster than trees ($O(\log n)$) and table or list ($O(n)$). Note that, when the Bloom filter is implemented and $k$ will be a constant. Then both the insertion and query complexity will be $O(1)$.

\textbf{One-sided error}. Intrinsically, BF suffers from unavoidable false positive errors during a query, but no false negative errors. Specifically, if BF infers that an element $x$ is not in the set $S$, users can exactly trust the judgment. By contrast, if BF concludes that $x$ belongs to $S$, users cannot rule out the probability that $x\notin S$.

\subsection{Challenging issues of Bloom filter} \label{subsec:disadvantage}
BF is proven to be easy-deployable and practical for situations where space is limited and membership query is required with allowable errors. However, BF incurs intrinsic challenges. We uncover the challenges as follows.

\textbf{False positives}. Although the false positive rate can be controlled by careful setting of the parameters, the misjudged elements may lead to serious impact on the upper-level applications. For instance, when routing with BF, the leaked flows caused by misjudgements may burden or even block bandwidth-scarce networks (e.g., Wireless Sensor Network, Internet of Vehicles). To reduce the false positive rate, more space or more complicated operations are needed. Note that decreasing the false positive rate by extending the length of BF incurs marginal effect. That is, with the same increment of $m$, the decrement of false positive rate will be more and more indistinctive. Another strategy is to remedy the misreported elements (e.g., using whitelist). However, picking up the false positives from the query results can be challenging for a large-scale dataset.

\textbf{Implementation}. Despite the feature of easy-deployable, when BF or its variants are employed, some implementation ingredients (e.g., memory access, space utilization, computation overhead) may be the bottleneck of the applications. To query an element, the accessed bits can be arbitrarily located in the bit vector, which results in $O(k)$ memory access. For a standard BF design, roughly only half of the bits are set to 1. Besides, high-performance hash functions (e.g., RSA, SHA-1, MD-5) calls for complex computation process which may be not possible for low-end or light-weight hardware. The above measurements get even worse when the dataset contains a large number of elements.

\textbf{Elasticity}. Note that, the parameters (including $m$, $n$ and $k$) are predefined, and the employed hash functions cannot be changed once selected. The bits are not permitted to be changed once they have been set as 1s. Consequently, the BF can only successfully represent a static set. Once been implemented, the number of bits can be neither extended nor shrunk adaptively. Besides, the internal logical relationship (e.g., distance, similarity, precedence, etc) between elements will be annihilated. For example, when representing a multicast tree, the BF only programmes the nodes into the bit vector. Therefore, the generated BF cannot tell the children or father of any node in the multicast tree directly. 

\textbf{Functionality}. The original BF was designed to support fast membership query and only offer two operations to the users, i.e., insertion and query. More complicated operations (e.g., deletion and decoding) or other types of queries (e.g., multiplicity) are not enabled. When an element $x$ is removed from $S$, the bits $BF[h_1(x)], \cdots, BF[h_k(x)]$ cannot be reversed as 0s. The $k$ 1s may also represent the membership of other elements, thus changing them from 1s to 0s may lead to false negatives of other elements. Moreover, BF fails to tell the multiplicity of an element in a multiset (a set in which elements are permitted to have multiple replicas), and cannot report which set or sets an element belongs to, when representing multiple sets with a shared bit vector simultaneously. To realize more operations and enable more types of queries, the BF framework calls for additional design.

\section{Applications of Bloom filters}\label{sec:Applications}
Before detailing the optimizations and generalizations, we first summarize the applications of BFs in the area of communication and networking. Usually, BFs are employed to represent a given set of elements and support membership queries. We focus on the extensive applications of BFs in recent years since previous applications have been covered by existing surveys.

\subsection{Content caching}\label{subsec:caching} 
BFs are naturally helpful for caches and storages via representing the contents to cache and supporting constant-time membership queries.

Consider a generic system composed of a user, a main memory containing all the data, and a cache with a subset of the data. Usually, BFs are employed to represent the content stored in the cache. The access of a specific element $x$ is first directed to the BFs. If the BFs indicate that $x$ is an element in the cache, the access tries to read the element from the cache. Due to the potential false positive errors, $x$ may be not found in the cache, then the access will be routed to the main memory. If BFs judges that $x$ is not stored by the cache, the request will fetch $x$ from the main memory directly without accessing the cache. In this manner, BFs eliminate unnecessary cache reads \cite{Paradox_BF}. 

In reality, Akamai establishes BFs in its servers for efficient cache accesses \cite{ContentDlivery_1}. Extensively, Akamai records the accessed elements with BFs to select the elements which should be pushed into the cache. As a consequence, the cache hit-rate is guaranteed \cite{ContentDlivery_1}. Moreover, in a computing system with multiple cores, BFs are utilized to reduce the cache coherence cost \cite{cache_1}. Both local caches and the shared bus system have their BFs. With the BFs in the cache and the system interconnections, the filter mechanism screens out the unnecessary snooping messages that would be otherwise handled by each core. The BF for the bus system further reduces system-wide data broadcasts. As for web caching, BF is redesigned to represent cache content of each Internet proxy in a compact form \cite{WebCache_1}. Thereafter, the BFs are shared with other proxies in the web caching system. To further decrease the inter-proxy overhead, a BF only records the portion of a proxy's cache content that will be of interest to other proxies.

In the field of wireless communications, BFs are employed to speed up the cache lookups \cite{cache_2} and update the cache mechanism \cite{cache_3}. On-demand routing protocols for wireless ad hoc networks cache the discovered paths locally for subsequent routing operations. In this case, BF provides summaries of the cache content for testing cache membership and thereby averting negative lookups and easing the computational burden \cite{cache_2}. In large wireless networks, each node caches the public key of some nodes in the network. The cache space in each node is limited and is capable of storing a few public keys only. By representing these public keys as BF vectors, more keys can be stored and queried within constant time \cite{cache_3}.

\subsection{Packet routing and forwarding}\label{subsec:routing}
BFs are space-efficient so that they can be embedded in the header of a packet or implemented on chip to improve both wired and wireless networks. 

\subsubsection{BFs in wired networking}
In traditional wired networks, BFs are widely used to speed-up IP lookups, enable multicast, support named data forwarding, etc.  

High-speed IP address lookup is essential to achieve wire-speed packet forwarding in Internet routers \cite{Trie_search}. High-performance hardware such as ternary content addressable memory (TCAM) has been adopted to solve the IP lookups. However, this hardware requires vast investment and incurs non-trivial power consumption. Alternatively, on-chip BFs are utilized to record the IP lengths thereby enabling fast longest IP prefix matching with ordinary memories \cite{Trie_search}. Recently, the on-chip BF is further improved.  The primary idea is to explore the discrepancy in length distribution between the set of patterns and the set of prefixes of input text that are examined against the patterns \cite{Prefix_1}. Specifically, the stored prefixes are grouped according to their lengths and mapped into the BF vectors with variable numbers of hash functions. 

The in-packet BF naturally enables multicast routing by recording the nodes in the pre-calculated multicast tree \cite{inpacket_1}. For any node in the multicast tree, it checks the membership of its neighbours against the BF in the incoming packet. Then the node forwards the packet to the neighbours which pass the membership test. Most-recently, endeavours have been made to improve the BF-enabled multicast routing in terms of scalability \cite{Multicast_2} \cite{Multiclass_BF} \cite{FP_Free_BF}, loop mitigation \cite{Multicast_1}, and flow leakage \cite{Multiclass_BF} \cite{FP_Free_BF}. Due to the limited bits in the header and the large number of nodes in the multicast tree, the in-packet BF incurs scalability to some extent. A possible solution is to split the multicast tree into multiple parts and represent them separately. Nikolavskiy et al. propose to represent the multicast tree as multiple trees with their splitting algorithms \cite{Multicast_2}. By contrast, as stated in \cite{FP_Free_BF}, the multicast tree can also be divided hierarchically so that nodes in each stage are represented by a BF. In the destination-oriented multicast, the in-packet BFs records the destination IP addresses instead of the multicast tree. The multicast routing is accomplished with the collaboration of the in-packet BFs and the local routers. In this framework, the forwarding loop can be hopefully eliminated if specific conditions are satisfied. More details can be found in \cite{Multicast_1}.   

In the above multicast routing strategies, the false positive errors of BFs lead to flow leakage in the network. Tapolcai et al. reduce the false positive errors by dividing the multicast tree as multiple layers and adjusting the BF length according to the number of nodes in each layer \cite{FP_Free_BF}. For in-switch BF based multicast routing, Li et al. alter the number of hash functions for each multicast group according to the probability that the switch is a node in that multicast tree \cite{Multiclass_BF}. These techniques are detailed in Section \ref{sec:ReducingFPP}. 

Moreover, in named data networking (NDN), packets forwarding decisions are driven by content names instead of IP addresses \cite{NDN_1} \cite{NDN_2} \cite{NDN_3}. BFs are natively helpful for the longest prefix matching of content names in NDN. The NameFilter \cite{NDN_2} leverages a two-stage BF-based scheme for NDN name lookups.  The first stage determines the length of a name prefix, and the second stage searches the prefix in a narrowed group of BFs based on the results from the first stage. Instead of using the two-stage BFs,  Quan et al. propose to split the name prefix into B-prefix followed by T-suffix \cite{NDN_3}. B-prefix is matched by BFs whereas T-suffix is processed by the small-scale trie. The length of B-prefixes (and T-suffixes) is dynamically throttled based on their popularity in order to accelerate the lookup. Thus, they achieve a lower false positive rate than NameFilter.

\subsubsection{BFs in wireless networking}
In wireless networking, BFs are mainly utilized to represent the routing table and support fast lookups in each node. 

In mobile ad hoc networks (MANETs), the nodes are free to move independently in any direction, and will therefore change their links to other devices. The self-organizing and infrastructure-less features disable the traditional protocols in MANETs. HRAN protocol \cite{adhoc_1} \cite{adhoc_2} proposes to store and spread topology information with BFs among the nodes to downsize the routing messages in the network. HRAN merges BFs to discover and maintain routes, rather than broadcast topology information when the network state changes. In the scenario of vehicular ad hoc networks (VANETs), BFs are applied to maintain and disseminate 2-hop neighborship information among the nodes. The using of BFs reduces the length of the beacon messages, thereby keeping channel load and packet collision probability considerably low \cite{adhoc_3}. Such a design enables various applications on top of neighborship information, including broadcast, routing, clustering, etc. 

In wireless sensor networks (WSNs), each node keeps a precise list of events that may be found through each neighbor \cite{WSN_1}. Event flooding-based routing protocols are enabled by querying these events. Here, BFs are used to represent these events and support fast queries \cite{WSN_1}. Hebden and Pearce propose to partition the sensors into a set of clusters where each node in a cluster can only exchange data with its cluster head \cite{WSN_2}. Cluster heads process member data and/or reports from other cluster heads, make routing decisions and forward the result to another cluster head or the sink node. A hierarchy of BFs were used by the sink, cluster head ports, and each cluster head itself to filter out unpromising transmissions \cite{WSN_2}. Particularly, in tree-structured data collection WSNs, packets are routed towards a sink node \cite{WSN_3}. Each node in the collection tree stores the addresses of its direct and indirect child nodes in its local BF. With such setting, the node's local BF identifies routes from the sink node to any other nodes.   

Especially, in wireless NDN, BFs are applied to improve the gossip algorithm \cite{NDN_1}. Gossip protocols guarantee robust data dissemination at the cost of a more aggressive bandwidth usage. To reduce the communication overhead, Angius et al. propose to record the pre-calculated path with an in-packet BF \cite{NDN_1}. As a result, unnecessary transmissions are avoided. We note that BFs are extensively utilized to navigate vehicles and mobile robots in the environment of wireless sensors \cite{Navigation_1}. The wireless sensors provide necessary information for the navigation. Thereafter, the derived global routing table is represented with a BF in each sensor. By querying the routing table, a vehicle can be navigated to any destination.

\subsection{Privacy preservation}\label{subsec:privacy}
BFs naturally anonymize the elements with the support of membership queries. Therefore, they are employed to preserve data privacy in various scenarios. 

Nowadays, location-based services (LBSs) bring us unprecedented convenience, as well as potential challenge to location privacy. Calderoni et al. divide the geographical areas into different classes and represent them with BFs \cite{Spatial_BF}. In this way, they hide the actual locations of users with the support of a fast location query. In cognitive radio networks, geo-location database-driven methods are proposed to identify vacant frequency bands for the secondary users without harm the primary users \cite{location_pravicy_1} \cite{location_pravicy_2}. In these methods, the secondary users are equipped with GPS devices to query the geo-location database. Then the database returns the available channels to the secondary users. To preserve the location privacy of secondary users, Grissa et al. propose to represent the locations with probabilistic data structures \cite{location_pravicy_1} \cite{location_pravicy_2}.   
 
Additionally, biometric data (iris, face, handshape, fingerprint, etc) is widely adopted in authentication mechanisms. However, any privacy leakage of these data may lead to severe security crises. Consequently, researchers suggest representing these biometric data with BFs and enable the authentication mechanisms based on membership queries. Specifically, Barrero et al. code the facial biometric templates with BFs \cite{Biometric_2}. Rathgeb et al. represent iris biometric template with BFs \cite{Biometric_4}. Hermans et al. further employ non-linear and non-invertible hash functions to map the biometric data to ensure the unlinkability of BF based representations \cite{Biometric_1}.   
 
In the context of record linkage analysis, items are searched and matched to identify records that refer to the same entity across different data sources. When data to be matched is deemed to be sensitive or private, the privacy preservation problem arises \cite{Linkage_Privacy}. Karapiperis and Verykio anonymize the records in the database by programming them as BFs and thereby enabling approximate matching \cite{Linkage_Privacy}.

Mobile social networks (MSNs) allow users to discover and interact with existing and potential friends to exchange information on a subject of common interest \cite{MSN_pravicy}. The MSNs are helpful for chatting, file-sharing, and web page pre-fetching applications. However, security and privacy issues in these applications remain severe. Oriero et al. employ BFs to represent users' interested topics and thereafter share these BFs among trusted friends \cite{MSN_pravicy}. The using of BFs leads to less communication and storage overhead. Most importantly, the actual interested topics are anonymized for privacy.  

In particular, call detail records (CDRs) that are generated by users of mobile devices and collected by telecom operators could potentially be used for the socio-economic development and well-being of populations \cite{Detail_privacy}. Therefore, it is necessary to mine CDRs while preserving the privacy of the individuals contained in this data. To this end, Alaggan et al. sanitize CDRs data with BFs and thus preserve users' privacy \cite{Detail_privacy}. The idea is to anonymize the users detected by cellular antennas with BFs. These BFs will be exchanged for further analysis. 

\begin{table*}[tbp]
\centering

\begin{footnotesize}
\caption{The purposes and gains of using BFs in diverse scenarios.} \label{table:gains} \vspace{-0.1in}
\begin{tabular}{c|c|c|c|c|c}\hline
Scenarios & Purposes & Gains  & Scenarios & Purposes  & Gains  \\ \hline
Caching  \cite{Paradox_BF} & Content summarization &CQ, AR & Akamai \cite{ContentDlivery_1} & Content summarization& CQ, AR  \\  \hline
Wireless ad hoc \cite{cache_2} & Content summarization &  CQ, AR&  Wireless net.  \cite{cache_3}& Content summarization &  CQ, AR\\  \hline
IP lookup \cite{Trie_search}\cite{Prefix_1}  & Representing routing table  & CQ, SS &  Multicast \cite{inpacket_1} &Representing multicast tree &  CQ, SS   \\  \hline
MANETs \cite{adhoc_1} \cite{adhoc_2} &Representing topology info. &CQ, SS & NDN \cite{NDN_2} \cite{NDN_3} & Representing content names & CQ, SS    \\  \hline
Wireless NDN \cite{NDN_1}   & Data dissemination &CQ, SS & WSNs \cite{WSN_1} & Routing packets & CQ, SS   \\  \hline
Cog. rad. net. \cite{location_pravicy_1} \cite{location_pravicy_2}   & Location anonymization & CQ, CA & LBSs \cite{Spatial_BF} & Location anonymization & CQ, CA    \\  \hline
Biometrics \cite{Biometric_2} \cite{Biometric_1}  & Biometrics info. anonymization &CQ, CA & MSNs \cite{MSN_pravicy}  &  Interested topics anonymization & CQ, CA    \\  \hline
Record linkage \cite{Linkage_Privacy} & Entry anonymization & CQ, CA & CDRs \cite{Detail_privacy} & Record anonymization & CQ, CA    \\  \hline
Smart grid \cite{security_1} & Representing CRL list & CQ, SS & & &   \\  \hline
\end{tabular} \vspace{-0.1in}
\end{footnotesize}
\end{table*}

Note that, several recent surveys and tutorials, which target at the privacy issues in diverse scenarios, also mention the usage of BFs explicitly \cite{COMST_info_centric} \cite{COMST_pitfalls} \cite{COMST_NSF} \cite{COMST_bitcoin_1} \cite{COMST_bitcoin_2}. In information-centric networking, BFs enable the DoS-resistant self-routing mechanisms, name obfuscation, and  access control \cite{COMST_info_centric}. The recent tutorial \cite{COMST_pitfalls} systematically reveals the shortcomings of anonymization with hash functions, yet fairly highlights that BF is a popular choice in designing privacy-friendly solutions. In the survey which aims at the security and privacy of U.S. National Science Foundation's future internet architectures,  M. Ambrosin et. al emphasize the BF-based method to defeat DoS attacks \cite{COMST_NSF}. Especially, in cryptocurrency such as Bitcoin, BFs are used to secure the wallet by letting the simplified payment verification clients only request matching transactions and merkle blocks from full nodes \cite{COMST_bitcoin_1} \cite{COMST_bitcoin_2}.

\subsection{Network security}\label{subsec:security}
The survey \cite{ExistingSurvey_3} published in 2013 has systematically summarised the application of BFs and their variants to address security problems in different types of networks. Specifically, in wireless networks, BFs are employed for authentication, anonymity, firewalling, tracebacking, misbehavior detection, replay attack detection, and node replication detection. In wired networks, various uses of BFs are found in the design of different security mechanisms, including string matching, IP tracebacking, spam filtering and e-mail protection, DoS and DDoS attacks detection, and anomaly detection. Herein, we briefly introduce a recent BF usage in the smart grid. 

In smart grid advanced metering infrastructure (AMI) networks, to revoke a certificate, the certificate revocation lists (CRLs) are maintained for each smart meters \cite{security_1}. However, both maintenance and access of these CRLs are challenging due to the large geographic deployment and scalability of the AMI networks. Therefore, Rabieh et al. propose BF based revocation schemes for AMI networks that can enable the meters to identify and nullify the false positives \cite{security_1}. Basically, one BF records the serial numbers of the revoked certificates in the AMI network and an additional BF represents the valid certificates. To fix the potential false positives of BFs, Merkle tree is established to verify the query results from BFs.

We find more extensive and novel applications in data dissemination in WSN \cite{Dissemination}, application discovery in classification \cite{Classification}, device-to-device communications \cite{Discovery}, etc. From the length concern, we omit the details here. As Mitzenmacher explains that: whenever you have a set or list, and space is an issue, a BF may be a useful alternative \cite{ExistingSurvey_2}. Next, we detail the existing optimization techniques of BFs with the following four sections by analyzing each related variant. The readers can also continue to Section VIII directly for comprehensive analysis and comparison for all the covered BF variants.

\subsection{Gains of using BFs}
 From a general perspective, we highlight the potential gains of using BFs in these applications as follows. 

\textbf{Constant-time query (CQ)}. As stated in Section \ref{subsec:characteristics}, the time-complexity of membership query of BF is constant, which is much faster than trees ($O(\log n)$) and table or list ($O(n)$). This feature leads to higher query throughput.

\textbf{Space saving (SS).} BF represents the information of each element with the bits in its vector, instead of storing the content of the element like other data structures. The required space cost is determined by the target false positive rate and the number of elements to represent while being independent to the size of each element. BFs introduce space efficiency and space constant features into their applications. Therefore, BFs are both storage-friendly and communication-friendly, so that on-chip and in-packet implementations are often possible. 

\textbf{Access refinement (AR).} BFs summarize elements as a bit (or cell) vector and filter unnecessary accesses of the elements. Intuitively, the access requests are directed to the BFs first. If BFs indicate the requested elements are stored by the memory, the requests will further check the memory and fetch the requested elements. Otherwise, the requested elements are beyond this set and the access will be denied accordingly. This refinement is extremely helpful in the scenarios of caching and storage systems.  

\textbf{Content anonymization (CA).} The actual content of each element is programmed as 1s in the vectors by BFs. Therefore, BFs naturally anonymize the elements with the support of membership queries. The applications are able to hide the real content while releasing the BFs for membership queries. This ability is meaningful to privacy preservation and data security.

Especially, Table. \ref{table:gains}  summarizes the major purposes and gains of using BFs in the aforementioned scenarios. All these systems represent elements with BF and benefit from constant-time membership query for sure. Additionally, for diverse design purpose, the gains of using BFs may be different. For content caching, BF works as content summarization and refines unnecessary access of the cache memory. For packet routing and forwarding, BFs enable the routing strategies and provide space efficiency so that they can be embedded into packet headers. For privacy preservation, BFs anonymize data and support fast queries simultaneously. Lastly, in our example of using BFs for network security, BFs save space for the representation of CRLs.

\section{Reduction of false positives} \label{sec:ReducingFPP}
\begin{figure}[t]
\centering
\subfigure[False positive rate.]
{\includegraphics[width=1.71in,scale=0.5]{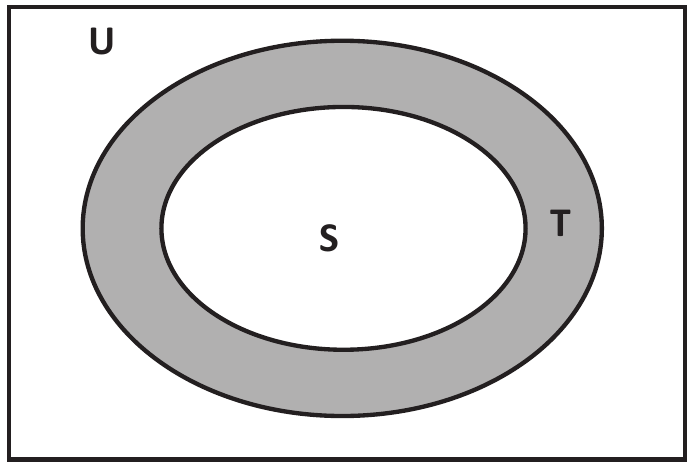}} \vspace{-0.07in}
\subfigure[False positive proportion.]
{\includegraphics[width=1.71in,scale=0.5]{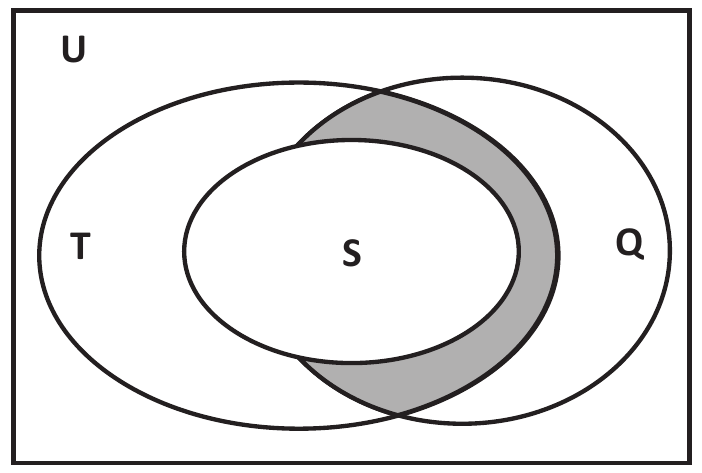}} \vspace{-0.07in}
\caption{The false positive rate and false positive proportion in Bloom filters, shown by the filled areas.}
\label{fig:FPRFPP} \vspace{-0.1in}
\end{figure}

Before detailing the related variants, we formulate the false positive proportion (FPP) and false positive rate (FPR) as follows. FPP is an empirical concept that calculates the ratio between the occurrence of false positive errors and the total query times. In contrast, FPR is a theoretical term used to quantify the probability of any query result incurs a false positive error. Usually, FPP can be employed as the point estimation of FPR. FPR is a fixed value and determined by the BF framework itself, while FPP is variable and fluctuates around the value of FPR. As shown in Fig. \ref{fig:FPRFPP}, let $U$ be the universal set, $S$ denote the set represented by a BF, $Q$ denote the elements to be queried and $T$ record the elements which return positive for the membership query, respectively. Then the elements which lead to false positive errors can be derived as $F\mathrm{=}T\mathrm{-}S$. With the above notations, the false positive rate for all the query of elements in $U$ can be calculated as $f_r\mathrm{=}|F|/|U|\mathrm{=}|T\mathrm{-}S|/|U|$. Pair-wisely, the false positive proportion for the query of elements in $Q$ is $f_p\mathrm{=}|(T\mathrm{\cap} Q) \mathrm{-}(S\mathrm{\cap} Q)|/|Q|$.

Therefore, to reduce the number of false positives in an instance of BF, we have methods to lower FPR and techniques to control FPP. Reducing FPR means to decrease $T$ and thus $F$ directly.  A brute-force solution to reduce FPR is to increase the number of bits or cells in the vector, with the penalty of more space overhead. But this method doesn't work in space-scarce situations. Reducing FPR is more general to be extended to diverse scenarios, while decreasing FPP is more specific and case-dependent. For example, for specific instance of BF with given $Q$, it is possible to select hash functions which lead to lower FPP. Both the two design philosophies are reasonable. 

The intrinsic reason of false positive is that BF arbitrates the membership of an element based on the presence or non-presence of the corresponding 1s, without concerning which element (or elements) fills these hash bits. The bits in the vector, however, may suffer from hash collisions, which may finally cause false positive query results. One may reduce false positives from the following aspects. First, prior knowledge can be helpful when setting the parameters of BF to reduce FPs. Second, the one-sided error property can also identify FPs. Given a main BF and multiple attached BFs, the negative query results in the attached BFs can prone some false positive results of the main BF. Third, if the non-zero bits are reset as 0s, the potential false positive errors will not happen. But this will introduce the possibility of false negative errors instead. Fourth, the hash functions are optional and one may select the hash functions which lead to the least false positives. At last, if the elements are inserted differentially, the hash collisions will not impose false positives to the query result. The details are shown in the following five subsections, respectively.

\subsection{Reducing FP with prior knowledge}
Note that, some prior knowledge is helpful to identify some false positives of BF queries and accordingly decrease the resulted FPs. For example, if the elements to be queried ($Q$) are known as prior knowledge, it is possible for the BF users to select the hash functions which yield the least FPs. A typical scenario where $Q$ is known is multicast routing based on BF.  Actually, BF enables both in-switch routing and in-packet routing schemes for multicast. For in-switch routing, each interface in the switch uses a BF to maintain the multicast groups it joins. By contrast, in-packet routing embeds a BF in the header of each packet to encode the multicast tree information. The false positives will result in traffic leakage via leading flows to additional nodes which are not in the multicast tree. Multi-class BF \cite{Multiclass_BF} and False-positive-free multistage BF \cite{FP_Free_BF} are proposed to reduce the FPs of in-switch and in-packet BFs, respectively. Additionally, in the case of web caching, Rottenstreich et al. uncover the Bloom paradox and suggest to insert and query elements selectively based on their priori membership probability. Moreover, the Optihash \cite{optihash_BF} scheme selects the hash functions which generate the lowest FPP in the context of PSIRP (Publish/Subscribe Internet Routing Parading) systems.

\textbf{Multi-class BF.} For in-switch multicast routing,  Li et al. point out that, the presence probability of a multicast group on a certain switch interface can be evaluated via joint consideration of the multicast group size and the data center topology \cite{Multiclass_BF}. With this insight, they propose Multi-class BF, a variant of BF to minimize the traffic leakage \cite{Multiclass_BF}. 
They observe that an element with a higher presence probability indicates a lower impact of its false probability on the expected number of falsely matched elements. Therefore, Multi-class BF proposes to employ different number of hash functions for each element to minimize the expected number of falsely matched elements. Specifically, in a switch interface, the groups with higher presence probability are encoded with fewer hash functions, while the groups with lower presence probability should be programmed with more hash functions. In this way, the multicast groups with lower presence probability must check more bits when querying their memberships. In other words, Multi-class BF removes some elements which may lead to false positives with high probability from set $T$, by setting stricter conditions to pass the queries. 

\textbf{False-positive-free multistage BF (FPF-MBF).} As for in-packet multicast routing, Tapolcai et al. devote to solving the scalable forwarding problem, especially the scalability of the multicast trees \cite{FP_Free_BF}. To this end, the multistage BFs \cite{Multistage_BF1} \cite{Multistage_BF2} are employed to record the links in each stage of the multicast tree. A multicast tree of $v$ hops is represented by $v$ BFs, and the $v^{th}$ BF contains only the links residing at $v$ hop-distance from the source node. When leaving the source, the multicast in-packet BF header consists of $v$ stage filters, which then shrinks as the packet travels along the tree. To this end, each stage BF consists of two parts, i.e., the first $\gamma$ bits to indicate its stage, and the later $b$ bits to record the memberships of links in this stage. 

Literature \cite{FP_Free_BF} proposes to build FPF-MBF, based on the prior knowledge of the number of elements to be included ($\rho$) and the number of elements to be excluded from the filter ($\zeta$). 
In effect, FPF-MBF reduces the generated false positive proportion from two aspects. On one hand, unlike the traditional in-packet BFs which record all links in a multicast tree, each stage filter only encodes the links in the corresponding stage. On the other hand, given the value of $\rho$ and $\zeta$, the length of each stage filter is adjusted to further decrease the false positive proportion. Essentially, the design philosophy of FPF-MBF is to reach the goal that $(T\mathrm{-}S)\mathrm{\cap} (Q\mathrm{-}S)\mathrm{=}\emptyset$, where $Q$ is the known set of elements to be queried. As a result, all queried elements will not be matched false-positively, and it performs like with a false positive proportion of 0.

\textbf{Bloom paradox.} We consider a generic system composed of a user, a main memory containing all the data, and a cache with a subset of the data. A BF is employed to record the data in the cache. When the user reads a piece of data, he checks the BF first. If the query result is negative, the user will access the main memory directly; otherwise, it goes to the cache instead. However, the positive results can be true or false. For the false positives, the user still needs to access the main memory. Let $M$, $C$ and $T$ be the elements in the main memory, the elements in the cache, and the elements in main memory that lead to positive result when querying the BF, respectively. We have $|T|\mathrm{=}|C|\mathrm{+}f_p\mathrm{*}(|M|\mathrm{-}|C|)$, where $f_p$ is the false positive proportion of the BF. The probability that a positive query result is false positive can be evaluated as:
\begin{equation}
 \frac{|T|\mathrm{-}|C|} {|T|}\mathrm{=} \frac{f_p\mathrm{*}(|M|\mathrm{-}|C|)} {|C|\mathrm{+}f_p\mathrm{*}(|M|\mathrm{-}|C|)}.
\end{equation}
Given $|M|\mathrm{=}10^{10}$, $|C|\mathrm{=}10^{4}$ and $f_p\mathrm{=}10^{\mathrm{-}3}$, the probability that a positive query result is false positive is almost $1\mathrm{-}10^{\mathrm{-}3}$. That is, among the positive results of the BF for the various elements, most of them are false positives. This phenomena is called Bloom paradox \cite{Paradox_BF}.

With the prior knowledge, i.e., the value of $|M|$, $|C|$ and the parameter setting of BF, one may calculate the condition of Bloom paradox. Rottenstreich et al. suggest selective BF insertion, as well as selective BF query \cite{Paradox_BF}. When inserting (querying) an element, if the element satisfies the Bloom paradox condition, it will not be inserted (queried). Namely, in the scenario of Bloom paradox, the users should not trust the BF such that they will survive the impact of false positives.

\begin{figure}
  \centering
  \includegraphics[width=3.2in]{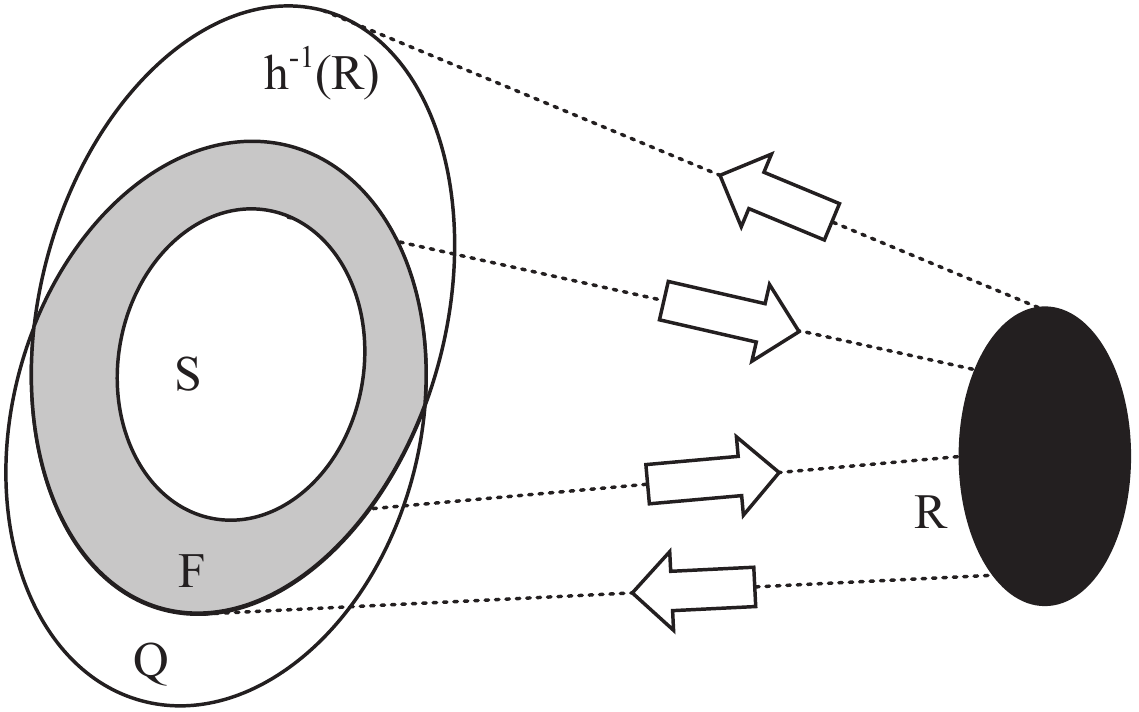}\\ \vspace{-0.05in}
  \caption{The theory of Optihash \cite{optihash_BF}. Its basic insight is to choose a set of transformed hash values which leads to the smallest volume of $F$.}\label{fig:Optihash} \vspace{-0.15in}
\end{figure}

\textbf{Optihash.} Optihash \cite{optihash_BF} is designed to extend the BF in the context of PSIRP (Publish/Subscribe Internet Routing Parading) which is a new redesign of the whole Internet architecture as far as the physical layer \cite{PSIRP}.  The in-packet BFs are employed as encoding to identify routes and links between nodes. The optihash is a bit array $\textbf{w}$ which consists of three parts: a bit array $\textbf{v}$ and two integer parameters $\alpha$ and $\beta$. Only one hash function $h$ is initialized to map each element into the bit array $\textbf{v}$. As depicted in Fig. \ref{fig:Optihash}, $S$ is a set of elements ($S$ may be a multiset) encoded in the optihash, the hash function $h$ maps the elements in $S$ to a generally smaller set $R$. The inverse $h^{\mathrm{-}1}$ maps the set of hashes $R$ to a set of elements which is bigger than $S$. Let $Q$ denote the set of elements to be queried, then the set of false-positive elements $F$ can be derived as $F\mathrm{=}(h^{\mathrm{-}1}(R)\cap Q)\mathrm{-}S$.

Optihash proposes to generate a family of hash values according to the value of $\alpha$, $\beta$ and the employed hash function $h$. Generally, there will be $2^{m_\alpha}2^{m_\beta}$ new sets of hash values, where $m_\alpha$ and $m_\beta$ are the number of bits to store $\alpha$ and $\beta$, respectively. Consider that all elements in the set $Q$ are known, the user can employ the set of hash values which lead to the smallest volume of $F$. In this particular manner, optihash decreases the number of elements which cause false positives, at the cost of complicated computation.

\subsection{Reducing FP with the one-sided error}
As stated in the subsection \ref{subsec:characteristics}, BF suffers from unavoidable false positive errors during query, but no false negative errors. Namely, the conclusion drawn by BF that an element $x$ is not a member of $S$ is 100\% correct. This one-sided error characteristic, indeed, is helpful to identify some false positives from the query results \cite{Trie_search} \cite{Trie_BF} \cite{Cross_checking_BF} \cite{Complement_BF}.

For example, it has been recently studied that the search performance of trie-based algorithms can be significantly improved by adding a BF \cite{Trie_search} to record the nodes in the trie. In such algorithms, the number of trie accesses can be greatly reduced because BF can determine whether a node exists in a trie without actually accessing the trie. However, the false positives of BF bring unnecessary trie accesses. Fortunately, Mun et al. notice that arbitrary node (except the source node) in a trie can only exist if its ancestors are also in the trie \cite{Trie_BF}. Therefore, they propose to use more BF queries to reduce the false positive proportion of a BF in trie-based algorithms.

When the BF returns positive upon querying an element $y$, to check it is a false positive or not, the ancestor node of $y$, i.e., node $z$ will also be queried. If the BF indicates that $z$ is not a node of the trie, obviously $y$ is a false positive node. By contrast, if the query result of $z$ is also positive, $y$ is a trie node with higher probability. But $z$ may also be a false positive node. To further test the trueness of the positive result for $z$, one may query the ancestor of $z$ additionally. For any set whose elements share a strong internal dependency, the same strategy can be introduced to recognize unnecessary false positives. As for more general sets, the following three variants of BF are proposed base on the one-sided error characteristic.

\begin{figure}
  \centering
  \includegraphics[width=3.2in]{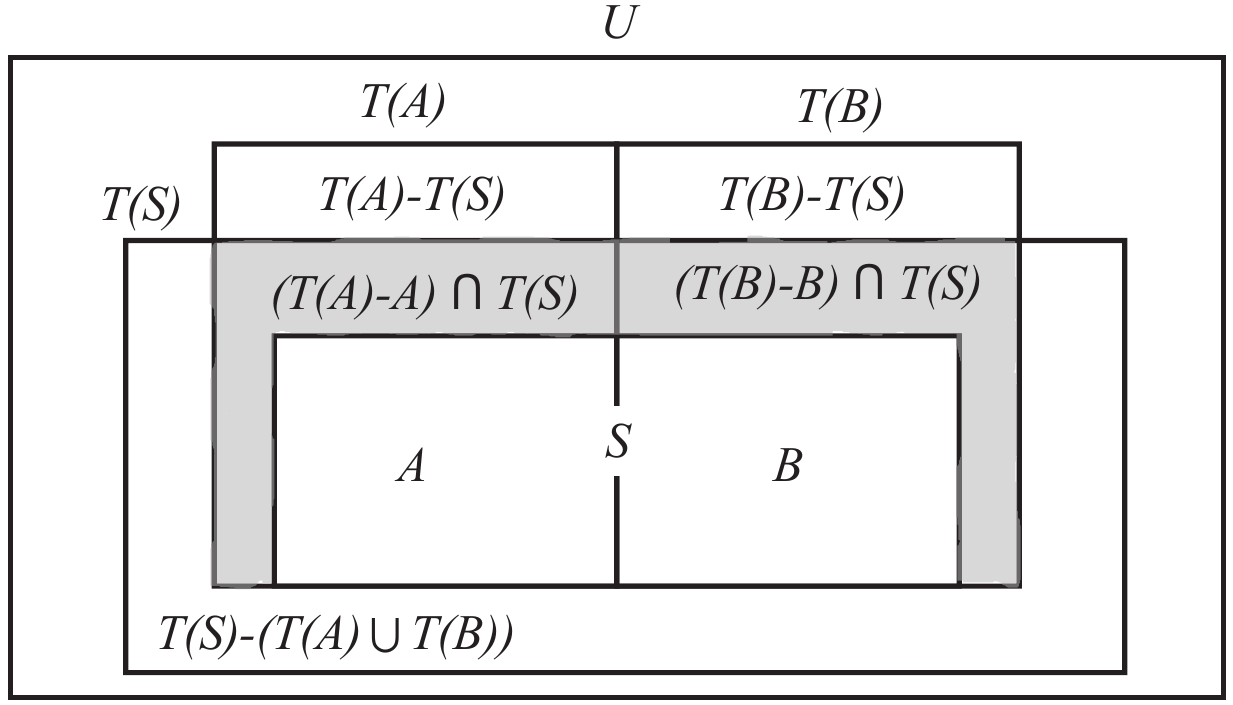}\\ \vspace{-0.05in}
  \caption{Set diagram with $T(A)$, $T(B)$ and $T(S)$ for cross-checking BFs and the main BF, with the assumption that  $T(A)\cap T(B)=\emptyset$ \cite{Cross_checking_BF}. }\label{fig:Cross-checking} \vspace{-0.15in}
\end{figure}

\textbf{Cross-checking BF.}  Lim et al. propose a new architecture to reduce the false positive proportion of BF \cite{Cross_checking_BF}. The architecture consists of a main BF which is programmed for all the elements in the set $S$, and multiple cross-checking BFs which are responsible to record the elements in several disjoint subsets of $S$. As depicted in Fig. \ref{fig:Cross-checking}, consider the set $S$ as the union of two disjoint subsets $A$ and $B$, i.e., $S\mathrm{=}A\mathrm{\cup} B$ and $A\mathrm{\cap} B\mathrm{=} \emptyset$. In this case, the BF for the entire set $S$ is the main BF, while the BFs for the set $A$ and $B$ are cross-checking BFs to check the false positives of the main BF. A positive result of the main BF for element $x$ will be recognized as false positive, if both of the cross-checking BFs return negative query results for $x$. The underlying basis is that the negative query results of BF are 100\% correct, due to the one-sided error characteristic. 

Theoretically, as shown in Fig. \ref{fig:Cross-checking}, the main BF and the two cross-checking BFs all suffer from false positives. Let $T(S)$, $T(A)$, and $T(B)$ denote the set of elements which return a positive result when querying upon the main BF and the two cross-checking BFs, respectively. By querying the cross-checking BFs, the elements in $T(S)\mathrm{-}(T(A)\mathrm{\cup}T(B))$ will be identified as false positives. Therefore, with the help of cross-checking BFs, the global false positive rate will be decreased from $\frac {|T(S)\mathrm{-}S|} {|U|}$ to:
\begin{equation}
  \frac {|((T(A)\mathrm{-}A)\cap T(S))\cup ((T(B)\mathrm{-}B)\cap T(S))|} {|U|}.
\end{equation}
The specific value of the false positive rate is further given in \cite{Cross_checking_BF}. Undoubtedly, the cross-checking BFs requires more queries and more space overhead.

\textbf{Complement BF.} When BF is associated with an off-chip hash table,  the hash table can verify the positive results of the BF for $S$. To reduce the access frequency of the hash table, complement BF  \cite{Complement_BF} is introduced. Unlike the cross-checking BFs which split the set $S$ as multiple disjoint subsets, Lim et al. propose to divide the union set $U$ as two independent subsets, i.e., $S$ and its complement set $S^C$  \cite{Complement_BF}. Typically, a main BF is programmed according to the elements in $S$, and another BF (complement BF) is initialized to record the elements in the set $S^C$. 
The complement BF helps to identify some false positive errors since the elements not in $S^C$ must belong to $S$.

With this insight, Table \ref{tab:complementBF} presents the truth table of a query with joint consideration of the main BF and complement BF. Note that, the two BFs will never return negative query result simultaneously, since $U\mathrm{=}S\mathrm{\cup} S^C$ and any element not in $S$ must belong to $S^C$. Only if both the main BF and the complement BF return positive when querying an element $x$, the membership of $x$ will be further checked by the off-chip hash table. Consequently, the frequency of hash table access can be significantly decreased from $T(S)$ to $T(S)\mathrm{\cap} T(S^C)$. It has been proven that the probability of both BFs producing positives converges the summation of the false positive rate of each BF \cite{Complement_BF}. However, in the case of large-scale complement set $S^C$, the complement BF may cost vast on-chip memory.

\textbf{Yes-no BF.} Unlike the previous proposals which remove the elements from the set $T$ directly, yes-no BF keeps track not only the elements belonging to the set $S$, but also the elements which generate false positives \cite{Yes_No_BF}. The yes-no BF is composed of two parts, i.e., the yes-filter which encodes the set elements, and the no-filter which stores the elements which generate false-positives. The $m$ bits in yes-no BF are consequently split into two parts, $p$ bits for the yes-filter, and $r\mathrm{\times} q$ bits for the $r$ no-filters each of $q$ bits. The yes-filter performs just like a normal BF and record the membership of elements in $S$. In contrast, the no-filter for the set $S$ tracks the elements which cause false-positives in the yes-filter. 
Note that, an element can only be stored in one of the $r$ no-filters. When querying, the yes-no BF will conclude that an element $x$ belongs to $S$, if and only if the corresponding bits in the yes-filter are all 1s and none of the $r$ no-filters indicates $x$ is a false positive. 
In Yes-no BF, the positive results given by the yes-filter are further tested by the no-filters. Consequently, the number of elements in $T$ can be decreased, but at the risk of false negatives. To reduce the potential false negatives, additional queries must be conducted.

\doublerulesep 0.1pt
\begin{table}[tbp]
\centering
\begin{footnotesize}
\caption{Truth table of query with both main BF and complement BF.} \label{tab:complementBF}\vspace{-0.05in}
\begin{tabular}{lcccc}
\hline
Main BF for $S$ & 1 & 0 &1 &0 \\ \hline
Complement BF & 0 & 1 & 1& 0 \\ \hline
Conclusion & $x\in S$ & $x\notin S$ &Hash table &Not exist \\ \hline
 \end{tabular}
\end{footnotesize}\vspace{-0.17in}
\end{table}

\subsection{Reducing FP via bit resetting}
Intrinsically, the false positives appear because BF checks the $k$ 1s without distinguishing the elements which are hashed into these positions. As a result, an element can be false-positively matched because its corresponding bits are set to 1s by other elements. Thus, several proposals try to reduce the false positive errors by modulating the bit vector directly. Note that the retouched BF \cite{Retouched_BF} and generalized BF \cite{Generalized_BF} are designed for standard BF, while multi-partitioned counting BF \cite{Multi_partitioning_CBF} are proposed to reduce the false positive rate of CBF \cite{Counting_BF}.

\textbf{Retouched BF.} Retouched BF is proposed as an extension to make the BF more flexible by permitting the removal of selected false positives at the expense of generating random false negatives \cite{Retouched_BF}. To do so, the selected bits are cleared by resetting from 1 to 0. The authors present two strategies for the clearing. The randomized bit clearing resets a certain number of bits from the bit vector randomly. Theoretical analysis demonstrates that, after executing the randomised bit clearing process, the increment of false negatives and the decrement of false positives are very close \cite{Retouched_BF}.

By contrast, another strategy called selective clearing is designed to reset the bits which trigger false positives. This can be realized since the system can learn a portion of false positives through previous queries. Four algorithms are proposed to trade-off the false positives and false negatives. When removing an element, there are $k$ candidate bits to reset. Thereafter, Retouched BF have: 1) the random selection algorithm randomly selects a bit amongst the $k$ candidates; 2) the minimum FN selection resets the bit which generates the least increment of false negatives; 3) the maximum FP selection chooses the bit which enables the maximum false positive decrement; and 4) the ratio selection prefers resetting the bit which minimizes the generated false negatives while maximizing the false positives removed. With the above strategies, Retouched BF reduces FPR at the cost of false negatives.

\textbf{Generalized BF.} Generalized BF \cite{Generalized_BF} employs two groups of hash functions, i.e., $\{g_1, \cdots g_{k_0}\}$ and $\{h_1, \cdots h_{k_1}\}$. When inserting an element into the bit vector, the $k_0$ bits mapped by the hash functions $\{g_1, \cdots g_{k_0}\}$ are reset as 0, while the $k_1$ bits derived by the hash functions $\{h_1, \cdots h_{k_1}\}$ are set to 1 (break the tie by keeping reset). When querying an element $x$, generalized BF checks whether the bits corresponding to the positions $\{g_1(x), \cdots g_{k_0}(x)\}$ are all 0 and the bits $\{h_1(x), \cdots h_{k_1}(x)\}$ are all 1. If at least one bit is inverted, generalized BF returns a negative query result; otherwise, it returns a positive result.

False positive will happen if the $\{g_1(x), \cdots g_{k_0}(x)\}$ are all 0, and if the bits $\{h_1(x), \cdots h_{k_1}(x)\}$ are all 1, due to the insertion of other elements in the set $S$. Moreover, it is possible that an element $x\mathrm{\in} S$ may not be reported as a member of $S$, resulting in a false negative. False negative will occur if at least one of the $\{g_1(x), \cdots g_{k_0}(x)\}$ bits is set as 1 by the $k_1$ hash functions for another element $y$, or one of the  $\{h_1(x), \cdots h_{k_1}(x)\}$ bits are reset as 0 by the $k_0$ hash functions. Certainly, generalized BF offers rigid constraints to pass the membership query, thus reducing the false positive rate. But it calls for reconsideration when put generalized BF into real use, due to the non-negligible false negatives.

\textbf{Multi-partitioned Counting BF (MPCBF).} CBF \cite{Counting_BF} extends BF by allowing insertions and deletions to support dynamic datasets. To reduce the number of memory accesses, MPCBF divides the $m$ cells into $l$ words so that each word can be fetched in a single memory access \cite{Multi_partitioning_CBF}. To insert an element $x$, an extra hash function $h_e$ is employed to map $x$ into one of the $l$ words. Thereafter, the $k$ hash functions $\{h_1(x), \cdots, h_k(x)\}$ map $x$ into $k$ cells in the selected word. This strategy, however, leads to more false positive errors. To reduce the false positives, MPCBF suggests to reconstruct each word as a hierarchical structure.

\begin{figure}
  \centering
  \includegraphics[width=3.2in]{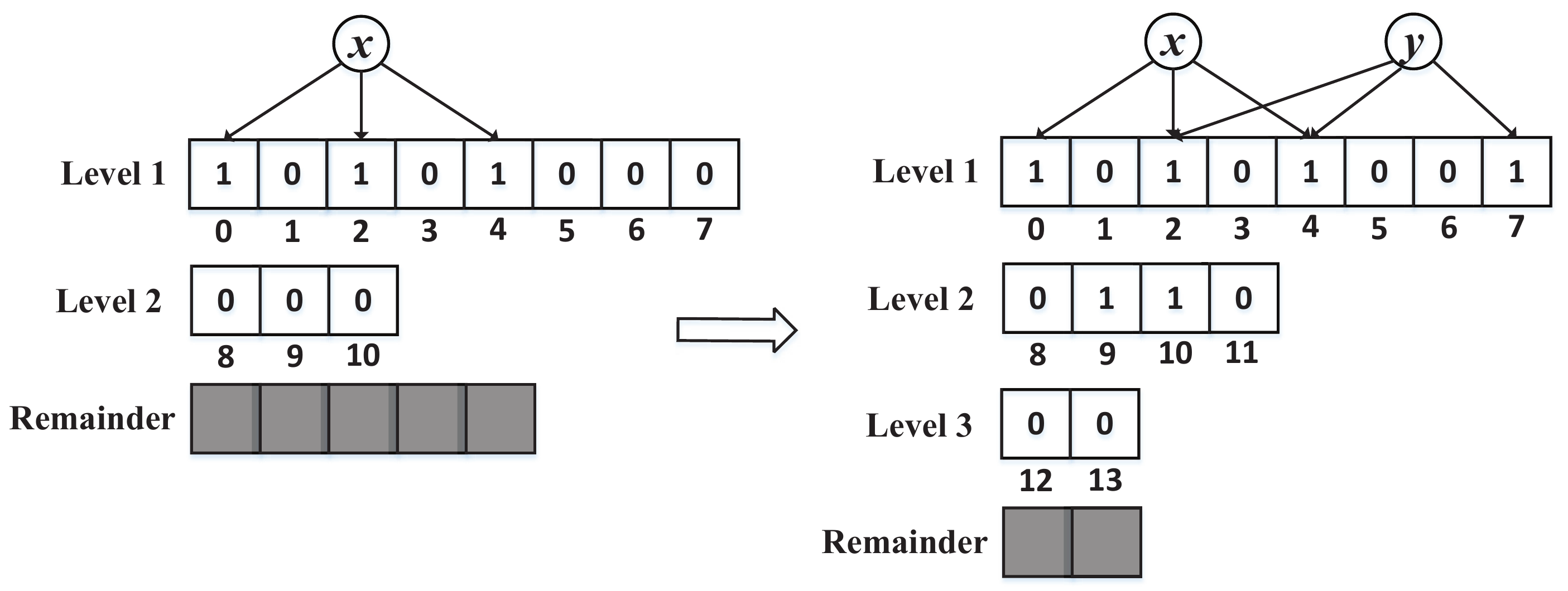}\\ \vspace{-0.05in}
  \caption{Hierarchical structure in a 16-bit word of MPCBF  \cite{Multi_partitioning_CBF}. Initially, the number of hash functions $k\mathrm{=}3$, and level 1 has 8 bits which have been initialized as 0. }\label{fig:MPCBF} \vspace{-0.15in}
\end{figure}

As depicted in Fig. \ref{fig:MPCBF}, MPCBF allocates the bits in each word as multiple levels. The basic principle for constructing the hierarchical structure is as follows. Whenever an element is inserted into the word, $k$ bits must be set from 0 to 1. And whenever a bit is set from 0 to 1, an empty bit should be added into the next level and initialized as 0. This is realized by using a function $popcount(i)$ which computes the number of ones before position $i$ at the hierarchy level that bit $i$ belongs to. 
Note that, only level 1 is utilized for membership query. Therefore, the membership in this word can be recorded with 8 bits, instead of 4 cells (suppose that each cell in CBF has 4 bits). As a result, the false positive rate can be significantly decreased at the cost of a little more computation overhead.

\subsection{Reducing FP with selected hash functions}
False positive errors are not avoidable for BF and its variants, due to the potential hash collisions. Typically, the false positive rate of BF is proportional to the number of 1s in the bit vector. Based on this observation, endeavours have been made to select proper hash functions for the elements. A typical work is to introduce ``the power of two choices'' into the design of BF \cite{Power2choice}. The basic idea is to employ two (or more) groups of hash functions, and the insertion will utilize the group of hash functions which increase the least 1s in the bit vector. During membership query, if the element passes the checking of any group of hash functions, the query result will be positive; otherwise, negative. The cost is more hash computation and lookups. 

Moreover, a partitioned hashing scheme \cite{Hash_partition} is then proposed to further optimize the hash function selection, based on the balls and bins theory. Before insertion, the elements in set $S$ is divided into $g$ independent groups by hashing their keys. Given $H$ hash functions, each group of the elements will be mapped into the bit vector $t$ times for test, where $k\mathrm{\leq} t\mathrm{\leq} \binom k H$. In each test, every element will be hashed $k$ times. Thereafter, the set of hash functions which increase the least 1s in the bit vector will be selected. Then, the group of elements will be mapped into the vector with the selected $k$ hash functions. Note that, the hash functions intra a group are independent, but can be dependent inter groups. That is, a hash function is allowed to be shared among multiple groups of hash functions. A greedy algorithm is designed to speed up the selection process. The conducted experiments indicate that the partitioned hashing scheme results in as much as a ten-fold increase in accuracy over standard BF.

Selecting the hash functions is a computation-intensive task, especially when the number of candidate hash functions is large, there are in total $\binom k H$ possible combinations. Besides, for dynamic datasets, this kind of methods call for re-selection of the hash functions when the dataset changes. Therefore, selecting $k$ hash functions from many candidates suits for the situations where the query accuracy must be guaranteed, while the computation is not an issue and the dataset is static.

Most recently, Kiss et al. propose EGH filter to replace the $k$ hash functions $\{h_1,\cdots, h_k\}$ with the $k$ simple functions $\{\hat{h}_1,\cdots, \hat{h}_k\}$ generated based on $k$ prime numbers. Intuitively, EGH filter supports the Bloom filter operations and additionally guarantees false positive free operations for a finite universe when a restricted number of elements stored in the filter \cite{EGH_filter}. In other words, given a finite universe set $U$ with $|U|$ elements, an EGH filter vector with $m$ bits will not suffer from any false positive errors if at most $n_t$ elements are stored. The essence of the solution is to use the Chinese Remainder Theorem \cite{ChineseRemainder} and solve a combination group testing (CGT) problem \cite{CGT} by finding a solution to a system of linear congruences. Note that there are strong constraints between the parameters $m$, $|U|$, $n_t$ and $k$. Basically, $n_t ^{|U|}$ should be less than the product result of the first $k$ prime numbers. Then the value of $m$ will be set as the summation result of the first $k$ prime numbers. With the above parameter setting, EGH filter performs the insertion and query just like the standard BF.  Furthermore, the EGH filter can be extended to support deletion and listing of the recorded elements. 

Definitely, the used functions in EGH filter are deterministic, fast and simple to calculate, enabling a superior lookup performance compared to BFs. However, the generated false positive free zone is relatively small. For example, an EGH filter with $m\mathrm{=} 2,127$ and $k\mathrm{=}34$ only guarantees a false positive free zone which covers 20 elements from a set with 562 elements. Still, when the number of  record elements is larger than the threshold $n_t$, EGH filter may incurs false positive errors. Besides, to maintain the false positive free zone, the bpe in EGH filter is  higher than Bloom filter. In the above example, the bpe of EGH filter is $\frac m {n_t} \mathrm{=} \frac {2,127} {20}\mathrm{=}106.35$ which is much higher than standard BF. Therefore, EGH filter is advisable when false positives should be completely avoided, the universe set is finite and the set is small while the available space is relatively large.

\subsection{Reducing FP by differentiated representation}
CBF \cite{Counting_BF} also suffers from false positives, since the counters fail to differentiate the elements mapped into it effectively. Unlike the MPCBF which optimizes the utilization of bits in the counter, variable-increment counting BF \cite{VIncrement_CBF} and fingerprint counting BF \cite{Fingerprints_BF} seek the ways to reduce FPR with differentiated representations of elements.

\begin{figure}
  \centering
  \includegraphics[width=3.2in]{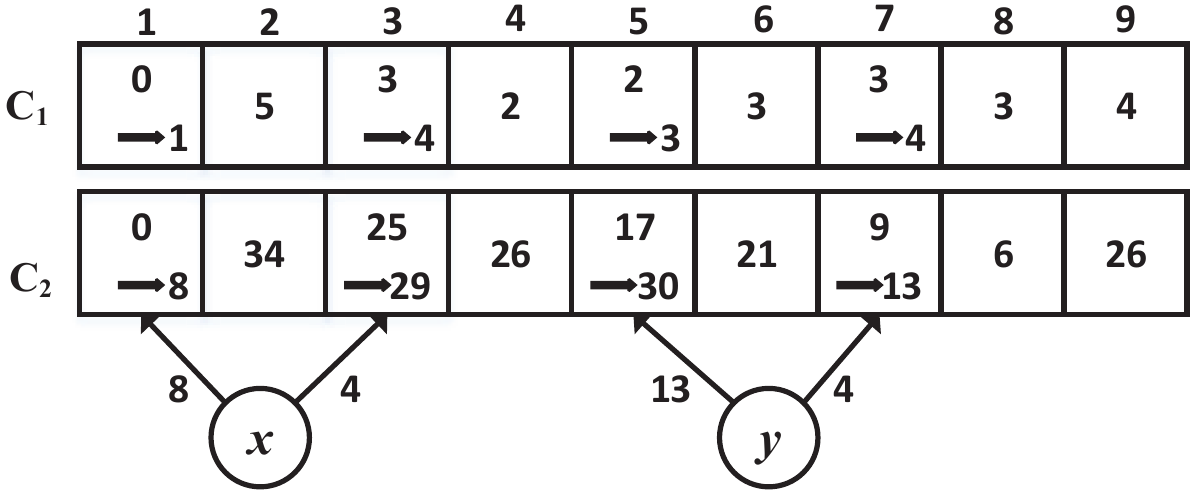}\\ \vspace{-0.05in}
  \caption{An illustrative example of VICBF insertion with $k\mathrm{=}2$ and the base set $D\mathrm{=}\{1,4, 8,13\}$ \cite{VIncrement_CBF}.}\label{fig:VICBF} \vspace{-0.15in}
\end{figure}

\textbf{Variable-increment counting BF (VI-CBF).} Unlike CBFs, when inserting an element, the counters of VI-CBF are incremented by a hashed variable increment instead of a unit increment \cite{VIncrement_CBF}. Then, to query an element, the exact value of a counter is considered, not just its positiveness. The specific insertion operation is shown in Fig. \ref{fig:VICBF}. VI-CBF consists of two counter vectors: the first counter vectors to record the number of elements hashed into this position ($C_1$), and the second counter vectors to provide a weight sum of these elements with diverse increments ($C_2$). A family of hash functions $G\mathrm{=}\{g_1,\cdots,g_k\}$ are employed to select the $k$ increments from the set $D$. Note that $D\mathrm{=}\{v_1, v_2, \cdots, v_u\}$ is a set of integers such that all the sums $v_{i_1}\mathrm{+} v_{i_2}\mathrm{+}\cdots \mathrm{+}v_{i_l}$ with $1\mathrm{\leq}i_1\mathrm{\leq} \cdots \mathrm{\leq}i_l\mathrm{\leq}u$ are distinct. Thereafter, $k$ hash functions $\{h_1,\cdots,h_k\}$ map the element into $k$ cells of VI-CBF. The $k$ counters in $C_1$ are increased by 1, while the counters in $C_2$ are updated with the $k$ selected increments, respectively.

To query an element $x$, VI-CBF first checks the $k$ counters $C_1[h_1(x)], \cdots, C_1[h_k(x)]$. If any counter is 0, obviously $x\mathrm{\notin} S$. If $C_1(i)$ is small, VI-CBF considers the exact values in both counter vectors. In this case, no more than $u$ elements were hashed into these cells. Thereafter, VI-CBF can deduce the employed increments in  the value of $C_2(i)$. If $v_{g_i(x)}$ is contained in $C_2[g_i(x)]$ ($i\mathrm{\in} [1,k]$), then $x\mathrm{\in} S$ with high probability; otherwise, $x\mathrm{\notin} S$. Lastly, if $C_1(i)$ is large, VI-CBF holds that this cell is not useful and examines other cells for possibly eliminating the membership of $x$. By updating the counters with diverse increments, VI-CBF effectively distinguishes the elements mapped into a cell. It was shown that the VI-CBF has an improved false positive rate than the CBF for fixed number of bits per element (bpe) although it requires more bits per counter allowing having a smaller number of counters.

\textbf{Fingerprint counting BF (FP-CBF).} Different from the VI-CBF, FP-CBF labels the elements with unique fingerprints \cite{Fingerprints_BF}. To be specific, each cell in FP-CBF consists of two fields, i.e., the fingerprint field and the counter field. The fingerprint has $a$ bits to store the fingerprints mapped into this cell. Note that the fingerprint of an element is generated by employing a hash function $h_{fp}(x)$ to map element $x$ into the range [0, $2^a$]. The counter field counts the number of elements with $c$ bits. To insert an element $x$, the $k$ hash functions map $x$  into the cells in positions $h_1(x), \cdots, h_k(x)$. Thereafter, in these cells, the fingerprint field is updated by executing the XOR operations between the existing fingerprint and $h_{fp}(x)$. By contrast, the counters are increased by 1. To delete an element $y$ from the FP-CBF, the corresponding $k$ counters are decreased by 1, while the fingerprint fields are updated by XORing the existing fingerprints in each counter with $h_{fp}(y)$.

To query an element $x$, if any counter in the corresponding $k$ cells is 0, $x\mathrm{\notin}S$. For all counters that have a value of 1, FP-CBF checks whether the fingerprint field is different from $h_{fp}(x)$. If so, $x\mathrm{\notin}S$. Otherwise, if the above two checks are passed, FP-CBF believes $x\mathrm{\in}S$ with high probability. That is, FP-CBF recognizes the false-positively matched elements which have at least one counter value of 1. This is realized by checking the fingerprint in the cell with the fingerprint of the queried element. In effect, the similar design philosophy is also achieved in the literature \cite{ICBF} by Luo et al in order to synchronize two given multisets. The dedicated encoding, subtracting and decoding operations are designed to identify the different elements between the multisets. So that only the different elements are transmitted to save bandwidth. However, imposing fingerprint field to the cells requires more memory.

\subsection{Summary and lessons learned}
As a simple summary of this section, numerous variants are proposed to reduce the false positive errors of BFs with novel intuitions. They remove or recognize the potential false positives by using prior knowledge \cite{Multiclass_BF} \cite{FP_Free_BF} \cite{optihash_BF} \cite{Paradox_BF}, selecting optimal hash functions  \cite{Power2choice} \cite{Hash_partition}, generating multiple BFs \cite{Yes_No_BF} \cite{Cross_checking_BF} \cite{Complement_BF} and queries \cite{Trie_BF}, resetting the bits \cite{Retouched_BF} \cite{Generalized_BF} \cite{Multi_partitioning_CBF}, or differentially representing the elements \cite{VIncrement_CBF} \cite{Fingerprints_BF}.  Among these variants, FPF-MBF \cite{FP_Free_BF}, Optihash \cite{optihash_BF}, strategies to lessen Bloom paradox \cite{Paradox_BF}, and Selected hash \cite{Power2choice} \cite{Hash_partition} control the FPP by carefully choosing the hash functions. Other variants, on the contrary, try to decrease the FPR directly. Both of these two design philosophies are reasonable and functional. However, all the above strategies impose either additional memory cost or complicated computation process. In reality, the users may trade-off the impact of false positives and the introduced cost of reducing FPs. 

\section{Optimizations of implementation measurements}  \label{sec:Implementation} 
BF is a lightweight and easy-deployable data structure. But its performance can be further improved. To this end, the existing desgins consider four practical measurements, including computation complexity, memory access, space efficiency, and energy consumption.  

\subsection{Computation optimization} 
The major computation overhead of BF stems from two folds, i.e., the computation of hash functions and the judgment process for a query. BF requires multiple independent hash functions, while well-designed hash functions are computation-intensive, e.g., MD5, SHA-1. Other hash functions, e.g., perfect hash, locality-sensitive hash, are even more complicated to calculate. 
To lessen the computation overhead due to hash functions, state-of-the-art techniques try to generate multiple independent hash values with only one or two hash functions  \cite{LessHash} \cite{OneHash_BF}. 

\textbf{Less hashing, same performance.} Kirsch and Mitzenmacher \cite{LessHash} use two pseudorandom hash functions $h_1(x)$ and $h_2(x)$ to generate additional hash functions. Specifically, the $k$ hash functions will be calculated as: $g_i(x) \mathrm{=} h_1(x) \mathrm{+} i\mathrm{\times} h_2(x) \ mod\ m$, where $0\mathrm{\leq} i \mathrm{\leq}k\mathrm{-}1$ and $m$ is the number of bits in BF. It has been soundly proved that using the generated hash functions imposes no any increase in the asymptotic false positive rate, based on the balls-and-bins analysis \cite{LessHash} \cite{LessHash1}. Since then, this kind of method has been widely utilized in practice to accelerate BF and its variants.  

\textbf{One Hash BF (OHBF).} Unlike the above strategy which needs two hash functions as seeds, OHBF \cite{OneHash_BF} is more ambitious and generates $k$ hash values with only one hash function. OHBF points out that the hash mapping consists of two stages, i.e., the hash stage which maps the input element into a machine word (e.g., 32 bits, 64 bits), and the modulo stage which maps the generated word into a given range via modulo. Consequently, OHBF divides the $m$-bit vector as $k$ parts unevenly, such that $m\mathrm{=}m_1\mathrm{+}\mathrm{\cdots} \mathrm{+} m_k$, where $m_i$ is the length of the $i^{th}$ part. Thereafter, the generated word will modulo with $m_1,\mathrm{\cdots}, m_k$ respectively to derive the location of the element $x$ in each part. Similar with the variants which partition the bit vector into multiple segments,  OHBF also slightly damages the randomness of the hash function. The reason is that the result of a hash function is limited in a dedicated range, rather than the overall bit vector. As a consequence, more hash collisions will be triggered thereby leading to a higher false positive rate. Fortunately, when $m$ increases, the gap of false positive rate between the standard BF and the partitioned variants will degrade gradually.

On the other hand, another thinking is to speed up the hash computation, sequent programmes and check operations. A general strategy is to parallelize the computations. In scenarios where the bit (or cell) vector is divided into multiple segments, these segments can be accessed simultaneously thereby the thereafter computations can be parallelized. This kind of variants include Space-code BF \cite{Space_Code_BF}, Dynamic BF \cite{Dynamic_BF}, Dynamic BF array \cite{DBF_Array}, Par-BF \cite{Par_BF}, BloomStore \cite{BloomStore}, Cross-checking BF \cite{Cross_checking_BF}, One hash BF \cite{OneHash_BF}, Bloom-1 \cite{Bloom_1}, OMASS \cite{OMASS}, Parallel BF \cite{Parallel_BF}, etc. The parallelism strategy can efficiently reduce the response time to $1/\xi$, where $\xi$ is the number of parallelized instances. Note that, some variants with multiple segments, on the contrary, cannot be parallelized, since they are designed for sequent checking, e.g., Yes-no BF \cite{Yes_No_BF}, Bloomier filter \cite{Bloomier_BF}, Complement BF \cite{Complement_BF}, and Multi-partitioned counting BF \cite{Multi_partitioning_CBF}. Specifically, the Yes-no BF \cite{Yes_No_BF} consists of both yes-filter and no-filter. Only the elements which pass the check of yes-filter will be further checked by the no-filter. Similarly, the Complement BF for the complement set $S^{C}$ only need to be checked if the element $x$ has passed the check of the main BF for set $S$. By contrast, the Bloomier filter \cite{Bloomier_BF} and Multi-partitioned counting BF \cite{Multi_partitioning_CBF} cannot be parallelized because the bit vectors in them are constructed recursively. 

In the following, we present two other variants which also enable the parallelism methodology with multiple divisions \cite{LoadBance_BF} \cite{Combinatorial_BF}, as well as one recent proposal which tries to parallelize the hash computation by employing acceleration technique named Single Instruction Multiple Data (SIMD) instructions \cite{UltraFast_BF}.

\textbf{Distributed Load Balanced BF (DLB-BF).} DLB-BF \cite{LoadBance_BF}  is designed for IP lookup (longest prefix matching). Usually, the IP prefixes have different lengths. Based on their lengths, the prefixes are categorized as diverse prefix groups, w.l.o.g., $g$ groups. DLB-BF employs $k$ equal-length BFs, as well as $k$ groups of hash functions, to represent these prefix groups. Note that, each group of hash function has $g$ independent hash functions in it. Each hash function in a group is responsible to a specific group of IP prefixes. Specifically,  the hash function $H_{i,j}$ ($1\mathrm{\leq} i\mathrm{\leq} g$, $1\mathrm{\leq} j\mathrm{\leq} k$) is responsible to map the $i^{th}$ group of IP prefixes into the $j^{th}$ BF. Under this framework, to encode a prefix with length $i$, the $k$ corresponding functions, i.e., $H_{i,1}, H_{i,2}, \cdots, H_{i,k}$, map the prefix into the $k$ BFs and programme the bits to 1s. Thus the membership information of the inserted IP prefix is dispersedly recorded in the $k$ BFs. To query an IP prefix $x$ with length $i$, DLB-BF checks the bits generated by the $k$ hash functions ($H_{i,1}, H_{i,2}, \cdots, H_{i,k}$) in the $k$ BFs. If all these bits are non-zero, the queried IP prefix is in the routing table and the associated forwarding port can be identified reasonably.

To speed up the query and computation, DLB-BF divides a single BF vector into multiple BFs and thus enables parallelism, i.e., the $k$ BFs can be accessed simultaneously and then the bitwise AND operation towards the $k$ bits will generate the query result. The problem is that the number of involved hash functions is $g\mathrm{\times}k$. The calculations of these hash functions can be challenging.


\textbf{Combinatorial BF.} Combinatorial BF \cite{Combinatorial_BF} considers the membership query of $S$ which has multiple groups. The query result should point out which group(s) the queried element belongs to, without the priori knowledge of the number of elements in each group. For simplicity,  Combinatorial BF supposes that an element only belongs to one of the $\gamma$ groups. For arbitrary element $x$, its group $g(x)$ has a $\Psi$-bit binary code $C(g(x))$. For each bit in the code, there is a corresponding group of $k$ hash functions. Therefore, $k\mathrm{\times}\Psi$ hash functions are required. To insert the element $x$, Combinatorial BF maps $x$ into the bit vector with the hash groups where the bits in $C(g(x))$ are 1s. For example, given $\Psi\mathrm{=}3$ and $k\mathrm{=}3$, if the element $x$ belongs to group 5, whose id code is 101, then the two groups of hash functions associated with the first and third bit of the code will be employed to map $x$. After that, to query the element $y$, Combinatorial BF checks all the $\Psi$ groups of hash functions. If all the $k$ bits in the bit vector associated with a group of hash functions are non-zero, the corresponding bit in the group code will be 1; otherwise, 0. In this manner, the group code can be determined reasonably. 

Besides of a large number of hash functions and memory access, the probability of misclassification in Combinatorial BF is high, since any false positive error for each bit will result in incorrect group code. An augment is to use fixed-weight group code, i.e., every group has the same number of 1s. Therefore, parts of the false positive errors can be recognized accordingly. The misclassification probability of Combinatorial BF is much larger than its false positive rate. Combinatorial BF also proposes a parallelism memory access strategy to speed up the query process. The idea is to partition the bit vector into $k$ chunks. The first hash function of each hash group will access the first chunks, and the second hash function of each hash group will map elements into the second chunk, so on and so forth. Therefore, the element can be inserted with $\theta$ memory accesses, where $\theta$ is the number of 1s in the group code $C(g(x))$.

\begin{figure}
  \centering
  \includegraphics[width=3.2in]{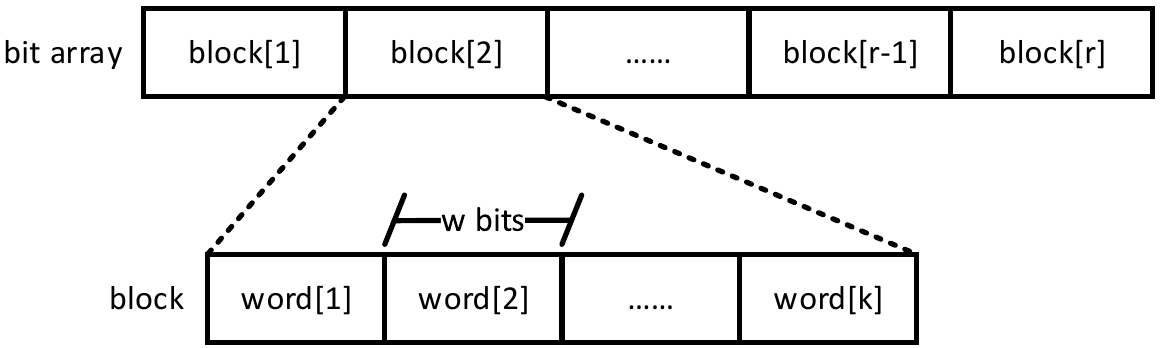}\\ \vspace{-0.05in}
  \caption{The framework of Ultra-Fast BF \cite{UltraFast_BF}.}\label{fig:UF_BF} \vspace{-0.15in}
\end{figure}

\textbf{Ultra-Fast BF (UFBF).} Ultra-Fast BF \cite{UltraFast_BF} tries to parallelize both the calculation of hash functions and check process directly. The UFBF vector is composed of $r$ blocks each of which has $b$ bits. Further, each block has $k$ consecutive words, thus $b\mathrm{=}k\mathrm{\times}w$, where $w$ is the length of a word. The overall number of bits can be calculated as $m\mathrm{=}r\mathrm{\times}k\mathrm{\times}w$. To insert an element $x$, a hash function $h_0$ selects one of the $r$ blocks randomly. Thereafter, $x$ will be mapped into the $k$ words with $k$ hash functions, respectively. Similar operations are executed to query an element. Based on this framework, the Single Instruction Multiple Data (SIMD) scheme is introduced to parallelize the calculation of the $k$ hash functions. The basic idea is to generate multiple hash values with one hash function using. Moreover, the membership query process can be naturally parallelized since the presence information is saved in $k$ separated words of one block. Consequently, the queries can be significantly accelerated. 

On the contrary, the SIMD requires supports of competent computing units, while lightweight devices (e.g., sensors, detectors) unable to run the instructions. Therefore, the SIMD instruction lacks of generality nowadays. Besides, the UFBF is proved to suffer from higher false positive rate, compared with the standard BF. 

\subsection{Memory access}
To accomplish the membership query, one has to read the $k$ bits, which may results in numerous memory accesses thereby hurting the performance of space-shared applications. Consequently, endeavours are made to reduce memory access times, such as Bloom-1 \cite{Bloom_1} and OMASS \cite{OMASS}. 

\textbf{Bloom-1.}  Bloom-1 \cite{Bloom_1} consists of $l$ words, each of which can be fetched from the memory to the processor in one memory access. Before inserting an element $x$, a string of hash bits is generated by a hash function. Thereafter, $\log_2 l$ hash bits are employed to select a word to record $x$, and $k\times \log_2 w$ hash bits map $x$ into the word by setting the corresponding bits from 0s to 1s. Therefore, $\log_2 l + k\times \log_2 w$ hash bits are required in total. To query an element, only one memory access is needed. If all the $k$ bits are non-zero, Bloom-1 concludes that element belongs to the set; otherwise, Bloom-1 infers that element is not a member of the set. Moreover, Bloom-1 can be further generalized as Bloom-$g$ by recording an element in $g$ words, rather than only 1 word. As a consequence, a lower false positive rate will be achieved, at the expense of $g$ memory accesses per query. 

\textbf{OMASS.} OMASS \cite{OMASS} focuses on the problem of set separation, i.e., identifying which sets an element $x$ belongs to. The existing approaches employing BFs in parallel. These schemes, however, incur multiple memory accesses which may be a bottleneck for certain applications. A solution is to divide each BF vector into multiple blocks, each of which is exactly a memory word size e.g., 32 bits or 64 bits. A global hash function is introduced to select a block when inserting an element. To resolve the set separation problem, this solution also calls for $s$ memory accesses, where $s$ is the number of queried sets. To further improve the strategy, one may overlay these BFs and share the same memory. This insight, however, leads to high false positive rate. Therefore, OMASS proposes to further equally divide each block as $k$ sub-blocks, where $k$ is the number of employed hash functions to map elements. The $j^{th}$ sub-block is responsible to the $j^{th}$ hash function. Moreover, to eliminate the interferences due to memory sharing, OMASS isolates the values generated from hash functions in a sub-block for diverse sets by letting $h_{ij}(x) \mathrm{=} (h_{1j}(x) \mathrm{+} (i\mathrm{-}1))\ mod\ b$, where $1\leq i\leq s$ and $b$ is the length of each sub-block. In this manner, the insertion of $x$ in one set will not affect the false positive rate when checking $x$ against the other $s\mathrm{-}1$ sets. Therefore, only one memory access is needed to tackle the set separation problem, without increasing of false positive rate.

Both Bloom-1 \cite{Bloom_1} and OMASS \cite{OMASS} record elements with given length of words, so that only one memory access is required for membership query. OMASS further eliminates the interference of elements which belongs to different sets but recorded in a shared word. A common shortcoming of them is that, when the number of elements in a dataset changes, they must be rebuilt. The lack of scalability makes them incapable of representing dynamic sets.

\subsection{Space efficiency} 
Space efficiency is always a problem whenever storing the BF locally or disseminating the BFs among hosts in a distributed system. Typically, to realize the minimum false positive rate, the bit utilization in a standard BF is only 50\%. Therefore, multiple works have been conducted to tackle this issue \cite{Compressed_BF} \cite{Compacted_BF} \cite{d_left_hash}   \cite{additional_hash_BF} \cite{Matrix_BF}. 

\textbf{Compressed BF.} Compressed BF \cite{Compressed_BF} optimizes the transmission overhead when the filters are exchanged in networks. Generally, the Compressed BF will be decompressed for real use after the transmission. The probability of each bit to be set as 1 under the optimal setting of BF is 1/2, which offers no compression gain at all. By contrast, under the constraint of number of bits to be sent after compression $z$, the number of bits $m$ of the array in the uncompressed form can be larger. With this insight, Compressed BF employs fewer hash functions $k$ yet larger number of bits $m$ to guarantee smaller false positive rate with less bits to transmit than standard BF. Alternatively, with the same number of bits to transmit, Compressed BF realizes lower false positive rate than standard BF. Specifically, the false positive rate is reduced as $0.5^{z/n}$, where $z$ is the length of bit vector after compression. However, both the compression and decompression algorithms consume more processing time and require additional computing and memory resources. Lightweight devices  may lack the capability to execute these complicated algorithms.

 \textbf{Compacted BF.} Similar with Compressed BF, Compacted BF \cite{Compacted_BF} reduces the length of bit vector before transmission to save bandwidth. In contrast, Compacted BF abandons the time-consuming compression algorithms but proposes a new pattern to condense the bit vector. Specifically, Compacted BF splits the original bit vector into $k_0$ blocks, each of which has $b$ bits. Thereafter, the original bit vector is translated as an array of $n$ indices and each index contains $m_0$ bits. Index $i$, denoted by $CmBFV[i]$, corresponds to the value of the $i^{th}$ bit positions in $block_1$, $block_2$, $\cdots$, $block_n$ in the original BF. For the $i^{th}$ index, the principles of deriving the index are follows: 1) if there is no 1 in the $i^{th}$ bits of all blocks, $CmBFV[i]$ is set as 0; 2) if only one 1 exist and this 1 appear in the $i^{th}$ bit of $block_r$, then $CmBFV[i]$ is set as $r$; 3) if all the bits or more than half of the bits are 1s, $CmBFV[i]$ will be set as $2^{m_0}\mathrm{-}1$; 4) in the case that less than half bits but more than one bit are 1, the Compacted BF randomly selects a block $block_s$ which contains 1, and set $CmBFV[i]$ as $s$. Thus, the $b\times k_0$ original bit vector will be compressed as a $n\times m_0$ Compacted BF.   
  
After transmission, the Compacted BF will be interpreted as a normal bit vector for later query. However, interpretion leads to both false positive and false negative errors. When the 1 in the original bit vector is interpret as '0', a false negative error happens. The false positive rate and false negative rate can be controlled by adjusting the parameters. But in some situations, the false negative is not allowed. Moreover, Compacted BF may not be robust due to the randomness in rule 4. 

\textbf{$d$-left counting BF (dlCBF).} For a static set $S$, one can employ a perfect hash function to map each element in $S$ into a hash table without any collision. But for a dynamic set, this scheme is not advisable, since each insertion or deletion of element leads to the reconstruction of the hash table. The CBF, although supports insertion and deletion smoothly, suffers from high false positive rate and space overhead. To this end, dlCBF \cite{d_left_hash} proposes to replace the general hash functions in CBF with a $d$-left hash function which is reported as ``almost perfect hash function'' \cite{almost_perfect} \cite{almost_perfect1}. dlCBF splits the cell vector into $d$ subtables, each with $n/d$ buckets, where $n$ is the total number of buckets. Each bucket resides $c$ cells, and each cell has a fingerprint field and a counter field. The fingerprint for an element $x$ consists of two parts, i.e., the first part corresponds to the index of the bucket which $x$ is placed in, and the second part is a remainder of $x$.   

Specifically, to insert an element $x$, a hash function maps $x$ as a bit string. The generated bit string is divided into $d\mathrm{+}1$ segments. The first $d$ segments offer candidate positions to record the fingerprint of $x$. The last segment is treated as the remainder of $x$. Thereafter, the remainder of $x$ will be stored in the cell with least load among the $d$ candidates (breaking ties to the left). To query $x$, all the $d$ cells will be checked. If the remainder of $x$ is found, dlCBF affirms the presence of $x$; otherwise not. Deletion is enabled by clearing the cell or decreasing the count field in the corresponding cell. Additionally, the augmented fingerprint creation scheme and the random permutations are introduced to handle the fingerprint collisions, so that mis-deletions will not happen. Compared with CBF, dlCBF achieves nearly 50\% space saving with the same false positive guarantee, and two magnitude reduction of false positive rate with the same space scale.

\textbf{Memory-optimized BF.}  Ahmadi et al. argue that the regular Bloom filter stores items from a set $k$ times $k$ memory locations that are determined by the $k$ addresses stored in the bit-array structure. The elements are stored quite redundantly. Based on this insight, an additional hash function is introduced to selected a cell among the $k$ candidate cells to store $x$ \cite{additional_hash_BF}. Therefore, the element will be stored only for once, and other $k\mathrm{-}1$ duplicates will not be necessary anymore. Consequently, the space usage will be highly improved. The experimental results indicate that this scheme results in much fewer hash collisions than the standard BF. However, the decrease of hash collisions never implies the reduction of false positive rate. It doesn't change the fact that there is still an optimal number of hash functions to realize minimum false positive rate.  

\textbf{Matrix BF.}  A Matrix BF is a bit matrix in which each bit can be set or reset to detect copy-paste contents in a literature library \cite{Matrix_BF}. The matrix BF consists of $N$ rows each of which records the contents in one document. Every row has $m$ bits and acts as a BF to support insertion and query, with $k$ shared hash functions. Before mapping, each document is divided into sub-strings by the Chunking Unit, and then feed the hash functions to set the $k$ corresponding bits to 1s. To detect the degree of copy-paste between any pair of documents, matrix BF just executes the bitwise AND operations between the two rows. Thereby, the similarity is measured by counting the number of 1s in the resultant bit array, rather than calculating the cosine or Jaccard similarity. If the number of 1s is larger than a predefined threshold, matrix BF believes the two documents are similar. Document-level scalability and deletion are enabled by simply inserting and deleting a row in the matrix, respectively. Matrix BF jointly supports a trade-off among accuracy, speed, space and privacy protection of the system. 

\begin{figure}
  \centering
  \includegraphics[width=3.2in]{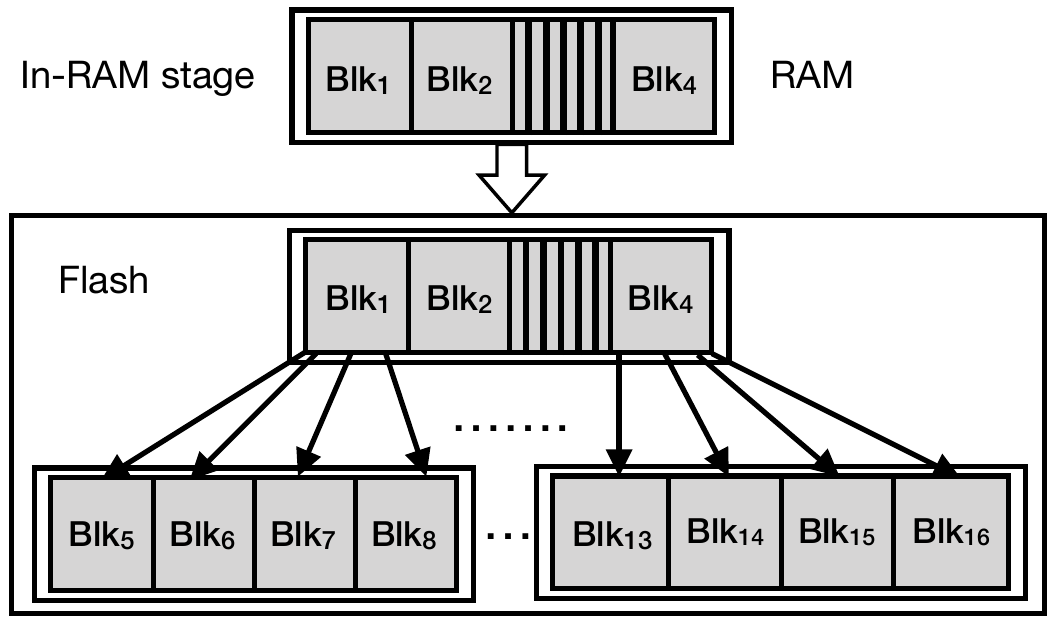}\\ \vspace{-0.05in}
  \caption{An example of Forest-structured BF \cite{Forest_structured_BF} with $\lambda\mathrm{=}b\mathrm{=}4$. }\label{fig:Forest_BF} \vspace{-0.15in}
\end{figure}

\textbf{Forest-structured BF  (FBF).} Usually, BF is saved in RAM whose space is a scarce resource. But once the BF exceeds the RAM size, secondary memory e.g. flash-based SSD, will be an alternative. Forest-structured BF \cite{Forest_structured_BF} consists of a collection of sub-BFs, each of size that exactly equals flash page size. $\delta$ sub-BFs are packed as a block. The blocks, thereafter, are organised as a forest structure. The highest layer contains $\lambda$ blocks. Each block (except for the ones at the lowest layer) has $\varsigma$ children ($\varsigma \geq 2$). By default, if the RAM is enough to save the blocks, FBF will only use the RAM; otherwise, the FBF will be moved into SSD and the RAM will be utilized as a buffer space when inserting elements. As depicted in Fig. \ref{fig:Forest_BF}, to query an element $x$, two hash functions are employed to select a block ($blk\_id$) and a sub-BF ($page\_id$) respectively. If the sub-BF fails to identify the existence of $x$, FBF will check the corresponding child of the block. The query algorithm will be terminated when $x$ is found or the lowest level of child has been checked. For instance, given $\lambda\mathrm{=}\varsigma\mathrm{=}4$, $blk\_id\mathrm{=} 2$, and $page\_id\mathrm{=}10$, FBF checks the sub-BF 10 of Block 2 in the first level. If $x$ is not found, the query algorithm will check the sub-BF 10 in Block 10 (Block 10 is the second child of Block 2) in the second level, so on and so forth. Therefore, at most $l$ sub-BFs will be checked, where $l$ the number of levels in FBF.  Besides, each sub-BF only requires one flash read.

However, the overall false positive rate can be enlarged due to the cascaded structure of FBF. Given the false positive rate of each sub-BF as $f_r$, the overall false positive rate can be calculated as $1\mathrm{-}(1\mathrm{-}f_r)^l$, where $l$ the number of levels in FBF. With the growth of levels, the overall false positive rate will be increased quickly.

\subsection{Energy saving} 
To query the membership of any element, all the $k$ corresponding bits in the BF vector must be checked. But the fact is that, for the elements which lead to negative query results, the zero bits will be located by a subset of the $k$ bits. This observation indicates the potentiality to lessen the calculation and energy consumption for negative query results. Therefore, several literature propose to check only part of the $k$ bits to save energy \cite{Pipelined_BF1} \cite{Pipelined_BF2} \cite{FullyPipelined_BF}  \cite{Energy_Efficient_BF}. Besides, different implementation environments also lead to diverse energy consumption \cite{L_CBF}. 

\textbf{Pipelined BFs.} Pipelined BF \cite{Pipelined_BF1} \cite{Pipelined_BF2} divides the $k$ hash functions into two stages, i.e., stage 1 and stage 2. The $k_1$ hash functions in stage 1 are always activated, while the $k_2\mathrm{=}k\mathrm{-}k_1$ stage 2 hash functions will be employed only when the queried element passes the checks of the stage 1 hash functions. For arbitrary query, the probability to employ stage 2 hash functions is approximately $(1\mathrm{-}e^{\frac {-kn}{m}})^{k_1}$. In the worst case (the element is either a member of $S$ or a false positive error), Pipelined BF will check all the $k$ bits, just like the BF does. The false positive rate of Pipelined BF is the same as BF. The average power saving ratio (compared with standard BF) can be calculated as:
\begin{equation}\label{equ:saving_ritio}
\frac {k_2+(k_1-k)(1-e^{\frac {-kn}{m}})^{k_1}}{k}.
\end{equation}
The two-stage Pipelined BF is further generalized as a fully Pipelined BF by regarding each hash function as a stage \cite{FullyPipelined_BF}. A queried element is progressed to the next stage only when the previous hash function produces a match. By doing so, the power saving ratio is further increased as: 
\begin{equation}\label{equ:saving_ritio1}
1-\frac 1 k \times \sum_{i=1}^{k} \rho^{i-1},
\end{equation}
where $\rho$ is the ratio of 1s in the bit vector. Pipelined BFs, however, introduce much higher latency to each query, since the multi-stage check scheme slows down the query process.

\begin{figure}
  \centering
  \includegraphics[width=3.2in]{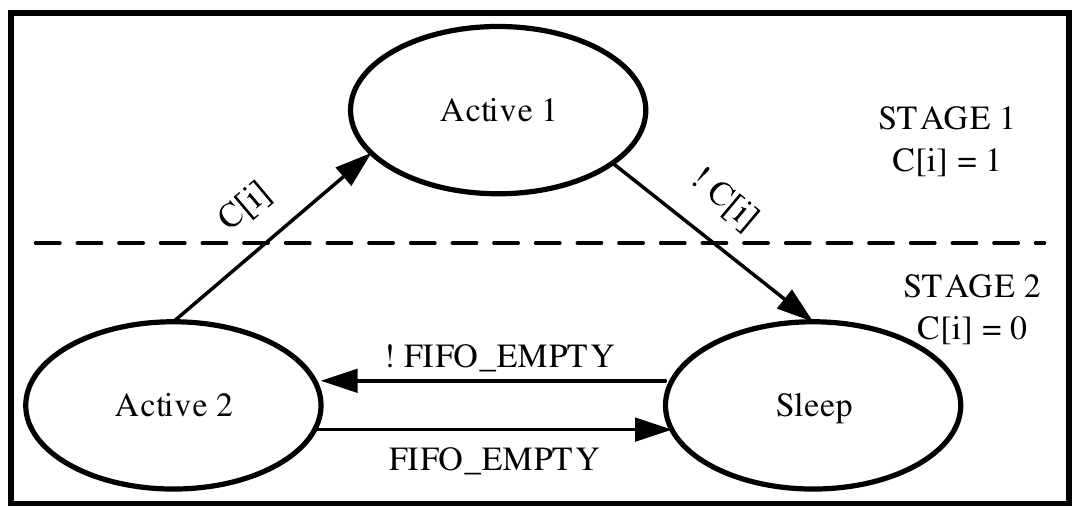}\\ \vspace{-0.05in}
  \caption{Hash function states in EABF \cite{Energy_Efficient_BF}. }\label{fig:EA_BF} \vspace{-0.15in}  
\end{figure}

\textbf{Energy efficient BF (EABF).} EABF \cite{Energy_Efficient_BF} augments the two-stage query scheme by adjusting the state of stage 2 hash functions adaptively. Specifically, $k_1$ hash functions are maintained as stage 1 for a tolerable false positive rate; the remained $k_2\mathrm{=}k\mathrm{-}k_1$ hash functions are allowed to move out to stage 2 for lower power consumption or moved back to stage 1 for faster response, according to the incoming workload. A FIFO buffer is introduced to cache the elements to be queried by stage 2 hash functions in the next time clock. As depicted in Fig. \ref{fig:EA_BF}, the hash function can adaptively and automatically adjust its state according to the value of the control bit $C[i]$ and the content in FIFO buffer. The flexibility of state migration makes sure that EABF acts as the two-stage scheme in \cite{Pipelined_BF1} \cite{Pipelined_BF2} for energy saving, as well as a regular BF in busy-hour for fast query speed. The adaption control policy, as well as the multi-stage design of EABF, is proposed for better performance. In real implementations, EABF needs a complex control circuit for each hash function. The complexity will be further amplified in the multi-stage sense.

\textbf{L-CBF.} Safi et al. rely on the following simple observations on CBF: (1) the actual count sequence used in a CBF is not important, and (2) externally, the users only care whether a counter is ``zero'' or ``non-zero'' \cite{L_CBF}. With the above insight, they proposed L-CBF \cite{L_CBF}, a hardware implementation of CBF, which implements an array of up/down counters while avoiding the overheads associated with using arithmetic counters. Specifically, L-CBF \cite{L_CBF} uses up/down linear feedback shift registers (LFSRs) which offer a better latency, power and complexity trade-off than other non-arithmetic counters. Additionally, L-CBF introduces an extra bit per counter to label the state of counters, i.e., the bit will be updated only when the count changes from or to zero  \cite{L_CBF}. This strategy significantly speeds up the query process, because the membership query only cares whether the corresponding counters are all non-zero or not, but never concerns the exact values in these counters. Experimental results indicate that L-CBF significantly outperforms the SRAM based implementation in terms of both query speed and energy-saving. The penalty is 3.2 $\times$ space occupation.

\subsection{Summary and lessons learned}
According to the above statement, the optimization space of computation is two-dimensional: reduce the amount of computation or speed up the computation. To reduce the amount of computation, it is advisable to employ computation-friendly hash functions or generate $k$ independent hash values with fewer hash functions  \cite{LessHash}\cite{OneHash_BF}. On the other hand, to speed up the computation process, parallelization can be a choice, e.g., DLB-BF \cite{LoadBance_BF} and Combinatorial BF \cite{Combinatorial_BF}.  As for memory access, the bit vector can be divided into multiple disjoint segments, thereby each query only needs to access one segment, such as Bloom-1 \cite{Bloom_1} and OMASS \cite{OMASS}. This will significantly reduce the memory access frequency. The space efficiency is not only about the utilization of bits in a BF (e.g., Compressed BF \cite{Compressed_BF}, Compacted BF \cite{Compacted_BF},  dlCBF \cite{d_left_hash}, Memory-optimized BF \cite{additional_hash_BF}, and Matrix BF \cite{Matrix_BF}), but also the global space utilization in a system where both fast RAM and secondary memory (e.g., SSD) are jointly employed  \cite{Forest_structured_BF}.  Lastly, for negative query results, when the query algorithm encounters the first zero bit, the algorithm should be terminated and return a negative result. This observation enables the possibility to reduce the energy consumption of membership queries against BFs. This kind of designs include Pipelined BF \cite{Pipelined_BF1} \cite{Pipelined_BF2}, EABF \cite{Energy_Efficient_BF}, and  L-CBF \cite{L_CBF}.

In a real implementation, the performance of BFs can be further strengthened by optimizing the computation cost,  reducing memory access, improving space efficiency, and decreasing the energy consumption. In scenarios where these metrics are sensitive, the above BF variants are possible candidates.

\section{Representation of diverse sets} \label{sec:SetDiversity} 
As depicted in Fig. \ref{fig:logic}, in the framework of BF, the input sets can be diverse and have their own features. Therefore, customized variants and techniques are proposed to record the elements under diverse scenarios. For example, if the dataset is a multiset, alterations are needed to record the multiplicity of each element; if the dataset is dynamic, the BF capacity should be expanded or shrunk on demand;  if the elements have different weight, BF should ensure the high-weight elements incurs lower FPR. Besides, the elements may be not independent but maintain a logical, temporal or spatial relationship. How to represent these special datasets with BF effectively is also challenging.  Consequently, in this section, for eligibility, we categorize the associated proposals as multiset, dynamic dataset, dataset with weighted elements, key-value system, sequence data and spatial data. 

\subsection{Multisets} 
We note that several literature mixes the definition of multi-set and multiset. In this survey, we regard the term multiset as a mathematical term and represents a dataset which permits the coexistence of multiple replicas of any element. Several parameters can be employed to characterize the features of a multiset. Let $x$ be an element of a multiset $A$. The multiplicity of $x$ is denoted by $m_{A}(x)$, which denotes the number of instances of $x$ in $A$. Note that a multiset is a generalization of a set. A simple set is a special case of a multiset where all elements only appear at most once. For example, in networks, a data flow consists of multiple packets with same source and destination IP addresses. Thus a flow can be regarded as a multiset element and the associated multiplicity is the number of packets in the flow. The major challenge of representing a multiset is that not only the existence of  elements, but also the associated multiplicities should be recorded correctly. To this end, Space-Code BF \cite{Space_Code_BF}, Spectral BF \cite{Spectral_BF}, Loglog BF \cite{Loglog_BF}, and Invertible Counting BF \cite{ICBF} record the multiplicities with the counters in each cell, at the cost of more space overhead. By contrast, Adaptive BF \cite{Adaptive_BF} and Shifting BF \cite{Shifting_BF} set additional bits to 1s to indicate the multiplicity of each element. Especially, Space-code BF \cite{Space_Code_BF} and Loglog BF \cite{Loglog_BF} employs the maximum likelihood estimation and probabilistic counting strategy to estimate the multiplicity of each element, respectively.

\textbf{Space-Code BF (SCBF).} Space-code BF \cite{Space_Code_BF}  novelly employs $g$ groups of hash functions to map an element into a shared bit vector. Thereafter, the multiplicity of a queried element $x$ will be estimated by the number of hash groups which indicate the existence of $x$. The estimation of multiplicity is conducted by running a maximum likelihood estimation (MLE) procedure. For convenience, a MLE table is pre-calculated so that only a simple query against the table is needed during estimation. When the multiplicity of an element is very high, the accurate multiplicity estimation will not be possible since all the $g$ groups of hash functions will be employed. To this end, the Multi-Resolution SCBF is designed by enabling $r$ SCBF simultaneously. When inserting, the element $x$ will be inserted into the $i^{th}$ SCBF with probability $p_i$, where $p_1\mathrm{>}p_2\mathrm{>}\cdots \mathrm{>} p_r$. By doing so, the elements with low multiplicities will be estimated by filter(s) of higher resolutions, while elements with high multiplicities will be estimated by filters of lower resolutions. As for the query, a joint MLE procedure will derive the estimation of multiplicity. Also, the overall MLE estimation can be pre-calculated as a MLE table. The user can tune the parameters for better estimation accuracy.

Although sharing a bit vector among multiple groups of independent hash functions will save space, both the MLE procedures for each SCBF and the overall estimation are computation-intensive. In other words, the saving of space is achieved by the sacrifice of more computation resource. Therefore, SCBF may suit devices with scarce space but ample computation capacity.  Besides, the MLE may overestimate as well as underestimate the real multiplicity of an element.

\textbf{Spectral BF.} Spectral BF \cite{Spectral_BF} is proposed to support approximate multiplicity queries of multiset elements. To this end, Spectral BF extends the bit vector of BF as a vector of $m$ counters. When inserting an element, the corresponding $k$ counters are increased by 1. By decreasing the dedicated counters, Spectral BF also enables element deletions. This is similar to CBF. However, when querying, Spectral BF employs the minimum value among the $k$ counters as the estimator of the multiplicity of $x$. Consequently, Spectral BF further optimizes the data structure with the ``minimum increase'' and ``recurring minimum'' schemes. The minimum increase scheme prefers conservative insertion, i.e., only increase the smallest counter(s). This strategy increases the accuracy of multiplicity queries significantly. As for the recurring minimum scheme, it further initialize a secondary Spectral BF to store the elements with a single minimum. Therefore, the error rate of multiplicity query can be further decreased. 
 
The resulted space overhead of Spectral BF can be a slightly higher than that of the standard BF. Employing the minimum counter as the estimator of multiplicity is helpful to realize high query accuracy. However, this strategy can also be a barrier of accurate querying when a hash collision occurs. For instance, consider that the multiplicities of elements $x$ and $y$ are 3 and 2 respectively. If $x$ and $y$ share a common counter, then the deletion of $y$ will lead to incorrect multiplicity query of $x$. The reason is that the deletion of $y$ will decrease the minimum counter of $x$ from 3 to 1. Thus, the multiplicity of $x$ will be wrongly reported as 1 instead of 3. Therefore, the Spectral BF may report an overestimate or underestimated multiplicity for an element.

\textbf{Invertible Counting BF (ICBF).} Unlike Spectral BF, ICBF  \cite{ICBF} is designed to achieve fast and effective synchronization between two given multisets. To this end, ICBF extends the BF into a vector of cells, each of which consists of two fields, i.e., $id$ and $count$. Specifically, the $id$ field is responsible to record the identifiers of elements which have been hashed into the cell, while $count$ field memorizes how many elements have been encoded into that cell. Thereafter, the corresponding encoding, subtracting and decoding operations are proposed to search out the difference between two given multisets. Given a pair of multisets, $A$ and $B$, when encoding them as $ICBF_A$ and $ICBF_B$, each of the elements is mapped into $k$ cells via the $k$ independent hash functions. The task of subtracting is to subtract the different elements from $ICBF_A$ and $ICBF_B$. This is completed by executing XOR operations between the $id$ fields and minus arithmetic subtractions towards the $count$ fields. Then a novel algorithm is proposed to decode the elements from the subtracting result in a recursive manner. Moreover, together with the local \emph{id table}, the ICBF can inevitably decode the different elements from the ICBFs with high probability. For synchronization, the involved hosts only need to transmit the difference caused by diverse elements and the gap of multiplicities can be fixed with replicas. 

By recognising the different elements, ICBF achieves the minimum transmission cost to achieve approximate multiset synchronization.  However, ICBF suffers from both false positive and false negative errors when synchronizing the multisets due to hash collisions. Moreover, exchanging the ICBFs between hosts also occupies a part of the link bandwidth.

\textbf{Loglog BF.}  Loglog BF \cite{Loglog_BF} introduces the probabilistic counting strategy \cite{coungting_estimate} to estimate the frequency of each memorized multiset element. The probabilistic counting strategy \cite{coungting_estimate} hashes all elements in a set into a given range of binary strings and then estimates the cardinality of each element with the locations of the first 1s in the binary strings. In the framework level, Loglog BF extends BF from two folds: 1) the bits in BF is extended as fixed-length counters; 2) an extra parameter $d$ (the duplication factor) is introduced so that Loglog BF inserts an item into the counter vector with different prefixes. To insert an element $x$, a prefix range from 0 to $d\mathrm{-}1$ is attached, before $x$ is mapped into the counter vector with $k$ hash functions. That is, the information of $x$ is memorized in $k\mathrm{\times} d$ cells. For each of the selected cells, a random integer of $x$ (geometric distribution) will be generated to replace the existing value in the cell if the existing value is less than the random integer. To answer the multiplicity of $x$, the probabilistic counting strategy is applied to calculate the estimated value based on the counters in the $k\mathrm{\times} d$ cells with high precision. Further functions like join and compress towards Loglog BFs are also enabled.

Besides of the explicit false positive rate of membership query, Loglog BF also introduces additional calculation cost to the BF paradigm by introducing the probabilistic counting strategy. Both generating the random integer during inserting and deriving the estimation value of multiplicity will occupy the computing resources. Moreover, the probabilistic counting strategy calls for comprehensive calibration of the related parameters (the number of bits for each counter, the value of $d$) to reach its best performance. 

\textbf{Adaptive BF.} Adaptive BF \cite{Adaptive_BF} also tells the multiplicity of an element with the number of hash functions. Different with SCBF, Adaptive BF doesn't need multiple groups of hash functions, but $k\mathrm{+}N\mathrm{+}1$ hash functions, where $N$ is the maximum multiplicity for elements in the set $S$. When an element $x$ is inserted for the first time, the $k$ corresponding bits in the bit vector will be set as 1s, and an extra bit is set as 1 to indicate its current multiplicity is 1. Thereafter, the latter inserted replicas of $x$ will be recorded by employing an additional hash function to set a bit as 1. In this manner, the membership information of an element is represented by the $k$ bits, and the multiplicity of the element is kept by the number of latter programmed 1s. Similar to BF, Adaptive BF tackles the membership query via checking the $k$ corresponding bits in the bit vector. For multiplicity query, Adaptive BF counts how many 1s are set by the latter hash functions.

However, Adaptive BF cannot support deletion, since resetting 1s to 0s may lead to false negative membership query errors and incorrect multiplicity query results. Besides, Adaptive BF treats the 1s which represent membership information and the 1s that indicate multiplicity without any differentiation. As a result, the 1s in the bit vector to represent multiplicity information can increase the false positive rate of membership queries. By contrast, the 1s in the bit vector to represent membership information may lead to inaccurate multiplicity query results.  Users may implement the Adaptive BF with two separated bit vectors. One bit vector records the membership information with $k$ hash functions, and the second bit vector stores the multiplicity with the $N\mathrm{+}1$ hash functions. This isolation significantly eliminates the interference between the membership information and the multiplicity information. 

\begin{figure}
  \centering
  \includegraphics[width=3.2in]{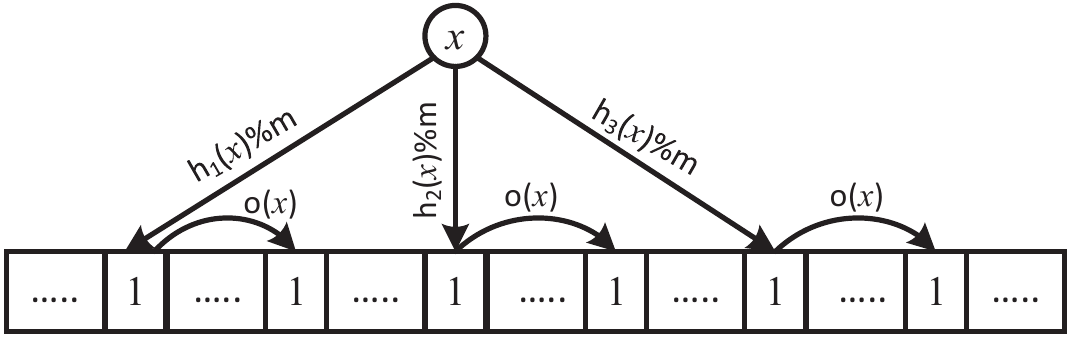}\\ \vspace{-0.05in}
  \caption{The construction of Shifting BF \cite{Shifting_BF}.}\label{fig:ShiftingBF} \vspace{-0.15in}
\end{figure}

\textbf{Shifting BF.} Shifting BF \cite{Shifting_BF} is an array of $m$ bits which are all initialized as 0s. It acts the same as BF to set $k$ bits to 1s with the $k$ independent hash functions, to store the existence information of an element $x$.  Additionally, Shifting BF employs the other $k$ bits to store the auxiliary information, e.g., multiplicity or affiliation. As depicted in Fig. \ref{fig:ShiftingBF}, the bits for auxiliary information are derived by an offset function $o(x)$. Based on the above framework, Shifting BF can support membership query for sure, but also customize other types of queries by adjusting the offset function. Taking the multiplicity query as an example, the offset function can be set as the multiplicity of the element. In the query phase, given the maximum multiplicity $C$ in the multiset, Shifting BF will check $C\mathrm{\times}k$ bits. The multiplicity of element $x$ will be the value of $i$ ($1\mathrm{\leq}i \mathrm{\leq}C$), if the $k$ bits in the locations $\{h_1(x)\%m\mathrm{+}i$, $\cdots$, $h_k(x)\%m\mathrm{+}i\}$ are all non-zero. If deletion is required, a counting version of Shifting BF can be generated by replacing each bit with a counter. Note that, under the framework of Shifting BF, the existence information and auxiliary information in the bit vector may interfere each other. 

As a summary of this subsection, the multiplicities of multiset elements can be represented by the counters in each cell vector (e.g., Space-Code BF \cite{Space_Code_BF}, Spectral BF \cite{Spectral_BF}, Invertible Counting BF \cite{ICBF}, Loglog BF \cite{Loglog_BF}) or setting additional bits to 1s in the bit vector (e.g., Adaptive BF \cite{Adaptive_BF}, Shifting BF \cite{Shifting_BF}). Note that there are always trade-offs in these variants. Recoding the multiplicity with counters generates additional space overhead, while setting the additional bits to 1s affects the accuracy of membership query. 

\subsection{Dynamic sets}
BF is designed for representation and membership querying for static datasets. However, in fact, distributed applications and online systems must deal with the dynamic datasets where the elements can join or leave dynamically. Although CBF \cite{Counting_BF} enables the deletion of elements reasonably, its capacity cannot be extended on demand. To overcome the obstacles of using BF in dynamic datasets, several compact variants are proposed. Variable length signatures \cite{Variable_length_BF} refers only part of the $k$ corresponding bits when inserting, querying, and deleting an element. Dynamic BF\cite{Dynamic_BF}, Scalable BF \cite{Scalable_BF}, Dynamic BF Array \cite{DBF_Array}, and Par-BF \cite{Par_BF} maintains multiple BF vectors (either homogenous or heterogeneous) in the memory and activate or merge these BFs on demand.

\textbf{Variable length signatures.} Especially, Lu et. al \cite{Variable_length_BF} focus on the flow deletion in the networking scenario. Obviously, flows in computer networks can be regarded as a time-varying set. Based on this insight, the variable length signature \cite{Variable_length_BF} scheme is put forward to enable straightforward deletion, query, aging and recovering of any flow. Different with the BF, variable length signature only sets $t\mathrm{\leq}k$ bits to 1s when inserting, and concludes that $x\mathrm{\in} S$ if at least $q\mathrm{\leq}t\mathrm{\leq}k$ bits are non-zero. To delete a flow $x$, at least $k\mathrm{-}d$ of its signatures are set to 0s, where $d\mathrm{<}q$ so that $x$ will not be declared as a member of flow set $S$. Moreover, the missing bits of a positive flow may be recovered by setting them to 1; thus, recovery strengthens or lengthens signatures. In practice, some flows stored by the bit vector may be out-of-date. Consequently, the aging mechanism is designed to set a part of 1s in the bit vector to 0s, in an either round-robin or random manner.

Note that, in reality, variable length signature enabled BF naturally incurs false positive errors. Additionally, false negative errors may also occur due to the aging operations. Fortunately, the value of $t$, as well as $q$ can be learned according to the flow distribution, so that the generated false positive and false negative errors can be significantly reduced. 

\textbf{Dynamic BF (DBF).}  The basic insight of DBF \cite{Dynamic_BF} is to reserve space for $s$ homogeneous BFs. Initially, only one BF is active and once the current BF is full, another untapped BF will be activated. According to the predefined upper-bound of false matching probability, the parameters of BFs, as well as the upper bound of recored elements in each BF, can be reasonably determined. The BF is full if the number of inserted elements reaches the upper bound. To insert an element, DBF first discovers an active BF and then set the corresponding $k$ bits as 1s. By doing so repeatedly, all elements in a dataset can be recorded. To answer the membership query of arbitrary element $x$, the DBF checks the BFs one by one. If in any BF, the corresponding $k$ bits are all non-zero, DBF believes $x\mathrm{\in} S$. If all the BFs report negative results, then DBF concludes $x\mathrm{\notin} S$. 

By contrast, deleting an existing element $x$ from DBF can be a little more complicated. First of all, DBF identifies the BFs in which all the $k$ bits for $x$ are non-zero. If only one such BF exists, DBF resets these bits as 0s. Otherwise, DBF quits the deletion operation. The reason is that, if there are multiple BFs indicate the existence of $x$, DBF cannot decide which is the right one. Therefore deleting $x$ from these BFs will lead to false negative judgments when querying other elements. Beyond deletion, DBF also enables the functionality to merge two active BFs via union operations. 

An obvious weakness of DBF is its high false positive rate. On one hand, the multiple BF mechanism increases the risk of false positive matching. Suppose the designed false positive rate of each BF in DBF is $f_r$, then the false positive rate of DBF will be calculated as $1\mathrm{-}(1\mathrm{-}f_r)^s$ which is definitely higher than $f_r$. On the other hand, the failures of deletion lead to false positive query results of the elements that should be deleted.

\textbf{Scalable BF.} Unlike DBF, Scalable BF \cite{Scalable_BF} consists of a series of heterogeneous BFs. That is, the parameters of each sub-BF may be differentiated. Specifically, to maintain a given false positive rate, the designed false positive rate of the successive sub-BFs are increased geometrically. To this end, the length and bit utilization of the sub-BFs are varied and well-designed. However, as a consequence, the sub-BFs cannot share the hash functions, which leads to more hash calculations. Besides, the sub-BFs with diverse lengths cannot be joined as one to save space. In the worst case, some sub-BFs may only record one element after the deletion of other elements, which is definitely not space-efficient.

\textbf{Dynamic BF Array (DBA).} DBA \cite{DBF_Array} dedicates to realize a scalable and space-efficient approximate data structure for storage systems. DBA consists of groups of BFs on demand. Within each group, there are $g$ homogeneous BFs. In this manner, DBA naturally support parallel queries, since the access of one group of BFs will not affect others. Therefore, the memory-access complexity for querying an element is $O(k\mathrm{\times}r/(2g))$ for a positive result and $O(k\mathrm{\times}r/g)$ for a negative result, where $r$ is the total number of BFs. As for deletion, the authors point out that large-scale storage systems always delete the out-of-date elements in off-peak time. Therefore, DBA  sets a predefined threshold for the number of stale elements in each BF and deletes these stale elements in a batched fashion.
 
DBA also suggests loading the entire DBA into the RAM so that the query can be quickly responded. To eliminate potential false positive errors, the positive responses of DBA will be further checked with the index of the dataset in the RAM. Basically, DBA is quite similar with DBF. The false positive rate of DBA is also $1\mathrm{-}(1\mathrm{-}f_r)^r$, where $f_r$ is the false positive rate of each BF, and $r$ is the total number of BFs.  The false positive rate can be controlled, but at the cost of more space overhead by prolonging the length of each BF or decreasing the number of BFs. What's worse, the batched deletion scheme is not feasible for the online application which requires to add or delete elements frequently and timely.

\textbf{Par-BF.} Par-BF \cite{Par_BF} argues that both Dynamic BF and Scalable BF are not compact enough, in terms of system performance and memory overhead. Dynamic BF cannot control the overall false positive, while Scalable BF suffers from higher memory cost and vast hash calculations. Like DBA, Par-BF is made up of sub-BF lists, each of which consists of multiple homogeneous sub-BFs. The parameters of each sub-BF are carefully designed to achieve the overall false positive rate guarantee. Par-BF supports parallelism of membership queries. To insert an element, the Par-BF locates the active sub-BF and set the corresponding bits to 1s. Deletion will be accomplished by resetting the corresponding bits to 0s in the sub-BF. Besides, the garbage collection scheme is also proposed to unite the unfilled sub-BFs for space recycling.

Actually, DBF \cite{Dynamic_BF}, Scalable BF \cite{Scalable_BF}, DBA \cite{DBF_Array}, and Par-BF \cite{Par_BF} all share the insight that, with the coming elements, the additional BFs can be activated dynamically. However, this strategy intrinsically calls for space reservation since the number of elements in the dynamic dataset is unknown in advance. Besides, the aggregated false positive rates of these data structures increase with the number of initialized BFs. Furthermore, the time-complexities of query and deletion of an arbitrary element are no longer $O(k)$ since they must travel all the sub-BFs to make a confident decision. 

\subsection{Weighted sets}
BF is designed for general representation of a dataset, with elegant support of membership query. In practice, the sets and queries in distributed systems can be highly skewed \cite{Weighted_BF}\cite{Popularity_Conscious_BF}. That is, some elements may be queried frequently, while some may not. With the query distribution and membership likelihood distribution as prior knowledge, the original BF framework can be further improved \cite{Weighted_BF} and \cite{Popularity_Conscious_BF}. 

\textbf{Weighted BF.}  Weighted BF \cite{Weighted_BF} proposes to adjust the number of employed hash functions for different elements, based on the query frequency and likelihood of being a member of set $S$. Intuitively, an element is assigned more hash functions if its query frequency is high and its chance of being a member is low. In particular, the number of hash functions for an element $x$ is jointly determined by the normalized query frequency of $x$ and the probability of $x$ to be a member of set $S$.
Weighted BF is a generalization of BF since when the frequency distribution and membership likelihood of every element is the same, the generated Weighted BF will be a regular BF. However, the problem is that, before querying an element $x\mathrm{\in} U$, the value of $k_x$ has to be calculated. If $x$ is queried frequently, $k_x$ will be recalculated repeatedly, which is definitely not efficient. Besides, the value of $k_x$ is not convergent. If an element only occurs in query but never appears to be a member of $S$, then $k_x \mathrm{=} \infty$; while  $k_x\mathrm{=} \mathrm{-}\infty$ when $x$ is a member of $S$ but has never been queried.

\textbf{Popularity Conscious BF.} Unlike Weighted BF, Popularity Conscious BF \cite{Popularity_Conscious_BF} predefines $k_{max}$ for the number of hash functions and thereafter profiles the problem of allocating hash functions to each element as a nonlinear integer programming problem. The target is to achieve the smallest overall false positive rate, with respect to the predefined parameters of BF. To this end, two approximate algorithms are proposed based on the concept of per-object importance score. Specifically, per-object importance score is calculated as $\frac{q'(i)}{p(i)}$, where $q'(i)$ is the probability that element $x_i$ occurs in a mis-matched query (false positive error), and $p(i)\mathrm{=} \frac{Pr(x_i\mathrm{\in} S)}{n}$ denotes the membership popularity distribution. Based on the insight that the optimal solution for the number of per-object hashes must follow the order of importance scores, a polynomial-time 2-approximation algorithm is designed. Further, to speed up the calculation, a $(2\mathrm{+}\varepsilon)$-approximation algorithm is proposed.  

We argue that both Weighted BF and Popularity Conscious BF might not be scalable from the time dimension. For online applications, both the query frequency and membership likelihood may vary from time to time. As a result, the value of $k_x$ calculated when inserting maybe not optimal in other time slots. Additional efforts are still needed to totally settle this problem so that this kind of variant is pratical. 

\subsection{Key-values}
Key-value (KV) store systems (e.g., Dynamo \cite{Dynamo}, Memcached \cite{Memcached}, Cassandra \cite{Cassandra}, Redis \cite{Redis}, BigTable \cite{BigTable}, etc.) handle numerous keys and corresponding values, with the supporting of key lookup and KV pair insertion. Usually, index structures (e.g., B-Tree\cite{Btree}, B+ Tree \cite{Bptree}, Hash Table\cite{HashTable}) are built for fast and deterministic access of the values. However, due to the space limitation, saving the entire index structure in RAM can be challenging when store large number of KV pairs. The BF is indeed space-efficient, but not designed for KV stores. The major obstacles to employ BF to memorize KV pairs come from two folds. First, the number of KV pairs are dynamic, hence the parameters of BF cannot be decided previously. Second, KV stores call for smooth deletion of KV pairs, but BF fails to support this functionality. To overcome this dilemma, BloomStore \cite{BloomStore}, kBF \cite{KBF}, and Invertible Bloom lookup table \cite{IBLT} are proposed reasonably.

\textbf{BloomStore.} BloomStore \cite{BloomStore} establishes multiple BloomStore instances in the KV store system. The space of keys is divided into multiple disjoint ranges, each of which is mapping into a dedicated BloomStore instance. Every BloomStore instance consists of four components, i.e., KV pair write buffer,  BF buffer, BF chain, and a number of data pages. KV pair write buffer and BF buffer are saved in RAM for fast access, while BF chain and the data pages are stored in the secondary memory (i.e., SSD, flash). The KV pair write buffer memories the incoming KV pair insertions temporarily. Once the write buffer is filled up, the content will be written into the secondary memory. In this manner, the number of write operations will be lowered. Correspondingly, the BF buffer represents the membership of the keys stored in the KV pair write buffer. Other KV pairs are represented by the Bloom chain (a series of BFs) in secondary memory. Note that each BF is responsible to memory the KV pairs in one flash page.

Based on the above framework, a parallel BF checking scheme is employed to speed up the query process. Notice that, the parallelization is possible because all the BFs in the chain are homogeneous and thus share the hash outputs. Apparently, BloomStore is naturally scalable and supports deletion by insert a \emph{null} value to the key in the corresponding BF bits. Sure, the aggregated false positive possibility is relatively high. 

\textbf{kBF.}  Instead of saving the KV pairs directly, kBF \cite{KBF} converts the values into fixed-length string bits with respect to given constraints. kBF consists of cells, each of which has two components, i.e., the counter to track the number of encodings into the cell, and the encoding field records either an original encoding, or the XOR results of the encodings that are mapped to this cell. To insert a KV pair, the value part is first converted to be a string bits and denoted as encoding. Thereafter, the keys and corresponding encodings are inserted into the kBF for later query, update, deletion, join and compress operations. Moreover, the original KV pairs can be inversely decoded from the generated kBF cells. In fact, for a KV pair, if any counter in the corresponding cells is 1, the encoding field will be exactly the encoding string of the value. Otherwise, complicated decoding algorithm is employed. The basic insight is to construct a BF which records all the encodings, and then execute the XOR operation of the encodings literately to check whether the XOR results contains the encoding to be decoded. A distributed version of kBF is also suggested for cloud computing scenarios.

Intrinsically, the conversion from values to encodings formalizes the disordered values as fixed-length bit strings, so that the later update, deletion, join and compress operations can be enabled reasonably. However, the cost is also very high. First, the time-complexity of the proposed decoding algorithm is $O(N^2)$, which is unacceptable for delay-sensitive online applications. The decoding algorithm also suffers from non-decodable cells and may fail to derive the original values from the intersection encodings. Second, the conversion scheme lacks of scalability, since the upper bound of the encodings must be predefined according to the number of KV pairs. Third, kBF calls for additional secondary BFs to speed up the searching of encodings or decode the encodings inversely. Fourth, kBF may suffer from false negatives, if all the $k$ corresponding cells for a KV pair fail to tell the encoding of its value.

\textbf{Invertible Bloom lookup table (IBLT).} IBLT \cite{IBLT} supports not only insertion, deletion, and lookup of key-value pairs, but also allows a complete listing of the pairs it contains with high probability, as long as the number of contained key-value pairs is below a designed threshold. Conceptually, IBLT consists of $m$ cells, each of which contains the following three fields: 1) $count$ field which counts the number of KV pairs that mapped into this cell; 2) $keySum$ which records the sum of all keys that mapped into this cell; and 3) $valueSum$ which is the the sum of all values that mapped into this cell. Note that IBLT treats both keys and values as integers for simplicity. To insert a KV pair $(x,y)$, $k$ hash functions are employed to map this pair into $k$ disjoint cells. In each of the corresponding cell, the $count$ field will be increased by 1; $keySum$ will be updated as $keySum\mathrm{+}x$; and $valueSum$ will be calculated as $valueSum\mathrm{+}y$. The deletion will be accomplished by executing inverse calculations of insertion. To query a key $x$, the IBLT will return the associated value $y$ or ``null'' or ``not found''. In the $k$ corresponding cells of $x$, if a cell which only records the key $x$ is found, IBLT will return the value. If any cell is empty or the $count\mathrm{=}1$ but the $keySum\mathrm{\neq} x$, the answer will be ``null''; otherwise, IBLT will return ``not found''. The result ``not found'' means all the $k$ cells store multiple KV pairs and IBLT cannot tell $(x,y)$ is among them or not. 

Additionally, IBLT is capable of listing all the KV pairs successfully with high probability. The algorithm searches out an anchor cell which only records one KV pair, and then returns the associated key and value. Thereafter, the selected pair will be deleted from the cell vector to expose more anchor cells. Recursively, all the KV pairs may be decoded. Definitely, IBLT can only list part of the KV pairs, when no such anchor cells can be found.

\subsection{Sequence sets and spatial sets}
Except for the above types of datasets, there are also other specific datasets with special characteristics. Recently, $k$-mers (substrings with equal length derived from the full sequence) are employed to support high-level algorithms \cite{Kraken}\cite{Velvet} towards the sequence data, e.g., DNA or RNA sequence. However, a single sequence may generate a massive number of $k$-mers, and thereby causing storage challenges. Consequently, BF is introduced to record the $k$-mers and support fast membership query. Due to the characteristics of sequence data, the application of BF can be further improved \cite{k-mer_BF} \cite{Trie_BF0}. Moreover, in the context of location based applications, spatial data should be represented efficiently for query. Spatial BF \cite{Spatial_BF} tries to represent the geographical areas with priorities. 

\textbf{$k$-mer BF.}  Pellow et al. \cite{k-mer_BF} point out that the internal dependency between the mapped subsequences can be utilized to reduce the false positive rate of membership query. Specifically, two adjacent $k$-mers share $k\mathrm{-}1$ common characters. Therefore, the presence information of neighbours can be employed to further judge the membership of the query substring. In this manner, some false positive errors will be identified via additional queries of the neighbours. For instance, consider a sequence TAAGCCA and it is stored as 4-mers, i.e., TAAG,  AAGC, AGCC, and GCCA. When querying AGCC, if the BF returns positive for AGCC, while reports negative for GCCA, then the BF experiences a false positive error with high probability. Note that BF never suffers from false negative errors, and it is 100\% true that an element $x\mathrm{\notin} S$ if the query result is negative. $k$-mer BF offers two optional schemes, i.e., one-sided $k$-mer BF and  two-sided $k$-mer BF. The one-sided $k$-mer BF only checks the presence of a single overlapping neighbour. By contrast, the two-sided $k$-mer BF which checks the neighbours from both directions. Surely, checking more neighbours will further lower the FPR.
 
$k$-mer BF seems compact enough to represent sequence sets. However, the functionality of $k$-mer BF is one-fold only, i.e., membership query. Many complicated algorithms call for comprehensive operations upon these $k$-mers, e.g., traverse, deletion, permutation, inversion, etc. $k$-mer BF fails to enable these operations, which in turn narrows its usefulness.  

\textbf{Spatial BF.} Spatial BF \cite{Spatial_BF} considers the location information with the concern of privacy protection. To this end, the locations of the users are divided into different interest areas. According to the Manhattan distance, the areas are categorized as diverse classes. Before inserting any element, each cell of the  Spatial BF is initialized as 0. Then $k$ hash functions map the element $x$ (which belongs to class $L$) into the cells, and set the values as $L$. By sorting the locations with increasing order, the locations with greater class labels will be inserted later.  Consequently, the locations with greater class labels will be correctly memorized, even if a hash collision occurs. Thereafter, the constructed Spatial BF will be implemented in location-aware applications based on the designed private positioning protocols.

Spatial BF is a variant of BF to represent spatial data which consists of many geographical areas. However, abstracting the spatial data as simple interest areas is oversimplified and thereby failing to support high-level applications such as navigation. Besides, the location of a user is highly time-dependent due to his or her motions. Spatial BF fails to track the motions if a user always moves in one class of areas.

\subsection{Summary and lessons learned}
According to different abstraction of the dataset, diverse variants are proposed to match the characteristics and functionality requirements of the elements. Thanks to these novel proposals, the BF paradigm can be extended to represent and support multiform operations towards multisets \cite{Spectral_BF} \cite{ICBF} \cite{Loglog_BF}, dynamic sets \cite{Variable_length_BF} \cite{Dynamic_BF} \cite{Scalable_BF} \cite{DBF_Array} \cite{Par_BF}, skewed datasets \cite{Weighted_BF} \cite{Popularity_Conscious_BF}, KV pairs \cite{BloomStore} \cite{KBF}, sequence sets \cite{k-mer_BF} \cite{Trie_BF0}, spatial data \cite{Spatial_BF}, etc. These variants may lack of generality, but function well in their own domains. In practice, the users of BFs can customize their BF variants to represent their datasets with diverse features. This malleability further extends the usage of BFs in various contexts.

\section{Functionality enrichments}\label{sec:Functionality}
Standard BF only supports membership query, since the bit vector only indicates the membership information of each element. Therefore, from the framework of BF, with the input set, the output functionalities can be enriched in diverse scenarios. 

\subsection{Element deletion}
In real systems, an element may be deleted from the dataset and other elements may be added. A general drawback of standard BF is that it fails to delete elements from the bit vector, since resetting the bits from 1 to 0 may cause false negative errors to other elements. Besides of the CBF \cite{Counting_BF}, Deletable BF  \cite{Deletable_BF} and Ternary BF \cite{Ternary_BF} are proposed to enable the deletion of arbitrary element.

\textbf{Deletable BF (DlBF).} In the bit vector of BF, hash collisions may happen, so that the deletion of an element may lead to false negative errors to other elements. Based on this insight, Deletable BF \cite{Deletable_BF} tracks the positions where hash collisions happen when inserting elements. Then only the bits in collision-free areas will be reset from 1 to 0 when deleting an element. DlBF divides the $m$ bits in the bit vector into $r$ regions, each of which has $m'/r$ bits, such that $m'\mathrm{=}m\mathrm{-}r$. Before the $r$ regions, the first $r$ bits in the bit vector are employed to identify the corresponding region is collision-free or not. Note that, the first $r$ bits are all initialized as 0s. If a hash collision happens in the $i^{th}$ region, the $i^{th}$ bit in the first $r$ bits will be set to 1. Both insertion and query will be accomplished just like the standard BF. When deleting, the element $x$ will be mapped into $k$ bits. Then the algorithm only reset the 1s located in collision-free regions. The element $x$ will be successfully deleted if at least one of the $k$ corresponding bits is reset to 0. However, if all the $k$ bits locate in conflicted areas, the element is declared as non-deletable. Let $p_0=(1-1/m')^{kn}$ and $p_1=kn/m'(1-1/m')^{kn-1}$ denote the probability that a given cell is set to 0 and 1 only once after inserting $n$ elements. An arbitrary element can be deleted successfully with approximate probability $(1-(1-p_c)^{m'/r})^{k}$, where $p_c \mathrm{=} 1\mathrm{-}p_0\mathrm{-}p_1$ is the probability that a given cell has at least one hash collision.

\textbf{Ternary BF (TBF).} Ternary BF \cite{Ternary_BF} allocates the minimum number of bits to each counter and thereby implies more counters. Each counter in the proposed TBF has three different values: 0, 1, and X. If 2 or more elements are mapped into a counter, the value is programmed as X and the counters with value X are not referenced in queries. Compared with the 4-bit CBF, a TBF vector has larger number of counters using the same amount of memory, since each counter of the TBF occupies less bits. 

Note that the counters with X are not increased when inserting an element, nor decreased when deleting an element. They are also not referenced when querying an element. Consequently, TBF derives the \emph{indeterminable} elements for membership query and \emph{undeletable} elements for deletion. To be specific, when querying an element $x$, if the corresponding $k$ counters of $x$ are all X, then $x$ is identified as \emph{indeterminable}. In contrast, when deleting element $y$, if the corresponding $k$ counters of $y$ are all X, then $y$ is considered as \emph{undeletable}. By labelling the counters where hash collisions occur, and defining the \emph{indeterminable} and \emph{undeletable} elements, TBF significantly reduces both the false positives for query and false negatives due to misdeletions. Namely, TBF is a conservative design. It prefers correct answers or operations, while leaving the controversial counters (the counters with value X) alone. Moreover, to lessen the number of  \emph{indeterminable} or \emph{undeletable} elements, the value of a counter can be adjusted as: 0, 1, 2 and X, such that more elements can be queried or deleted. Surely, doing so may increase the false positive rate.

Besides of CBF \cite{Counting_BF}, Deletable BF \cite{Deletable_BF} and Ternary BF \cite{Ternary_BF}, other variants which have a counter filed in the cells (e.g., SCBF \cite{Space_Code_BF}, Spectral BF \cite{Spectral_BF}, ICBF \cite{ICBF}, IBLT \cite{IBLT}, etc) also enable the deletion operation successfully. Consider that deletion is not the prime functionality or goal of these variants, we present them in other sections.

\subsection{Element decay}
For online services and applications, new data will come and stale elements will banish. Therefore, to represent a dataset more efficiently, the BF should also support decay operation, such that those stale information can be eliminated to leave space for the coming elements. The difference between deletion and decay is that, decay is a proactive operation and will be executed by the BFs periodically or aperiodically, while deletion is an inactive function that is called by the users. Consequently, during deletion, the BFs know exactly which element should be removed. In contrast, when decaying, the BFs don't acknowledge which and how many elements will be removed from the vector. Stable BF \cite{Stable_BF}, Temporal Counting BF \cite{TCBF}, Double buffering \cite{Double_Buffering}, $A^2$ buffering \cite{Two_Active}, and Forgetful BF \cite{Forgetful_BF} are designed to enable decay of the recorded elements.

\textbf{Stable BF.}  Stable BF \cite{Stable_BF} is proposed for duplicate elimination of streaming data flows which arrive continually. Stable BF is defined as an array of cells with integer range from 0 to $Max$. The number of bits for each cell is $b$ such that $Max\mathrm{\leq}2^b\mathrm{-}1$. To check whether a recently arrived element $x$ is a duplicate or not, $x$ is mapped into $k$ cells in Stable BF. If all of the $k$ cells are non-zero, $x$ is considered as a duplication; otherwise not. To insert an element into Stable BF, $P$ randomly selected cells are decreased by 1. Thereafter, the corresponding $k$ cells for the element is set as $Max$. In this way, the whole cell vector will be stable, i.e., after numerical iterations, the fraction of zeros in the Stable BF will become fixed, irrespective of the initial state. In the framework of Stable BF, the number of cells which are 0s after $N$ iterations is a constant when $N$ is large enough. 
Note that, Stable BF provides both false positive and false negative errors. Evicting the stale information in a random fashion certainly results in false negative errors. However, the duplicate detection will be handled with constant time, so that Stable BF suits the scenarios of streaming data or other series datasets well.

\textbf{Temporal Counting BF (TCBF).} TCBF \cite{TCBF} is an extension of CBF. Upon inserting an element, the $k$ corresponding counters will be set as a given value called initial counter value (ICV), rather than increased by 1. If a hash collision occurs, TCBF will remain the existing value of the counter. The decay mechanism is to constantly decrease all the counters' values with a rate given by the decay factor (DF). By tuning the values of ICV and DF, TCBF enables differentiated decay granularity for different elements. Additionally, TCBF offers two kinds of merging schemes for two TCBF vectors, i.e., additive merging and maximum merging. Additive merging generates a new TCBF vector whose counters are the sum of the counters in the original vectors. By contrast, the maximum merging set the counters as the maximum value of the counters in the original vectors. The membership query will check the $k$ corresponding cells, and the caused false positive rate is the same as standard CBF. 

Note that, TCBF predefines the ``time-to-live'' of each element in the vector by assigning the ICV value. However, false negatives will occur when one of the $k$ counters remains of a smaller value than its ICV. For example, $x$ is mapped into cells 1, 5 and 9 in the TCBF vector, and the ICV and DF are 7 and 1 respectively. But cell 5 has already been set as 2 by a former element. In this case, the counters in cells 1, 5 and 9 will be 7, 2 and 7, respectively. Therefore, after two evictions, the membership query of $x$ will be negative since cell 5 has been degraded as 0. A possible repairment is to keep the maximum value among the existing value and the inserted ICV, if hash collisions occur. This scheme, despite eliminating false negatives, imposes more false positive errors for the membership query.

\textbf{Double Buffering.} Double buffering \cite{Double_Buffering} offers an active/standby scheme to evict the stale elements in a first-in-first-out manner. The memory is physically divided into two independent parts to maintain an active  BF and a warm-up BF, respectively. For a coming element $x$, if $x$ is already recorded by the active BF, and the active BF is more than 1/2 full, then $x$ will be inserted into the warm-up BF. On the contrast, if $x$ hasn't been recorded by the active BF, $x$ will be inserted into the active BF. After that, if the active BF is more than 1/2 full, $x$ will be additionally inserted into the warm-up BF.  Once the active BF is filled up, double buffering switches the active BF and warm-up BF, and flushes the generated warm-up BF. With such a design, the active BF stores all the recent data and the warm-up BF always records a subset of the active BF. An essential drawback of this scheme is that it doubles the memory space to store the same number of elements, as well as the memory access of inserting a new element.

\textbf{$A^2$ Buffering.} $A^2$ buffering \cite{Two_Active} divides the memory equally as two buffers which implement two BFs denoted as $active1$ and $active2$, respectively. Novelly, $active1$ stores the recently inserted elements, while $active2$ records previously inserted elements. When $active1$ becomes full, $active2$ is flushed and the two BFs switch their roles. Basically, a queried element $x$ will generate a positive answer if it passes the membership check in either $active1$ or $active2$. Specifically, $x$ is first queried against $active1$. Only when $active1$ returns a negative answer, $active2$ will be accessed and queried. If $active2$ returns a positive result, the element $x$ will be inserted into $active1$. Suppose the false positive rate of each BF is $f_r$, the global false positive rate is $f\mathrm{=}1\mathrm{-}(1\mathrm{-}f_r)^2$. With this scheme, the users can update BFs with recently used data. $A^2$ buffering stores twice as many elements as double buffering in the best-case scenarios, and as many elements as double buffering in the worst-case scenarios.

\textbf{Forgetful BF.} The most-recently proposed Forgetful BF \cite{Forgetful_BF} maintains several active BFs simultaneously, including one future BF, one present BF and one or multiple past BF(s). A new-arrival unrecorded element will be inserted into both the future BF and the present BF. Forgetful BF periodically refreshes the BFs to evict the stale elements. Specifically, 1) the oldest past BF is dropped for eviction; 2) the current present BF is turned into the newest past BF; 3) the current future BF is degraded as present BF, and lastly 4) a new, empty future BF is added. The membership query proceeds by checking for membership in the most recent window of time (the future BF) until the oldest window of time (the oldest past BF) is reached. If the membership check passes in any window of time (two consecutive BFs return positive), the Forgetful BF terminates the query process and return positive. Inherently, since each element is inserted twice, the elements that only pass one membership query are identified as false positives. Moreover, to control the false positive rate, Forgetful BF proposes to adjust its number of past BFs and the refresh frequency. However, Forgetful BF also has problems with a large number of memory accesses inherited from the structure of multiple active BFs.  

Generally speaking, the decay can be enabled by changing the values of bits in a single BF vector (e.g., Stable BF \cite{Stable_BF}, Temporal Counting BF \cite{TCBF}) or flushing the oldest BF among multiple BF vectors (e.g., Double buffering \cite{Double_Buffering}, $A^2$ buffering \cite{Two_Active}, and Forgetful BF \cite{Forgetful_BF}). Decay results in false negative errors thereby triggering the reinsertions of some elements. 

\subsection{Approximate membership query} \label{subsubsec:AMQ}
As an opposite of accurate membership query,  approximate membership query is to answer the query ``Is $x$ close to an element in $S$?'' when the closeness is measured by a dedicated metric. This kind of query can find its applications in databases, networking, image processing, social networks, and biological scenarios. The existing proposals usually employ the locality-sensitive hash family \cite{locality_hashing} \cite{locality_hashing1} \cite{Multigranularity_LBF} to map the elements so that the close elements will be stored in neighbouring or same cells with high probability. 

\textbf{Distance-sensitive BF.} Distance-sensitive BF \cite{DS_BF} employs local-sensitive hash functions based on the distance to map elements into the vector. Distance-sensitive BF is an array (denoted as \emph{DSBF}) of $k$ disjoint arrays, each of which has $m'$ bits, such that the total number of bits is $m\mathrm{=}k\mathrm{\times} m'$. Each hash function is responsible to one dedicated array. Let $\mathcal{H}$: $U\mathrm{\rightarrow} V$ be a $(p_L,p_H)$-distance sensitive hash function family, and $\mathcal{H'}$: $V\mathrm{\rightarrow} [m']$ be a weakly pairwise independent hash family. Before programming the arrays, a group of hash functions $\{h_1,\cdots, h_k\} \mathrm{\in} \mathcal{H}$, as well as another group of hash functions $\{h'_1,\cdots, h'_k\} \mathrm{\in} \mathcal{H'}$, are selected independently. If $V\mathrm{=}[m']$, $g_i\mathrm{=}h_i$; otherwise, $g_i\mathrm{=}h_i \mathrm{\circ} h'_i$; Thereafter, for $x\mathrm{\in} S$, $DSBF_i[g_i(x)]$ ($i\mathrm{\in} [1,k]$) will be set as 1. As for query, Distance-sensitive BF refers to the bit locations of an element $x$ in the vector, to test whether $x$ is close to some elements in $S$ or not. Let $\mathcal{B}(x) \mathrm{=} \sum_{i\in[1,k]} DSBF_i[g_i(x)]$. If $\mathcal{B}(x)$ is larger than a predefined threshold $t$, $x$ is far away from all elements in $S$; otherwise, $x$ is close to some elements in $S$. 

Practically, the Distance-sensitive BF is complex to be implemented. Firstly, the local sensitive hash functions are hard to define for distance metrics. Secondly, the configuration needs more computation cost since two families of hash functions are involved. Thirdly, this paradigm lacks of scalability. The parameters for the current dataset are customised and cannot be generalized to other datasets. Besides, Distance-sensitive BF is constructed based on the assumption that $S$ is static, which further restricts its scalability. Therefore, a lightweight and more scalable solution may still be required.

\begin{figure}
  \centering
  \subfigure[False positives from multidimensional inconsistency.] {\includegraphics[width=1.7in,scale=0.5]{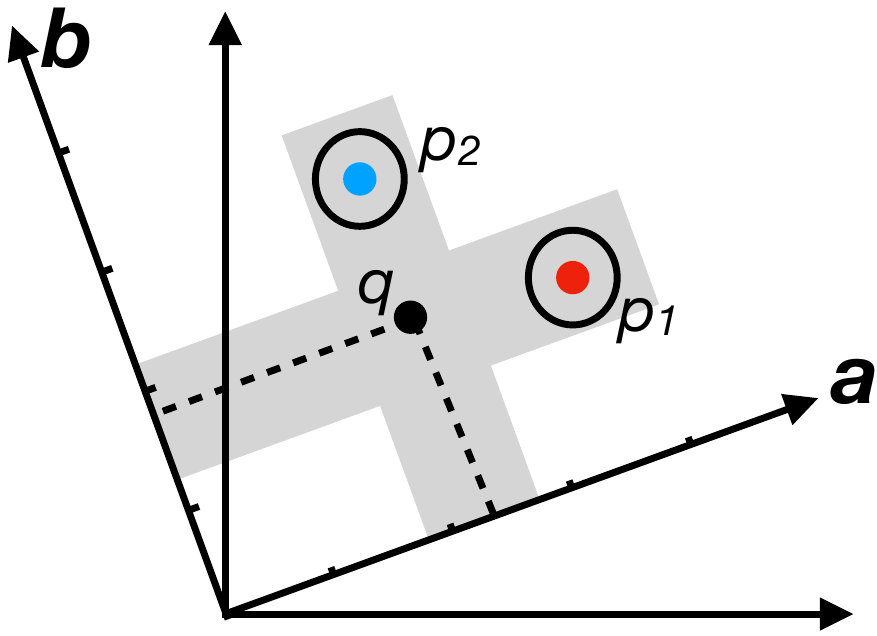}}
  \subfigure[False negatives from multidimensional checking.] {\includegraphics[width=1.7in,scale=0.5]{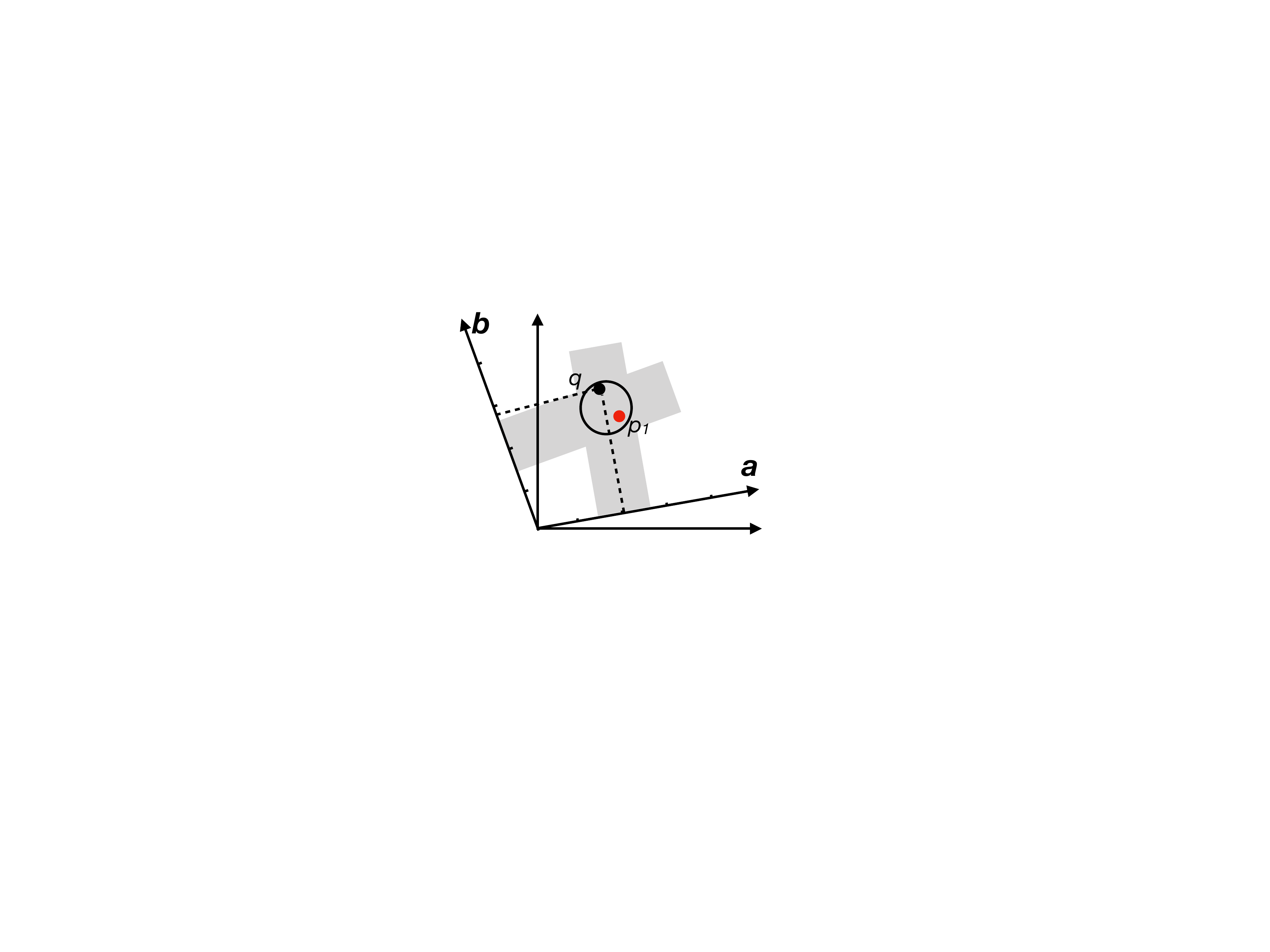}}
  \caption{An example of false positive and false negative errors in LSBF \cite{LSBF}.} \vspace{-0.15in}
  \label{fig:LSBF}
\end{figure}

\textbf{Locality-sensitive BF (LSBF).} LSBF \cite{LSBF} also replaces the independent uniform hash functions in BF with locality-sensitive hash functions, thus support the approximate membership query. Unlike Distance-sensitive BF, LSBF concludes that $x$ is close to an element in $S$ only when all the corresponding bits are non-zero. This design further restricts the condition to pass the check. Intrinsically, the FPs are partially inherited from the BF framework. Another inexplicit reason of FPs is the lack of multidimensional attributes of elements. As depicted in Fig. \ref{fig:LSBF} (a), the queried point $q$ is projected on two dimensions, i.e., vectors $\textbf{a}$ and $\textbf{b}$. LSBF fails to tell whether the approximate membership of $q$ in each dimension comes from an existing item or from multiple items. As a consequence, $q$ will be concluded as an approximate member, although $q$ is far away from both the points $p_1$ and $p_2$. By contrast, the reason for false negative errors is that, the locality-sensitive hash functions can map close elements into neighbouring cells with high probability but not 100\%. Therefore, in Fig. \ref{fig:LSBF} (b), the queried $q$ is close to point $p_1$, but the distance in dimension $\textbf{b}$ is too far so that one bit for $q$ in the bit vector will be far away from that of $p_1$ and maybe a 0. In this case, LSBF will return a negative query result.

LSBF proposes a verification scheme based on an additional BF to recognize part of the false positive errors, as well as an overflowed scheme to decrease false negative errors.
Besides the introduced computation and space overhead, the configuration of LSBF is also time-consuming, since the verification BF can only be established after all the elements have been inserted into the LSBF vector. The reason is that the neighbours of any bit may change from 0 to 1 due to later insertions. This limitation also indicates weak scalability of LSBF, since each coming element will cause the reconstruction of the verification BF. 

\textbf{Multi-granularity locality-sensitive BF (MLBF).} The using of locality-sensitive hash functions has been further improved in MLBF \cite{Multigranularity_LBF}, by enabling multiple distance granularities. MLBF contains a basic multi-granularity locality-sensitive BF (BMLBF) and a multi-granularity verification BF (MVBF). Specifically, the BMLBF is a vector of $m$ bits and constraints $s\mathrm{-}1$ virtual LSH BFs (VLBFs) which are not physically constructed but derived from the BMLBF (VLBF 0). One bit in VLBF $\theta$ ($1\mathrm{\leq} \theta \mathrm{\leq} s\mathrm{-}1$) covers $2^{i}$ bits in VLBF $\theta\mathrm{-}i$ for any $0\mathrm{\leq} i \mathrm{\leq} \theta$. If one of the locations in the VLBF 0 is set to 1, its covered virtual location in each of the VLBFs is regarded as 1 too. $L$ AND-constructions (to improve the locality-sensitiveness together with later OR-constructions), each of which has $k$ hash functions, are employed to map an element into BMLBF. Therefore, $L\mathrm{\times} k$ hash functions are needed. For each group of hash functions, the $k$ generated locations are concatenated together and then stored in the MVBF. Although the MVBF only stores the concatenations from the BMLBF, it can still support multi-granularity verification by constructing virtual verification BFs (VVBFs). The reason is that the address of a location at the coarser level is a prefix of the two addresses of its contained locations at the next finer level. 

To query an element $x$ in the $\theta$ level, MLBF checks the VLBF $\theta$ first, and then the MVBF.  If there is a group of hash functions such that both its BMLBF and MVBF return positive results, MLBF judges $x$ is close to some element in the set $S$. Despite the introduced AND-constructions and verification scheme, MLBF still suffers from both false positive and false negative errors. Moreover, MLBF is overcomplicated to construct and query. $L\mathrm{\times} k$ hash functions are required for the BMLBF and $k'$ hash functions are needed for the MVBF. 

Generally, the locality sensitive hash functions are employed to map the elements into the bit vector. Distance-sensitive BF \cite{DS_BF} infers an element $x$ is close to some element in $S$ if the number of 1s in the bit vector exceeds a predefined threshold. LSBF \cite{LSBF} and MLBF \cite{Multigranularity_LBF}, however, concludes $x$ passes the approximate membership query only if all the $k$ corresponding bits are non-zero. Besides, LSBF and MLBF also implement the verification schemes to further identify the false-positively matched elements. Especially, MLBF further enables the query and verification in different granularities.  

\subsection{Enrichment of BF semantics}
An inborn drawback of BF is that it only saves the membership information while lacking other features of the elements. Therefore, the enrichment of BF semantics is needed to support more types of queries. From the perspective of elements, the information saved in the bit (or cell) vector can be extended in either a scale-up or a scale-out manner. A scale-up scheme programmes not only the membership information, but also other features associated with the element. By contrast, scale-out proposals care about the internal relationship between elements. As shown below, Invertible BF \cite{IBF}, Bloomier filter \cite{Bloomier_BF}, and Parallel BF  \cite{Parallel_BF} try to store more information.

\textbf{Invertible BF (IBF).} With the assumption that each element has a unique binary identifier, an IBF \cite{IBF} vector for a dataset contains $m$ cells and each cell has three fields. The \emph{idSum} records the XOR results of the identifiers of elements that mapped into this cell. The \emph{count} counts the number of elements that mapped into this cell. IBF cell additionally employs a hash function $g(x)$ to map the identifiers of the corresponding elements into a fixed-length binary string. Thereafter, the XOR result of hash values $g(x)$ is saved as \emph{hashSum} field. The encoding, subtracting and decoding algorithms are designed to settle the set-reconciliation (or set difference) problem. Encoding algorithm represents the two sets as two IBF vectors, and then subtracting algorithm operates the two IBFs to generate a new IBF vector which only contains different elements. Lastly, the decoding algorithm tells which elements are unique for set $A$ and $B$. Note that, the identifier, the \emph{idSum} and \emph{hashSum} fields are all binary bit strings. Consequently, the simple XOR bit operations are implemented for both aggregating and removing elements.

IBF \cite{IBF} and IBLT \cite{IBLT} are similar and try to decode the elements from the cell vector inversely. Both of them select a cell which only records one element as an anchor and thereby return the elements one by one. The difference is that IBLT is designed for KV store while IBF suits general datasets well by turning the \emph{valueSum} field in IBLT as \emph{hashSum} field. Moreover, the \emph{hashSum} helps to filter the cells in which the $count\mathrm{=}$ 1 or -1 but saves more than one element.  These cells will not be chosen as an anchor. This is done by simply checking whether $g(idSum)\mathrm{=}hashSum$ or not. If $g(idSum)\mathrm{=}hashSum$, only the element whose identifier is exact \emph{idSum} is remained into this cell; otherwise, multiple elements are still stored in this cell and the $count=$ 1 or -1 is caused due to the subtracting operation. Especially, the length of IBF $m$ is decided as $\alpha d$, where $\alpha$ is a constant coefficient and $d$ is the number of different elements between $A$ and $B$. Although literature \cite{IBF} presents a strata estimator which tries the decoding iteratively until it successes, the estimation is still time-consuming and needs multiple rounds of communication. Furthermore, the both IBLT and IBF may fail to decode all the elements from the cell vector. 

\textbf{Bloomier filter.} The BF can be viewed as a boolean characteristic function of the represented set. With this insight, Bloomier filter \cite{Bloomier_BF} generalizes the boolean characteristic function to arbitrary functions. In other words, Bloomier filter expands the theory of BF by enabling the storage of not only membership information, but also the function values with the elements in $S$ as input. Consequently,  Bloomier filter also supports constant-time query of the function values that associated with the elements. Specifically, for the elements in $S$, Bloomier filter will return the function value; by contrast, for the elements that are not members of $S$, the query result will be $\perp$ which means an \emph{undefined} value. A near-optimal design for Bloomier filtering is proposed based on a cascading pipeline of BFs. The trick is to build the $i^{th}$ pair of BFs to represent the false positive elements in the $(i\mathrm{-}1)^{th}$ pair. For instance, $BF(A_0)$ and $BF(B_0)$ record the subset of elements (i.e., $A_0$ and $B_0$) whose function values will be 1 and 2 respectively. Then $BF(A_1)$ will be generated by recording the elements in $A_0$ but also pass the check of $BF(B_0)$. Similarly, $BF(B_1)$ will be generated by recording the elements in $B_0$ but also pass the check of $BF(A_0)$. Therefore, the elements which cause false positive errors will be reasonably recorded in the latter cascading BFs. The time-complexities for construction of Bloomier filter and query of a given element are $O(n \log n)$ and $O(1)$ respectively. Later researchers try to speed up the construction of Bloomier filter to linear time, or lessen space overhead at the cost of higher construction complexity.

\textbf{Parallel BF.} BF and its variants only support representation and query of single-attribute elements. Therefore, Parallel BF  \cite{Parallel_BF} denotes itself to represent multi-attribute elements. A typical scenario is that, in a database, a query always jointly consider multiple attributes of the entries. A naive solution to represent multi-attribute elements is to employ multiple BFs, each of which is responsible to one dedicated attribute. This method, however, leads to high false positive rate, since these BFs fail to store internal dependency of the attributes. Parallel BF is a matrix of CBFs, which consists of $\mathcal{P}$ sub-matrices to represent $\mathcal{P}$ attributes independently. For each sub-matrix, there are $k$ arrays of counters and $k$ hash functions are responsible to map an input into an array respectively. To insert an element with $\mathcal{P}$ attributes, each attribute is mapped by $\mathcal{P}$ groups of hash functions into $\mathcal{P}$ sub-matrices. The corresponding $\mathcal{P}\mathrm{\times}k$ counters are increased by 1. Additionally, a hash table is established to record the internal dependency of the attributes. In this way, the false positive errors from the sub-matrices can be significantly recognised. Typically, a function $\mathcal{F}$ is employed to calculate the verification value of an element based on the $\mathcal{P}\mathrm{\times}k$ positions in the matrix. In other words, the function $\mathcal{F}$ aggregates the attributes of an element together. With the generated verification value as input, the additional hash table offers a verification mechanism to the positive query results from the Parallel BF matrix. Compared to a single CBF, the Parallel BF matrix enlarges the false positive rate. But the majority of these false positive errors can be identified by the thereafter hash table. The representation of multi-attribute elements is realized at the costs of more hash computing, space occupation, and memory access. 

IBF \cite{IBF}, Bloomier filter \cite{Bloomier_BF}, and Parallel BF  \cite{Parallel_BF} are all the kind of scale-up extension in terms of semantic enrichment. IBF records the identifier of the elements and thereby enables inversely decoding. Bloomier filter injects the function values of the elements into the vector thus supports fast query of the function value. Parallel BF represents the multi-attribute of each element via multiple parallelized CBFs and an additional hash table. Some variants, on the other side, scale-out single elements and focus on the internal relationship. The traditional BF supposes by default that the elements in a dataset are independent. But the fact is that elements in a dataset can be intensively correlated. For instance, the nodes in a  multicast tree have a strict father-children relationship. The former mentioned Distance-sensitive BF \cite{DS_BF}, LSBF \cite{LSBF} and MLBF \cite{Multigranularity_LBF} in Section \ref{subsubsec:AMQ} can be viewed as the scale-out type semantic enrichment, since they try to maintain the distance between the elements with locality-sensitive hash functions. We believe other elegant designs are still needed in terms of other kinds of internal relationships.  

\subsection{Summary and lessons learned}
Besides of element insertion and query, real applications have a strong need for additional functionalities. For a dynamic set, deletion is needed. Therefore, CBF \cite{Counting_BF}, Deletable BF  \cite{Deletable_BF} and Ternary BF \cite{Ternary_BF} are proposed. For an online system, decay (or eviction) should be enabled to release the space for coming elements.  This functionality can be enabled by changing the values of bits in a single BF vector (e.g., Stable BF \cite{Stable_BF}, Temporal Counting BF \cite{TCBF}) or flushing the oldest BF among multiple BF vectors (e.g., Double buffering \cite{Double_Buffering}, $A^2$ buffering \cite{Two_Active}, and Forgetful BF \cite{Forgetful_BF}). Moreover, when the closeness is measured by a dedicated metric, the approximate membership query will be conducted to answer ``Is $x$ close to an element in $S$?''. In this case, the locality-sensitive hash functions are employed to map elements into the BFs \cite{DS_BF}  \cite{LSBF} \cite{Multigranularity_LBF}. Additionally, BF only saves the membership information but lacks other features of the elements. This disadvantage limits the functionality of BF. Therefore, the enrichment of BF semantics is required to support more complicated functionalities \cite{IBF} \cite{Bloomier_BF}  \cite{Parallel_BF}. When the users have additional requirements for other functionalities, they can alter the BF framework accordingly.

\begin{figure*}
  \centering
  \includegraphics[width=7.0in]{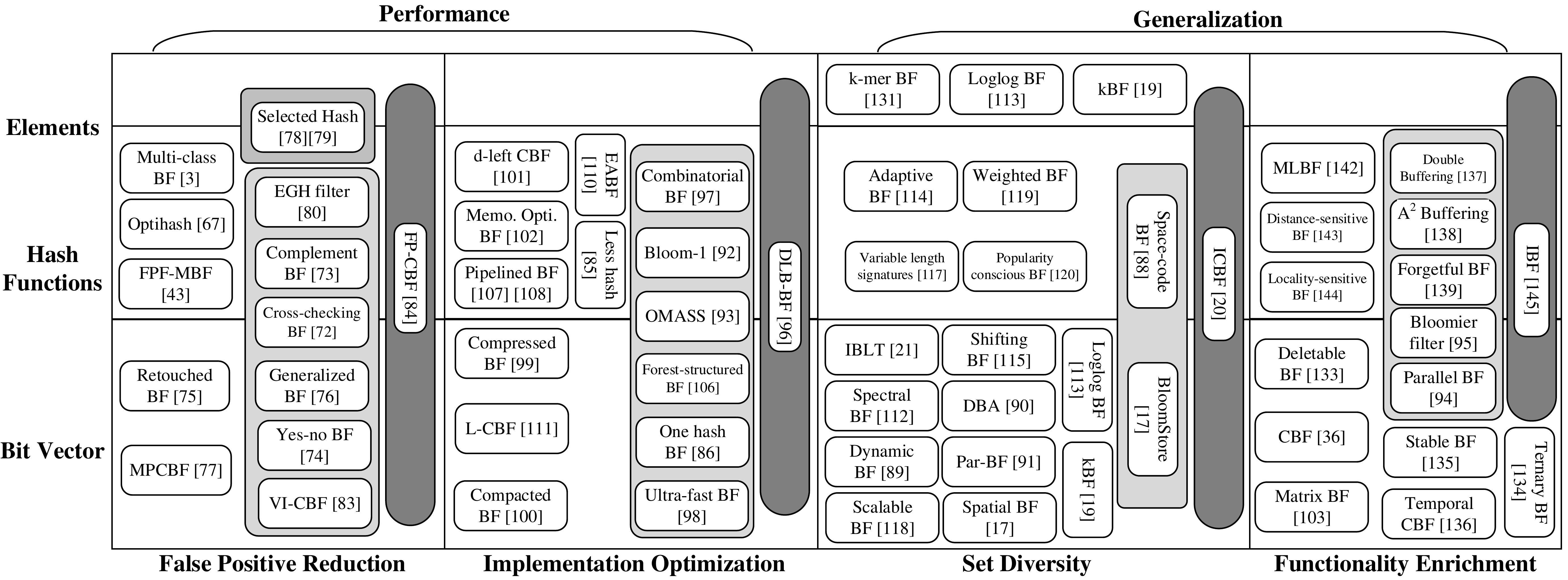}\\ \vspace{-0.05in}
  \caption{The taxonomy of the existing BF variants. The BF variants are classified in two main dimensions from the optimization perspective, i.e., performance and generalization. To improve the performance, dozens of variants devote themselves to reducing the false positives and easing the implementation. Besides, tens of variants generalize the BF framework in more scenarios by diversifying the input sets and enriching the output functionalities. We further analyze the optimization techniques towards the three components of BF, i.e.,  elements, hash functions, and bit vector, respectively. Obviously, many variants jointly optimize these components.}\label{fig:Taxonomy} \vspace{-0.15in}
\end{figure*}

\section{Classification and comparison}\label{sec:Analysis} 
In this section, we systematically analyze the optimization techniques introduced by the above BF variants. As shown in Fig. \ref{fig:Taxonomy}, the variants augment BF by operating its three components, i.e., elements, hash functions and bit vector, either separately or jointly. Consequently, we characterise the optimization techniques from three perspectives that refer to these three BF components.

\textbf{Elements}. The set $S$ in the framework of BF is a highly abstract term. In practice, a specific dataset may have its special feature. To better suit the real dataset, the users can augment the BF adaptively. For instance, recording the multiplicity of an element in a multiset calls for a redesign of the bit vector. Besides, before hashing, preprocessing operations (e.g., duplication, aggregation, extraction, outlier processing, dimension-reduction, etc) can be introduced to refine the entries. In the case of recording a set of text files, instead of hashing a complete text file, users may extract the keywords as the identification of a text file.

\textbf{Hash techniques}. Hashing the input elements into a given range contributes the major computation overhead of BF. To further speed up the BF or extend the usage of BF in devices with low computation capability, lightweight hash techniques can be considered. Moreover, BF assumes by default that the hash functions are independent. However, the randomness of hash functions is still an open theoretical problem. Therefore, designing or employing high-performance hash functions is critical to improve BF. 

\textbf{Bit vector}. The bit vector is the direct product of BF, and is responsible to record the membership of elements in a set. The bit vector can be queried, stored, transmitted, reproduced and updated in practice. From the angle of saving space overhead or communication overhead, one may compress, extend, shrink, segment or layer the bit vector. Moreover, in order to enable more functionalities or support complicated queries, each bit can be upgraded as a cell with multiple bits. 

\subsection{Techniques towards elements}
BF represents the membership of elements by the bits in the vector without differentiating them. As a consequence, BF cannot tell the 1s are caused by a dedicated element or not, thereby may result in false positive query results. A natural way to resolve this dilemma is to impose a unique fingerprint to each element. Alternatively, the elements may share common features which may help to improve the query accuracy by categorizing and then representing the elements in groups. Fig. \ref{fig:Taxonomy} indicates that the strategies which operate elements directly are usually aggregated with other techniques which augment the hash functions or (and) bit vectors.

\subsubsection{Imposing fingerprints to elements}
Generally, the fingerprint of an element is generated by mapping the element as a binary string with a hash function. Thereafter, the fingerprint can be stored directly in the vector. This kind of variants include kBF \cite{KBF}, FP-CBF \cite{Fingerprints_BF}, ICBF \cite{ICBF}, and IBF \cite{IBF}. Among them, kBF, FP-CBF, ICBF, and IBF store the fingerprint of element directly in the cells. To break the tie of collision where multiple elements are mapped into a shared cell, the XOR operations are introduced to aggregate the fingerprints together. In this case, deleting a fingerprint can be simply enabled by an additional XOR operation. kBF, FP-CBF, ICBF and IBF all associate the fingerprint with a counter in a cell, such that when the counter is 1, the stored fingerprint is exactly the fingerprint of the inserted element. Based on this insight, part of false positive matches can be identified via checking the fingerprint field. 

\subsubsection{Dividing elements into independent groups}
DLB-BF \cite{LoadBance_BF} and Partitioned hashing \cite{Hash_partition} divide the elements in the set into multiple groups, thereafter each group of elements is represented by a corresponding BF. This will effectively isolate the elements so that they will not affect each other upon query. To be specific, DLB-BF maps the IP prefixes with the same length into a dedicated BF. To query a prefix with a given length, DLB-BF only checks the corresponding BF. Similarly, the Partitioned hashing categorizes the elements into multiple groups with an additional hash function. Then each group of elements are stored in a BF. Before checking the $k$ bits in a BF, the queried element is first hashed by a hash function to decide which BF it should refer to. The advantages of the grouping strategy include false positive isolation and friendly memory access. On one hand, the elements in different groups will not interfere each other since they are physically isolated. On the other hand, the multiple BFs can be naturally accessed in parallel, and each query only requires the access of one BF. Therefore the query throughput will be increased reasonably. We also notice a special case, i.e., $k$-mer BF \cite{k-mer_BF}, which partitions the sequence data into $k$-mers without the above fingerprint nor grouping technique. The neighbouring mers can be utilized to identify incorrect matches of a given mer, during the querying process. 

An exception, which doesn't label elements with fingerprints nor group the elements, is Loglog BF \cite{Loglog_BF}. Loglog BF imposes a prefix to each element before each insertion. This is required by the probabilisitic counting strategy. 

\subsection{Techniques towards hash functions}
Hash functions are crucial ingredients of BF, and there are lots of proposals to augment this part. As depicted in Fig. \ref{fig:Taxonomy}, a dozen of the variants alter the hash functions directly. There are still tens of variants adjusting the hash functions together with other ingredients. We summarize these techniques from three aspects, i.e., leveraging the number of hash functions, optimizing the implementation of the  hash functions, and using advanced hash functions.

\subsubsection{Leveraging the number of hash functions}
For a static set, with the given number of elements $n$ and the bit vector length $m$, the optimal number of hash functions is $k\mathrm{=}\frac m n \ln 2$. But in reality, the number of employed hash functions can be more than $k$ or adjusted dynamically to achieve diverse targets. We roughly summarize the existing proposals which adjust the number of hash functions as three categories: 1) adding one extra hash function; 2) employing multiple groups of hash functions; 3) using hash functions on demand. We detail each of the categories as follows.

\textbf{One additional hash function.} Usually, an extra hash function is employed to randomly select an object from multiple available candidates. Typically, ICBF \cite{ICBF} and FP-CBF \cite{Fingerprints_BF} generates a fingerprint for each element with an extra hash function from a given range. Memory-optimized BF \cite{additional_hash_BF} chooses an accommodation cell for a file from all possible locations. OMASS \cite{OMASS} locates the block which stores an element via an extra hash function. Similarly, Forest-structured BF \cite{Forest_structured_BF} also derives the location of an element with extra hash functions. But Forest-structured BF calls for two extra hash functions, to locate the block and page respectively. Besides, IBF \cite{IBF} needs an extra hash function to calculate the $hashSum$ in  each cell to identify its purity.  
  
\textbf{Multiple groups of hash functions.} A dozen of BF variants require multiple groups of hash functions. Among them, Complement BF \cite{Complement_BF}, Cross-checking BF \cite{Cross_checking_BF}, Yes-no BF \cite{Yes_No_BF}, DLB-BF \cite{LoadBance_BF}, BloomStore \cite{BloomStore},  Bloomier filter \cite{Bloomier_BF}, Parallel BF \cite{Parallel_BF}, Combinatorial BF \cite{Combinatorial_BF} and FPF-MBF \cite{FP_Free_BF} establish multiple separated secondary BFs, thereby they need multiple groups of independent hash functions when  sharing the hash values are not allowed. If these secondary BFs have different length, sharing hash values is definitely not possible. Besides, to enable parallelized queries for different elements, each secondary BF must have its own hash group. But in scenarios where secondary BFs are of equal length and parallelism is not required, the $k$ hash functions can be shared, e.g., Dynamic BF \cite{Dynamic_BF} and Dynamic BF Array \cite{DBF_Array}. The main cost of employing multiple groups of hash functions is the computation overhead. 

On the contrary, Space-code BF \cite{Space_Code_BF} and Generalized BF \cite{Generalized_BF} utilize multiple groups of hash functions to set or reset a shared bit vector. The Space-code BF estimates the multiplicity of an element by counting the number of hash groups which result in non-zero bits. A passive impact of inserting an element into the bit vector for multiple times is that, the membership query may suffer from much higher false positive rate. Generalized BF employs two groups of hash functions, one for setting of bits and the other one for resetting of bits. As a consequence, the false positive rate can be decreased, with the penalty of false negative errors.

A more special case is VI-CBF \cite{VIncrement_CBF}, which introduces another group of hash functions to select the $k$ increments for an inserted element. The differentiated increments for elements, rather than a common 1, helps VI-CBF avoiding part of the false positive errors.

\textbf{Hash functions on demand.} The number of employed hash functions directly inference the performance of BF. Therefore, researchers integrate wisdom into this part and change the number of hash functions flexibly. Adaptive BF \cite{Adaptive_BF} records the multiplicity of an element with the number of hash functions (except for the $k$ hash functions for existence information). Weighted BF \cite{Weighted_BF}, Popularity conscious BF \cite{Popularity_Conscious_BF} and Multi-class BF \cite{Multiclass_BF} propose to adjust the number of employed hash functions for different elements. Typically, the elements which are more frequently queried or have a lower probability to be a member of $S$ should be associated with more hash functions. More hash functions implies stronger constraints to pass the check. The reason to do so is that the false positive errors happen to these elements impose larger increment of false positive proportion.

EABF \cite{Pipelined_BF1} \cite{Energy_Efficient_BF} and its same kind \cite{Pipelined_BF2} \cite{FullyPipelined_BF} adjust the number of activated hash functions dynamically for the purpose of energy saving. For the negative query results, they need not check all the $k$ hash functions, thus saving computation and energy. In other words, with these strategies, a negative result may be given instantly, but there will be a considerable delay for a positive result. If saving energy dominates the delay of a positive result, they are advisable choices. Note that the false positive rate of them will be the same as the standard BF.

The variant Variable length signatures \cite{Variable_length_BF}, novelly sets $t\mathrm{\leq}k$ bits of $h(x)$ to 1s when inserting, and concludes that $x\mathrm{\in} S$ if at least $q\mathrm{<}k$ bits of the $h(x)$ are 1s. The gaps between $t$, $q$ and $k$ enable the flexible insertion, deletion, and even decay of the recorded elements. Nevertheless, Variable length signature incurs both false positive and false negative errors.

\subsubsection{Optimization of hash implementation}
Due to the importance of hash functions in BF, tens of variants optimize hash functions from the perspective of implementation. To this end, Bloom-1 \cite{Bloom_1}, Less hash \cite{LessHash}, and One hash \cite{OneHash_BF} generate $k$ independent hash values with one or two hash functions. Specifically, Bloom-1 \cite{Bloom_1} divides a hash bits into $k$ parts, and each part is regarded as a hash value. By contrast, One hash \cite{OneHash_BF} modulos the machine word generated from the hash stage with $k$ diverse modulus to derive $k$ hash values for each segment of the bit vector. Less hash \cite{LessHash} reports a conservative method to generate as many hash values as the users want by using 2 independent hash functions, without any damage of the randomness. All of the these proposals can reduce the computation overhead of hash functions significantly.

For a given set, different hash functions lead to diverse false positive proportion. Therefore, Optihash \cite{optihash_BF} and Selected hash \cite{Power2choice} \cite{Hash_partition} try to pick an optimal group of hash functions which result in the lowest false positive proportion. Both Optihash \cite{optihash_BF} and Partition hashing \cite{Hash_partition} are computation-intensive since they need multiple rounds of tests or queries. A moderate method is to test two groups of hash functions and then select a better one for each element \cite{Power2choice}. This scheme achieves a significant reduction of FPP with an acceptable increase of computation. 

Especially, Ultra-fast BF \cite{UltraFast_BF} speeds up the calculation of hash functions via employing the SIMD technique. SIMD instructions parallelize the computation and thus only need $1/k$ time to generate $k$ hash values. SIMD, however, needs dedicated hardware and software environments, making this strategy lose of generality. 

\subsubsection{Application of advanced hash techniques}
Researchers introduce advanced hash techniques to improve BFs. $d$-left CBF \cite{d_left_hash} employs $d$-left hash functions to store the index of an element in a least-loaded cell (selecting the left-most one to break the tie of multiple equally loaded cells). The using of $d$-left hashing achieves nearly 50\% space saving with the same false positive guarantee, and two magnitude reduction of false positive rate with same space scale. Besides, the locality-sensitive hashing is also employed to replace the traditional hash functions. Locality-sensitive hashing maps two neighbouring elements into same or close locations in the vector with high probability. This characteristic enables Distance-sensitive BF \cite{DS_BF}, Locality-sensitive BF \cite{LSBF} and MLBF \cite{Multigranularity_LBF} to answer the query `` Is an element $x$ close to any element of $S$ or not?''. This kind of query, however, incurs both false positive and false negative errors, irrespective of the remedial designs, e.g., further verification and checking. The computation complexity of these locality-sensitive hash functions is also a barrier of real implementation. Note that, the EGH filter \cite{EGH_filter}, on the contrary, replaces the hash functions with some simple functions derived by prime numbers. By designing the parameters carefully, EGH filter guarantees a false positive free zone for a subset of elements.

We simply summarize this subsection here. Hash function is an essential ingredient of BF. The number of hash functions can be adjusted both proactively and passively. When there are multiple secondary BFs or one must select an object from multiple candidates, the number of hash functions has to be passively increased. By contrast, adjusting the number of employed hash functions proactively can achieve diverse design goals, i.e., energy-saving, false positive control, query acceleration. Moreover, the $k$ hash values can be generated by one or two hash functions with the guarantee of randomness. Meanwhile, given a set of elements, the performance of BF is hash-dependent. A proper group of hash functions can be selected by multiple rounds of tests and queries. Lastly, advanced hash techniques are available to enrich the functionality or improve the performance of BF directly. 


\subsection{Techniques towards the bit vector}
The bit vector stores the existence information of elements by setting the bits from initial 0s to 1s. Numerical variants extend the bit vector to enrich its semantics or impose additional operations on the bit vector for extra functionalities. There are two typical methodologies for semantic enrichment, i.e., a scale-up scheme which replaces each bit with a cell to store more information, and a scale-out scheme which implements more BF vectors for higher capacity. Also, two kinds of representative operations are popular to be executed upon the bit (or cell) vector, i.e., partition the vector into segments and changing the values of bits. Consequently, we organize the existing techniques towards the bit vector as the following four aspects. 

\subsubsection{Scale-up: beyond one single bit}
CBF \cite{Counting_BF} is the first variant which extends the bits of BF as fixed-length counters to seamlessly support element deletion. Thereafter, Spectral BF \cite{Spectral_BF} enables multiplicity query with the minimum value among the $k$ counters. Two variants of CBF, i.e., FP-CBF \cite{Fingerprints_BF} and VI-CBF \cite{VIncrement_CBF} leverage the fingerprint and differentiated increments for each inserted element to reduce the false positive rate, respectively. L-CBF \cite{L_CBF} speeds up the query of CBF from the implementation perspective. EGH filter \cite{EGH_filter} also extends the bits as counters to enable the deletion functionality.

IBLT \cite{IBLT}, IBF \cite{IBF} and ICBF \cite{ICBF} further extend a bit as a multi-field cell to store the necessary information for invertible decoding. The fundamental insights of their recursive decoding algorithms are similar. They first identify a pure cell which only stores one element. After recording the element, it will be deleted from the cell vector to hopefully expose more pure cells. The algorithm will be terminated until no pure cells can be searched out. In each cell, the three variants all have a counter to record the number of elements mapped into it. Especially, each IBLT \cite{IBLT} has $keySum$ and $valueSum$ field to aggregate the KV pairs mapped into the cell. Likewise, every IBF \cite{IBF} cell has two other fields, i.e., $idSum$ and $hashSum$ to record the information of stored elements. ICBF \cite{ICBF}, by contrast, has only one extra field, i.e., $id$. The differences between these three proposals stem from their functionalities. IBLT is designed for KV pairs, while IBF is targeted at general elements. ICBF, however, is proposed for multiset synchronization between two hosts, and an off-line id table is maintained to record the mapping between ids and elements. A common technique for them is the XOR, which can aggregate information together, delete elements with simple bitwise operations, and reveal the value or fingerprint of an element when only one single element stored in the cell. 

There are also other proposals to extend the bits, e.g., Bloomier filter \cite{Bloomier_BF}, Spatial BF \cite{Spatial_BF}. They are actually coupled with other features, so we analyze them in other chapters. As an opposite of extension, Compressed BF \cite{Compressed_BF} and Compacted BF \cite{Compacted_BF} compress the bit vector for bandwidth-friendly transmission. Compressed BF employs fewer hash functions so that the probability of each bit to be set as 1 is less than 1/2. Thereafter, the bit vector is compressed for transmission.  Compacted BF proposes a series of principles to encode the bit vector and recover it for later queries after transmission. The penalty of  Compacted BF is additional false negative errors due to its encoding principles.

\subsubsection{Scale-out: more BF vectors}
When one BF vector is not enough, or the users have multiple sets to represent, a natural way is to establish multiple BFs. Dynamic BF\cite{Dynamic_BF}, Scalable BF \cite{Scalable_BF}, Par-BF \cite{Par_BF} and DBA \cite{DBF_Array} are proposed for scalability and elasticity. The secondary BFs, either homogeneous or heterogeneous, can be added or merged on demand. Another concern for capacity extension is to partially migrate BF from fast but scarce RAM storage to slower yet massive flash memory. Forest-structured BF \cite{Forest_structured_BF} and BloomStore \cite{BloomStore} propose joint design with both RAM and flash. Specifically, Forest-structured BF \cite{Forest_structured_BF} first implement an in-RAM block which contains several BFs. Once all the in-RAM BFs are saturated, blocks will be added into the flash memory on demand and constructed as a forest architecture. Then the RAM will be employed as a cache for the most-recent arriving elements. On the contrary, BloomStore \cite{BloomStore} leverages the RAM as a buffer which only stores the most-recent elements for each BloomStore instance. Once the buffer is filled up, both the elements and the associated BF will be pushed into the flash. 

Complement BF \cite{Complement_BF}, Cross-checking BF \cite{Cross_checking_BF} and Yes-no BF \cite{Yes_No_BF} configure additional BFs to help identify some false positive errors from the main BF query results. Complement BF implements an additional BF to store the elements in the complement set of $S$ directly. Cross-checking BF divides $S$ into several independent subsets and establishes a BF for each of them separately. Yes-no BF simultaneously maintains a yes-filter for the elements in $S$, and no-filters for the false-positively matched elements. When the universe set $U$ is large while $S$ is small, Complement BF is not advisable. If the false positive queries are detectable, Yes-no BF will be a wise choice. Cross-checking BF is a pervasive scheme and it also leaves a trade-off space (between query accuracy and space cost) to its users.

Combinatorial BF \cite{Combinatorial_BF}, DLB-BF \cite{LoadBance_BF}, Space-code BF \cite{Space_Code_BF}, Matrix BF \cite{Matrix_BF}, and Parallel BF \cite{Parallel_BF} are all configured of multiple independent BFs. Combinatorial BF and DLB-BF have multiple groups of elements to represent, therefore they configure a BF for each group of elements. Parallel BF represents the multiple attributes of an element with multiple BF matrices. Space-code BF is functional with single BF vector. But with multiple BF vectors and executing the MLE for multiple times, Space-code BF generates a more accurate multiplicity estimation. In Matrix BF, each row of the bit matrix acts as a BF and record one specific document in a file library. These rows share $k$ hash functions and thereby supporting copy-paste detection between documents. Note that Bloomier filter \cite{Bloomier_BF} also has multiple BFs but it is none of the above kinds. The reason is that the BFs in Bloomier filter are nested and constructed recursively, rather than independent with each other. In Bloomier filter, false positives in the first level BFs will be recorded in the second level BFs. As a consequence, these BFs are accessed sequentially. 

Certainly, a common obstacle behind the scale-up and scale-out proposals is space overhead. For scale-out proposals which just configure multiple standard BFs, e.g., Dynamic BF, Scalable BF, Par-BF, DBA, DLB-BF, Matrix BF, Forest-structured BF and BloomStore, they lead to the same bpe (bites per element) as standard BF. Most of the scale-up proposals may result in higher bpe, with the gain of additional functionalities (e.g., deletion, invertible decoding, parallelism) or better performance guarantee (e.g., higher query accuracy). 

\subsubsection{The power of partition}

A simple partition method is to divide the bit vector into $k$ segments such that each hash function is responsible to one segment. This adjustment is a natural way to enable parallel access, at the cost of a slight increment of false positive rate. OMASS \cite{OMASS} stores the information of an element in a word-size block, such that a query can be responded with only one memory access. Bloom-1 \cite{Bloom_1} also stores the information of an element in one word, while the $k$ bits are chosen by only one hash function. The trick is to split the hash bits of an element into $k$ parts, each part select one position for the element from the word. Still, only one memory access is enough for a membership query. Ultra-fast BF \cite{UltraFast_BF} employs the similar scheme, but it has $k$ words in one block for the $k$ hash functions, respectively. As a consequence, Ultra-fast BF supports block-level parallelism and each query needs to access all the $k$ bits in the $k$ words of a selected block. The load of each block or word is determined by the global hash function which is responsible to select one word or block for an element. Ideally, the blocks or words are load-balanced. However, in reality, it is still possible that some blocks or words are overloaded while some are underloaded. We suggest that the load-balance friendly hash techniques, e.g., $d$-left hash, may be an effective solution.

MPCBF \cite{Multi_partitioning_CBF} also divides the bit vector into multiple words, and then goes further to establish a hierarchical structure for better utilization of these bits. The first level of bits are responsible for query, the later levels are leveraged for dynamical element insertion and deletion. Compared with CBF, MPCBF achieves lower false positive rate by logically increasing the value of $m$ in Equ. \ref{equ:fpr}. The partitions in OHBF \cite{OneHash_BF}  and Deletable BF \cite{Deletable_BF}, however, are not for memory access reduction nor better utilization. OHBF divides the bit vector into $k$ parts with different lengths, such that the machine word generated by one hash function can derive $k$ hash values with $k$ different modulus. Deletable BF, specially, splits the vector into $r\mathrm{+}1$ regions. The first region has $r$ bits, each of which indicates whether the corresponding region incurs hash collisions or not. When deleting, Deletable BF only resets the 1s in collision-free regions. Any of the associated $k$ bits is reset as 0 means a successful deletion of an element. In the situation where there are no collision-free regions, Deletable BF  \cite{Deletable_BF} declares that element is not deletable. 

\subsubsection{The game between 0s and 1s}
Another explicit way to adjust the vector is changing the values in the bits (or cells). Retouched BF \cite{Retouched_BF}, Generalized BF \cite{Generalized_BF} and Stable BF \cite{Stable_BF} allow to reset bits from 1s to 0s when inserting an element. Both Retouched BF \cite{Retouched_BF} and Generalized BF \cite{Generalized_BF} are nice trials to eliminate 1s thereby generating lower false positive rate if false negatives are permitted. Retouched BF proposes both random and selective clearing strategies, while Gneralized BF employs two groups of hash functions, i.e., one group for set and the other group for reset. Stable BF \cite{Stable_BF} targets at the dynamic dataset and maintains a fixed proportion of 1s in the bit vector, such that stale elements will be eliminated when inserting the new arrivals. Undoubtedly, all the three proposals incur false positive, as well as false negative errors. Resetting the 1s to 0s damages the one-sided error characteristic of BF. As a consequence, they offer no correctness guarantee for both the positive and negative query results. 

Ternary BF \cite{Ternary_BF} doesn't reset the bits to 0s, on the contrary, it records the bits which incur hash collision and marks their value as X. When querying an element $y$, if the corresponding $k$ counters of $y$ are all X, then $y$ is identified as \emph{indeterminable}. In contrast, when deleting element $z$, if the corresponding $k$ counters of $z$ are all X, then $z$ is considered as \emph{undeletable}. That is, Ternary BF refuses to answer a query which may be a false positive and rejects to delete an element which may cause a false negative. Temporal CBF \cite{TCBF} saves the ``time-to-live'' of an element in the $k$ corresponding cells. All the counters in the vector will be decreased periodically by one. The space overhead of this proposal will be an obstacle. Each cell must have enough bits for the maximum lifecycle. For a set of elements with highly skewed lifecycle, Temporal CBF is obviously not space-efficient. Spatial BF \cite{Spatial_BF} leverages the distinct values in the cells to label the elements in different geographical areas. Larger counter values indicate more central areas and will remain when conflicted with lower values in a cell. In this way, Spatial BF provides better accuracy to the queries of central areas. 

\doublerulesep 0.1pt
\begin{table*}[tbp]
\centering
\begin{footnotesize}
\caption{The key capabilities and complexities of BF and its variants. The capabilities include counting,  grouping, deletion, scalable, decay, parallelism and false negative (FN).  In contrast, we consider the complexities about insertion (Comp-I), query (Comp-Q), deletion (Comp-D) and memory access (Comp-M). } \label{table_feature}\vspace{-0.05in}
\begin{tabular}{lcccccccccccc}
\hline 
 \begin{turn}{-90}Structure$\ $ \end{turn}  & \begin{turn}{-90}Counting $\ $ \end{turn} & \begin{turn}{-90}Grouping $\ $ \end{turn}   & \begin{turn}{-90}Deletion $\ $ \end{turn}  & \begin{turn}{-90}Scalability $\ $ \end{turn}  & \begin{turn}{-90}Decay $\ $ \end{turn}  & \begin{turn}{-90}Parallelism $\ $ \end{turn}  & \begin{turn}{-90}FN $\ $ \end{turn} & \begin{turn}{-90}Comp-I $\ $ \end{turn}& \begin{turn}{-90}Comp-Q$\ $ \end{turn}  & \begin{turn}{-90}Comp-D $\ $ \end{turn} & \begin{turn}{-90}Comp-M $\ $ \end{turn} & \begin{turn}{-90}Context $\ $ \end{turn} \\ \hline
Standard BF \cite{BF}	&No &	No&	No&	No&	No&	No & No &	$O(k)$	& $O(k)$	&$--$	&$O(k)$ & General	\\ \hline
CBF \cite{Counting_BF}&	Yes&	No&	Yes&	No&	No&	No	&No & $O(k)$&	$O(k)$	&$O(k)$	&$O(k)$	& General\\ \hline
kBF \cite{KBF}&	Yes	&Yes&	Yes&	No&	No&	No&	Yes&	$O(k)$&	$O(k)$&	$O(k)$	&$O(k)$ & K-V store	\\ \hline
Loglog BF \cite{Loglog_BF}&	Yes&	No&	No&	No&	No&	No&	No&	$O(kd)$	&$O(kd)$	&$--$	&$O(dk)$ & Smart Grid \\ \hline
$k$-mer BF \cite{k-mer_BF} &	No	&No&	No&	No&	No&	No &	No&	$O(k)$	&$O(2k)\mathrm{-}O(3k)$&	$--$&	$O(2k)\mathrm{-}O(3k)$  & Biometric	\\ \hline
Adaptive BF  \cite{Adaptive_BF} &	Yes	&No&	No&	No&	No&	No &	No	&$O(k\mathrm{+}N\mathrm{+}1)$&	 $O(k\mathrm{+}N\mathrm{+}1)$&	$--$&$	O(k\mathrm{+}N\mathrm{+}1)$ & Networking\\ \hline
Weighted BF  \cite{Weighted_BF} &	No	&No	&No&	No&	No&	No & No &	$O(k)$&	$O(k)$&	$--$&	$O(k)$	& General	\\ \hline
Var. Len. Sin \cite{Variable_length_BF}&	Yes&	No&	Yes	&No&	No&	No &Yes &	$O(t)$&	$O(q)$	&$O(k\mathrm{-}d)$&	$O(k)$	& Networking	\\ \hline
Pop. cons. BF \cite{Popularity_Conscious_BF}&	No& 	No &	No &	No& 	No &	No&	No &	$O(k_i)$	&$O(k_i)$&	$--$&	$O(k)$ & General	\\ \hline
Spectral BF \cite{Spectral_BF}	&Yes	&No	&Yes	&No	&No	&No &No	&$O(k)$	&$O(k)$&	$O(k)$&	$O(k)$	& Networking	\\ \hline
Dynamic BF \cite{Dynamic_BF}&	No&	No&	Yes&	Yes	&No&	Yes &	Yes&	$O(k)$&	$O(sk)$&	$O(sk)$&	$O(sk)$ & Dynamic set\\ \hline
DBA \cite{DBF_Array} &	No&	Yes&	Yes&	Yes	&No&	Yes &	Yes&	$O(k)$&	$O(rk)$&	$O(rk)$&	$O(rk)$ & Dynamic set	\\ \hline
Par-BF \cite{Par_BF}	&No&	No&	Yes&	Yes&	No	&Yes	&No&	$O(k)$&	$O(k s_m)$&	$O(k s_m)$&	$O(k s_m)$ & Dynamic set	\\ \hline
Scalable BF \cite{Scalable_BF}&	No&	No&	No	&Yes&	No&	Yes&	No	&$O(k)$&	$O(lk)$&	$--$&$	O(lk)$ & Dynamic set	\\ \hline
Spatial BF \cite{Spatial_BF}	&No&	Yes&	No	&No	&No&	No&	No&	$O(k)$&	$O(k)$&	$--$&	$O(k)$ & Spatial data\\ \hline
Space-code BF \cite{Space_Code_BF}&	Yes&	No&	No&	No	&No	&No	&No&	$O(c k)$&	$O(lk)$&	$--$&	$O(lk)$ & Networking	\\ \hline
ICBF \cite{ICBF}&Yes&	No&	Yes&	No&	No	&No&	No&	$O(k)$&	$O(k)$&	$O(k)$	&$O(k)$ & General	\\ \hline
BloomStore \cite{BloomStore}&	No&	Yes&	No	&Yes	&No	&Yes&	No&	$O(k)$&	$O(xk)$&	$--$&	$O(xk)$ & Storage	\\ \hline
IBLT \cite{IBLT}&	Yes&	No	&Yes	&No	&No&	No&	No&	$O(k)$&	$O(k)$&	$O(k)$	&$O(k)$& K-V store	\\ \hline
Multi-class BF  \cite{Multiclass_BF}&	No&	No&	No&	No&	No&	No& 	No &	$O(k_i)$ & $O(k_i) $& $--$ & $O(k_i)$ & Networking	\\ \hline
Optihash \cite{optihash_BF}	&No&	No &	No&	No&	No&	No&	No&	$O(1)$&	$O(1)$&	$--$&	$O(1)$& Networking	\\ \hline
FPF-MBF \cite{FP_Free_BF}&	No&	No&	No&	No&	No&	Yes&	No&	$O(k)$&	$O(k)$&	$--$&	$O(k)$& Networking	\\ \hline
Retouched BF \cite{Retouched_BF}&	No&	No&	No&	No&	Yes &	No&	Yes&	$O(k)$&	$O(k)$&	$O(k)$&	$O(k)$& General	\\ \hline
MPCBF  \cite{Multi_partitioning_CBF}	&No	&No	&Yes	&No	&No	&Yes	&No	&$O(k\mathrm{+}1)$	&$O(k\mathrm{+}1)$&	$O(k\mathrm{+}1)$	&$O(1)$& General	\\ \hline
Ternary BF  \cite{Ternary_BF}&	No&	No	&Yes	&No&	No&	No&	No&	$O(k)$&	$O(k)$&$O(k)$&	$O(k)$& General	\\ \hline
Selected Hash \cite{Power2choice} \cite{Hash_partition}	&No&	No&	No&	No&	No&	No&	No&	$O(k)$&	$O(k)$&	$--$&	$O(k)$& General	\\ \hline
Complement BF \cite{Complement_BF}&	No&	No&	No&	No&	No&	No&	No&	$O(k)$	&$O(2k)$&	$--$	&$O(2k)$& General	\\ \hline
EGH filter \cite{EGH_filter}&	Yes&	No&	Yes&	No&	No&	No&	No&	$O(k)$	&$O(k)$&	$O(k)$	&$O(k)$& General	\\ \hline
Cross-checking BF \cite{Cross_checking_BF}&	No&	No&	No&	No&	No&	No&	No&	$O(k)$	&$O(\alpha k)$&	$--$&	$O(\alpha k)$& General	\\ \hline
Generalized BF \cite{Generalized_BF}&	No&	No&	No&	No&	Yes&	No&	Yes&$	O(k)$&	$O(k)$&	$--$&$	O(k)$& General	\\ \hline
Yes-no BF \cite{Yes_No_BF}&	No&	No&	No&	No&	No&	No&	No&	$O(k)$&	$O(2k)$&	$--$&	$O(2k)$& General	\\ \hline
VI-CBF \cite{VIncrement_CBF}&	No&	No&	Yes&	No&	No&	No&	No&	$O(2k)$&	$O(2k)$&	$O(2k)$&	$O(2k)$& General	\\ \hline
FP-CBF \cite{Fingerprints_BF}&	Yes&	No&	Yes&	No&	No&	No&	No&	$O(k)$&	$O(k)$&	$O(k)$&	$O(k)$& General	\\ \hline
Dist. Sens. BF  \cite{DS_BF}&	No&	No&	No&	No&	No&	No&	Yes&$	O(k)$&	$O(k)$&	$--$&$	O(k)$& General	\\ \hline
Loc. Sens. BF  \cite{LSBF}&	No&	No&	No&	No&	No&	No&	Yes&	$O(2k)$&	$O(2k)$&	$--$&$	O(2k)$& General	\\ \hline
MLBF  \cite{Multigranularity_LBF} &No &No&No&No&No&No & Yes &	$O(Lk\mathrm{+}k')$ &	$O(Lk\mathrm{+}k')$ &$--$ & 	$O(k)$ & General	\\ \hline
Deletable BF  \cite{Deletable_BF}&	No&	No&	Yes&	No&	No&	No&	No&$	O(k)$&	$O(k)$&	$O(2k)$&	$O(k)$& General	\\ \hline
IBF  \cite{IBF}&	Yes&	No&	Yes&	No&	No&	No&	No	&$O(k)$&	$O(k)$&	$O(k)$&	$O(k)$& General	\\ \hline
Stable BF	 \cite{Stable_BF} &No&	No	&No&	No&	Yes&	No&	Yes&	$O(k)$&	$O(k)$&	$O(k)$&	$O(k)$& Duplication	\\ \hline
Temporal CBF	 \cite{TCBF}&No	&No	&Yes&	No	&Yes	&No	&Yes	&$O(k)$	&$O(k)$	&$O(k)$&	$O(k)$& Networking	\\ \hline
Double Buffering  \cite{Double_Buffering} &No &Yes &No &No &Yes &Yes &Yes  &	$O(2k)$  &	$O(2k)$&$--$ &	$O(k)$& Networking\\ \hline
$A^2$ Buffering  \cite{Two_Active} &No &Yes &No &No &Yes	&Yes	 &Yes &$O(2k)$ &$O(2k)$&	$--$&$O(2k)$& Dynamic set	\\ \hline
Forgetful BF  \cite{Forgetful_BF} &No &Yes &No &No &Yes &Yes &Yes &	$O(2k)$ &	$O(pk)$ &	$--$&$O(pk)$& K-V store	\\ \hline
Bloomier filter  \cite{Bloomier_BF}&	No &	Yes&	No& 	No&	No&	No&	No&	$O(n\log n)$&	$O(\lambda k)$&	$--$&	$O(\lambda k)$& General	\\ \hline
Parallel BF  \cite{Parallel_BF}&	Yes&	Yes&	Yes	&No&	No	&Yes&	No&	$O(pk)$&	$O(pk$)&	$O(pk)$&	$O(pk)$ & Networking	\\ \hline
Less hash \cite{LessHash}	&No&	No&	No&	No&	No&	No&	No&	$O(k)$&	$O(k)$&	$--$	&$O(k)$& General	\\ \hline
One hash BF \cite{OneHash_BF}&	No&	No&	No&	No&	No&	No&	No&	$O(k)$&	$O(k)$&	$--$&	$O(k)$& General	\\ \hline
Compressed BF \cite{Compressed_BF}&	No&	No	&No&	No&	No&	No&	No&	$O(k)$	&$O(k)$&	$--$&	$O(k)$& General	\\ \hline
Compacted BF \cite{Compacted_BF}&	No&	No&	No&	No &	No&	No	&Yes	&$O(k)$&	$O(k)$&	$--$&	$O(k)$& General	\\ \hline
Combinatorial BF \cite{Combinatorial_BF}&	No&	Yes&	Yes	&No&	No&	Yes&	No	&$O(n_1 k)$&	$O(\gamma k)$&	$O(\gamma k)$	&$O(\gamma k)$ & Networking\\ \hline
Bloom-1 \cite{Bloom_1}&	No&	No&	No&	No&	No& 	No&	No&	$O(k)$&	$O(k)$&	$--$&$	O(1)$& General	\\ \hline
OMASS \cite{OMASS}&	No&	Yes&	No&	No&	No&	Yes	&No&	$O(k+1)$&	$O(k+1)$&	$--$&$O(1)$& General	\\ \hline
Forest-struc. BF \cite{Forest_structured_BF}	&No&	Yes	&No&	Yes&	No	&Yes&	No&	$O(k+2)$&	$O(lk)$&	$--$&	$O(lk)$& Storage	\\ \hline
DLB-BF \cite{LoadBance_BF}&	No&	Yes&	No&	No&	No&	Yes&	No&	$O(k)$&	$O(k)$&	$--$&$	O(k)$& Networking	\\ \hline
Ultra-fast BF \cite{UltraFast_BF}&	No&	Yes&	No	&No	&No&Yes&	No&	$O(k+1)$&	$O(k+1$)&	$--$	&$O(k)$& Implementation	\\ \hline
$d$-left CBF \cite{d_left_hash}&	Yes&	No&	Yes&	No&	No	&No&	No&$	O(d)$&	$O(d)$	&$O(d)$&$O(d)$& Implementation	\\ \hline
L-CBF \cite{L_CBF} &	Yes&	No&	Yes&	No&	No	&No&	No&$	O(k)$&	$O(k)$	&$O(k)$&$O(k)$& Implementation	\\ \hline
Mem. Opti. BF \cite{additional_hash_BF}&  No	&No&	No&	No&	No&	No&	No&	$O(k+1)$&	$O(k+1)$&	$O(k+1)$&	$O(k+1)$& Storage	\\ \hline
Energy effic. BF \cite{Energy_Efficient_BF}&	No&	No&	No&	No&	No&	No &	No	&$O(k)$&	$O(k_1)\mathrm{-}O(k)$&	$--$&	$O(k_1)\mathrm{-}O(k)$& Networking	\\ \hline
Pipelined BF \cite{Pipelined_BF1} \cite{Pipelined_BF2} &	No&	No&	No&	No&	No&	No &	No	&$O(k)$&	$O(k_1)\mathrm{-}O(k)$&	$--$&	$O(k_1)\mathrm{-}O(k)$& Networking	\\ \hline
Matrix BF \cite{Matrix_BF}&	No&	Yes&	No&	Yes&	No&	No &	No	&$O(k)$&	$O(k)$&	$--$&	$O(Nk)$& Copy dectection	\\ \hline 
Shifting BF \cite{Shifting_BF}&	Yes&	Yes&	Yes&	No&	No	&No& No	&$O(\lambda k)$&	$O(\lambda k)$&	$O(\lambda k)$&	$O(\lambda k)$& General	\\ \hline
 \end{tabular}
\end{footnotesize}\vspace{-0.17in}
\end{table*}

Double buffering \cite{Double_Buffering}, $A^2$ buffering \cite{Two_Active}, and Forgetful BF \cite{Forgetful_BF}, however, flush the stale BF(s) directly. In online scenarios, these variants always cache the recently inserted elements. Double buffering maintains an active BF and a warm-up BF in the memory. After the active BF is half full, the latter inserted elements will be also inserted into the warm-up BF. Once the active BF is filled up, double buffering switches the active BF and warm-up BF and flushes the generated warm-up BF. Note that, in Double buffering, only the active BF is responsible for membership query. The $A^2$ buffering, by contrast, implements two active BFs simultaneously. $active1$ stores the recently inserted elements, while $active2$ records previous inserted elements. When $active1$ becomes full, $active2$ is flushed and the two BFs switch their roles. The membership of an element will be confirmed if it passes the check of either $active1$ or $active2$. Especially, the Forgetful BF maintains several BFs simultaneously, namely, one future BF, one present BF and one or multiple past BF(s). During an update, the oldest past BF will be eliminated and an empty new future BF will be added. All the BFs in Forgetful BF are responsible for the membership query. 

Besides of the $k$ bits which represent the membership of elements, Adaptive BF \cite{Adaptive_BF} and Shifting BF \cite{Shifting_BF} programme additional bits as 1s to record more information associated with the elements. Adaptive BF sets additional $m(x)$ bits to represent the multiplicity of element $x$ with $m(x)$ hash functions. Shifting BF \cite{Shifting_BF} also enriches the semantics of the bit vector by additionally storing auxiliary information for diverse types of queries. For example, for association query, the cells or bits will record the set or group the element belongs to. The $k$ bits for auxiliary information are novelly selected by attaching an offset to the $k$ locations for existence information. However, the 1s for membership information and the 1s for additional information may interfere each other. The 1s in the bit vector to represent multiplicity information can increase the false positive rate of membership queries. By contrast, the 1s in the bit vector to represent membership information may lead to inaccurate multiplicity query results. To ease this dilemma, saving the two kinds of 1s in separated bit vectors may be a good idea.

We also noted that, in the last years, several methods to protect Bloom filters have been proposed. For example, Reviriego et. al propose to use a parity to detect soft errors \cite{protect_1}. Then upon an error detection, all the bits on that word of the error were set to 1 so that false negatives were avoided. The protection against transient errors was considered in \cite{protect_2} by for example performing a recomputation if all except one of the positions accessed had a value of one. These methods protect BFs against errors in the memories or the hash function calculation. 

\subsection{Qualitative comparison}
In this survey, we review the existing optimization techniques towards the three components of BF, from the perspective of both performance and generalization. To improve the performance of BFs, the variants focus on the reduction of false positives and implementation optimizations. Also, dozens of variants try to generalize BF framework by representing diverse type of sets from the input side and enabling more functionalities from the output side. As shown in Fig. \ref{fig:Taxonomy}, the variants augment one or more components to achieve their design goals. Furthermore, we qualitatively compare the capabilities and complexities of these variants and report the results in Table \ref{table_feature}. 

Note that, in the last column of Table \ref{table_feature}, we also highlight the contexts where these BF variants are proposed or applied. Obviously, a large portion (28 out of 60 variants) of the BF variants are designed for general usage -- to represent a static set of elements, support membership query and other functionalities. These variants redesign the standard BF framework to reduce the potential false positives and enable extensive functionalities. The rest of BF variants, on the other hand, are motivated by specific requirements in diverse contexts. In the networking field, Spectral BF \cite{Spectral_BF}, Multi-class BF \cite{Multiclass_BF}, Optihash \cite{optihash_BF}, FPF-MBF \cite{FP_Free_BF}, and Combinatorial BF\cite{Combinatorial_BF} are designed to improve in-packet or in-switch BFs for multicast forwarding; Adaptive BF \cite{Adaptive_BF} and Space-code BF \cite{Space_Code_BF} aim at traffic engineering; Double Buffering \cite{Double_Buffering}, DLB-BF \cite{LoadBance_BF}, EABF \cite{Energy_Efficient_BF}, and Pipelined BF \cite{Pipelined_BF1} are proposed for packet processing; Temporal CBF \cite{TCBF} speeds up the information sharing in human net; Parallel BF \cite{Parallel_BF}  supports multiattribute representation on network services. Besides, for online systems where elements may join and leave dynamically, several variants are proposed to realize capacity elasticity of BF, such as Dynamic BF \cite{Dynamic_BF}, DBA \cite{DBF_Array}, Par-BF \cite{Par_BF}, Scalable BF \cite{Scalable_BF} and $A^2$ Buffering \cite{Two_Active}. Other variants are also designed in given contexts, including kBF \cite{KBF}, IBLT \cite{IBLT}, and Forgetful BF \cite{Forgetful_BF} for KV store, Loglog BF \cite{Loglog_BF} in smart grid, $k$-mer BF \cite{k-mer_BF} for biometric data representation, BloomStore \cite{BloomStore}, Forest-structured BF \cite{Forest_structured_BF}, and Memory-optimized BF \cite{additional_hash_BF} for storage systems, Spatial BF \cite{Spatial_BF} for spatial data representation, Stable BF \cite{Stable_BF} for duplication,  Ultra-fast BF \cite{UltraFast_BF} and $d$-left CBF \cite{d_left_hash} for implementation of CBF, and Matrix BF \cite{Matrix_BF} for copy detection.

\subsubsection{Capability comparison} 
Generally, we consider six capabilities, i.e., counting, grouping, deletion, scalability, decay, and parallelism. Besides, we also highlight the variants which additionally incur false negative errors in Table \ref{table_feature}. 

\textbf{Counting.} Generally, the variants which have a counter field in each cell (e.g., CBF, Spectral BF, FP-CBF, IBLT, IBF, ICBF, etc) will be considered to support the counting capability and thereby multiplicity query. The counting capability is of great significance to extend the usage of BF from simple sets to more general multisets. There are also other alternatives of the counter field, e.g., using the number of hash functions \cite{Adaptive_BF}. Additionally, typical estimation algorithms (e.g., maximum likelihood estimation \cite{Space_Code_BF}, probabilistic counting \cite{Loglog_BF}) can also be leveraged to estimate the multiplicity.  A more special proposal is Shifting BF \cite{Shifting_BF} which regards the multiplicity of elements as auxiliary information. 

\textbf{Grouping}. Grouping answers the question which group(s) or subset(s) a queried element belongs to. This capability widens the application of BF from a single set to multiple sets (or groups) scenarios, e.g., IP lookup. Generally, the variants with multiple BF vectors or BF segments (e.g., DBA \cite{DBF_Array}, Matrix BF \cite{Matrix_BF}, BloomStore \cite{BloomStore}, Forest-structured BF \cite{Forest_structured_BF}, etc) naturally support the grouping capability via representing each group with a single BF vector. This strategy has to query the element against all the BF vectors or BF segments to draw a probabilistic conclusion. Especially, Combinatorial BF \cite{Combinatorial_BF} derives the group id for an element by leveraging multiple BFs. If the element passes the membership check of a BF, the corresponding bit in the output group id will be 1. Spatial BF \cite{Spatial_BF} distinguishes the areas with different integer labels. If we treat the labels as groups, Spatial BF enables grouping with differentiated accuracy for diverse groups. OMASS enables the grouping capability with shared bit vector for the multiple sets. To lessen the interference of bits for diverse sets, OMASS redesigns the hash functions to generate orthotropic hash values for a common element in multiple sets. Proposals like Bloomier filter \cite{Bloomier_BF}, Shifting BF \cite{Shifting_BF} and kBF \cite{KBF}, can store the group id as an associated value of the element. However, except for the proposals with multiple BFs or BF segments, other variants only resolve the grouping problem where an element exclusively belongs to exactly a single group. 

\textbf{Deletion and scalability}. Both deletion and scalability are crucial capabilities to enable the usage of BFs in dynamic datasets. Nevertheless, most of the existing proposals achieve one single capability, and only a few of them guarantee deletion and scalability simultaneously. CBF \cite{Counting_BF} and its variants, e.g., MPCBF \cite{Multi_partitioning_CBF}, VI-CBF \cite{VIncrement_CBF}, FP-CBF \cite{Fingerprints_BF}, Temporal CBF \cite{TCBF}, $d$-left CBF \cite{d_left_hash} are naturally deletable. Other proposals with a counter in each cell can also delete elements reasonably by decreasing the corresponding counters. The Deletable BF \cite{Deletable_BF} partitions the bit vector and only resets the bits in collision-free segments to delete elements. To achieve scalability and elasticity, DBF \cite{Dynamic_BF} and DBA \cite{DBF_Array} support both deletion of elements and introduction of additional vectors. Note that deleting the elements in DBF and DBA by resetting the bits from 1s to 0s may cause false negatives. A remedy is to replace the bits with cells like the CBF does, at the cost of nearly 4 times space overhead. The BFs in Matrix BF \cite{Matrix_BF} can be expanded but don't support element-level deletion. However, the BF-level deletion is permitted, since each BF is independent and records all the strings in a document. 

\textbf{Decay}. Decay means eliminating stale elements to better represent the newly-arrival ones. This capability is extremely important for online systems. Note that we distinguish the periodical decaying operations from deletion. When decaying the vector, we may not  acknowledge what and how many elements have been wiped out. By contrast, we are aware of exactly what elements will be deleted during deletion. Besides, decaying always leads to both false positive and false negative errors, while correct deletion imposes no impact to the existing elements. There are several variants designed to support decay, including Generalized BF \cite{Generalized_BF}, Retouched BF \cite{Retouched_BF}, Stable BF \cite{Stable_BF}, Temporal Counting BF \cite{TCBF}, Double buffering \cite{Double_Buffering}, $A^2$ buffering \cite{Two_Active}, and Forgetful BF \cite{Forgetful_BF}. Generalized BF, Retouched BF, Stable BF, and Temporal Counting BF reset the bits in a single BF vector with diverse strategies. By contrast, Double buffering, $A^2$ buffering, and Forgetful BF flush the oldest BF among multiple BF vectors to remove the stale elements in a batched manner.

\textbf{Parallelism}. Parallelism is helpful to accelerate the query process and thereby increase the throughput of query. Typically, the variants with multiple independent BFs or segments can be accessed in parallel, include DBF \cite{Dynamic_BF}, DBA \cite{DBF_Array}, Par-BF \cite{Par_BF}, Scalable BF \cite{Scalable_BF}, BloomStore \cite{BloomStore}, FPF-MBF \cite{FP_Free_BF}, MPCBF \cite{Multi_partitioning_CBF}, Parallel BF \cite{Parallel_BF}, Combinatorial BF \cite{Combinatorial_BF}, OMASS \cite{OMASS}, Forest-structured BF \cite{Forest_structured_BF}, DLB-BF  \cite{LoadBance_BF}, and Ultra-fast BF \cite{UltraFast_BF}. The parallelism of query process decreases the computation time to $1/k$ thus improves the query throughput significantly. 

\textbf{False negatives}. Certainly, all the variants of BF incur inherited false positive, despite the proposed optimization techniques. A dozen of the variants, additionally, suffer from false negative errors. These variants loss the one-sided-error characteristic. Therefore, no matter the query result is positive or negative, the users have to further check the correctness from an off-chip hash table (if they have). Generalized BF \cite{Generalized_BF}, Retouched BF \cite{Retouched_BF}, Stable BF \cite{Stable_BF} reset the bits to wipe the stale elements thereby decreasing the false positive rate. The decay strategies may delete legal elements hence leading to false negatives. Similarly, deleting elements from the standard DBF and DBA vectors via resetting of bits can also result in false negatives, if the reset bits experience hash collisions. The coding strategy in Compacted BF will also cause false negatives. Besides, the use of locality-sensitive hash functions may also lead to false negatives. Distance-sensitive BF \cite{DS_BF}, Locality-sensitive BF \cite{LSBF} and MLBF \cite{Multigranularity_LBF} employ locality-sensitive hash functions to maintain the distance between elements. However, it is still possible that two neighbouring elements are mapped into two distant cells or bits. As a consequence, false negative errors will be triggered for the approximate membership queries. Beyond the above major reasons, the overflow of counters in Variable length Signature \cite{Variable_length_BF}, as well as hash collisions in kBF \cite{KBF}, also result in false negative errors.

\subsubsection{Complexity comparison}
To achieve the above capabilities, the resultant penalty is higher computation complexity for insertion, query, and deletion, as well as higher complexity for memory access. The last four columns in Table \ref{table_feature} detail the considered complexities for each variant. Generally, the parameter $k$ denotes the number of hash functions for each BF vector, $n$ counts the number of elements in the set $S$. Other dedicated parameters can find their meanings in corresponding literatures. Note that, the complexities of several variants are non-deterministic, hence we show both the upper bound and lower bound. Due to space limitation, we will not elaborate the complexities one by one but spot several representative or special variants. Interested readers can refer to the references. 

Inserting an element means mapping the element into the bit or cell vector. Definitely, the insertion complexity is mainly determined by how many hash functions have been employed to map the element. For instance, Adaptive BF \cite{Adaptive_BF} records the multiplicity of an element with the number of hash functions, thereby the insertion complexity is $O(k\mathrm{+}N\mathrm{+}1)$, where $N$ is the highest multiplicity. The Bloomier filter \cite{Bloomier_BF} encodes all the elements recursively with multiple cascaded BFs. Accordingly, its insertion complexity is $O(n\log n)$, and query complexity is $O(\lambda k)$, where $\lambda$ is the number of BFs in Bloomier filter. 

Generally, the query complexity highly depends on the insertion complexity, e.g., CBF \cite{Counting_BF}, kBF \cite{KBF}, Ternary BF \cite{Ternary_BF}, Deletable BF \cite{Deletable_BF}, IBF \cite{IBF}, Stable BF \cite{Stable_BF}, Temporal CBF \cite{TCBF}, etc. By contrast, $k$-mer BF \cite{k-mer_BF} records the $k$-mers of a sequent dataset. For insertion, the $k$ hash functions map the sequence into $k$ bits and set them as 1s. When querying, besides of the queried sequence, one (one-sided $k$-mer BF) or both (two-sided $k$-mer BF) of the neighbouring sequences will also be queried, resulting in $O(2k)$ and $O(3k)$ query complexity, respectively. The other variants which must query against multiple BFs or segments to identify the membership of an element also need higher query complexity, e.g., DBF \cite{Dynamic_BF}, DBA \cite{DBF_Array}, Matrix BF \cite{Matrix_BF}, BloomStore \cite{BloomStore}, etc. In particular, the query complexities of EABF \cite{Energy_Efficient_BF}, Pipelined BF \cite{Pipelined_BF1} and its variants \cite{Pipelined_BF2} \cite{FullyPipelined_BF} are $O(k_1)$ ($k_1\leq k$) for negative or less accurate answers, while $O(k)$ for positive or more trustable results. 

As shown in Table \ref{table_feature}, the deletion operation (if supported) incurs the same complexity as a query operation since BFs have to query the existence of an element before actually deleting it. The only outlier is Variable length signatures \cite{Variable_length_BF}, which inserts an element by setting $q\mathrm{\leq}t\mathrm{\leq} k$ bits as 1s and declares element $x \mathrm{\in} S$ if at least $q$ bits are non-zero. Therefore, deleting an element must reset $k\mathrm{-}d$ ($d\mathrm{<} q$) bits as 0s, such that the element will not be recognized as a member. By contrast, the complexity of memory access quantifies the number of memory access for a query process. For variants in which the hash functions share the entire bit or cell vector (e.g., CBF \cite{Counting_BF}, Adaptive BF \cite{Adaptive_BF}, FP-CBF \cite{Fingerprints_BF}, IBF \cite{IBF}, IBLT \cite{IBLT}, ICBF \cite{ICBF}, etc.), arbitrary query has to read the $k$ target bits with $k$ memory accesses. The reason is that the $k$ target bits disperses randomly in the vector, and one memory access may not cover multiple target bits. MPCBF \cite{Multi_partitioning_CBF}, Bloom-1 \cite{Bloom_1} and OMASS \cite{OMASS}, however, only require exactly one memory access since the existence information for an element is stored in one block with a word-size length. 

\section{Summary and Open Issues}\label{sec:Summary}
Heretofore, tens of variants are proposed for diverse purposes, enriching the BF paradigm. Table \ref{table_feature} qualitatively summaries the key capabilities and complexities of BF and its variants. These variants leverage the component(s) in the basic framework to improve the performance or generalize the BF to more scenarios. Based on the review of current literature on optimization of BF, we now outline some open issues that should be considered in the future. 

\textbf{Implementation in extreme hardware}. BF is easy-deployable, however, implementing BFs in extreme hardware is still problematic. Firstly,  in light-weight hardware (wireless sensors, RFIDs), the memory, bandwidth, and computation unit are scare resources. As a consequence, deploying large-scale or complicated variants is not possible. To resolve this problem, one may pre-clean the dataset and only store the necessary elements, or alternatively, sacrifice the FPR and overload the bit vector to accommodate all the elements. As for the transmission, besides of shrinking the length of BF, one may reasonably utilize the bandwidth and send the vector when the channel is not busy. Secondly, newly advanced hardware (TCAM, CAM, programmable hardware like FPGA, ASIC), by contrast, bring great opportunities for the implementation of BF and its variants. Several existing proposals jointly leverage both RAM and flash memory to extend their capacity without significant damage of access speed. The TCAM or FPGA-embedded devices, on the contrary, will lose their superiorities if the BFs are deployed in a traditional way. This open issue calls for extra efforts. 

\textbf{Extension or downsizing at arbitrary scale}. The existing variants to support capacity extension or downsizing (e.g., DBF \cite{Dynamic_BF}, DBA \cite{DBF_Array}, Matrix BF \cite{Matrix_BF}, Scalable BF \cite{Scalable_BF}, etc.) can only be resized at the level of sub-BFs. That is, only an entire sub-BF can be added or deleted. However, storing only one element with an added sub-BF is uneconomical. By contrast, overloading a sub-BF results in unacceptable false positive rate. An ideal extension scheme, however, should have the ability to extend the BF with an arbitrary scale. This vision brings the flexibility of implementing BF at any scale. 

\textbf{Representation of inter-element relationships}. Basically, standard BF and most of its variants represents elements independently without any consideration of the inner relationship between them. Exceptions only include those variants which employ locality sensitive hash functions to map close elements into neighbouring or same bits, e.g., Distance-sensitive BF \cite{DS_BF}, LSBF \cite{LSBF} and MLBF \cite{Multigranularity_LBF}. Besides of the distance between elements, other kinds of relationships are still beyond the capability of the existing variants. For instance, the BFs are capable to record the nodes in a multicast tree, but fail to tell the parents and children of any node. Therefore, we envision new variants which not only record the membership of the elements, but also represent the inner dependency of the elements. Besides, researchers are trying to learn the proper location for elements in BFs (also other data structures), so that there are no false positive errors \cite{learned_BF_1} \cite{learned_BF_2}.

\textbf{Alternates of BFs}. Besides of optimizing the BF framework, another approach is to propose alternates with better performance. Recently, Cuckoo filter \cite{Cuckoo_filter}, Quotient filter \cite{Quotient_filter}, and their variants \cite{Simplification_Cuckoo}  \cite{Adaptive_Cuckoo}  \cite{Dynamic_Cuckoo} \cite{Consistent_CF}  \cite{Counting_Quntient} generate wide attention due to their similar or even more integrated functionalities as the BF framework. Unlike BFs, these hash tables store the fingerprints of elements directly. Cuckoo filter \cite{Cuckoo_filter}  is a redesigned hash table to support membership query based on the theory of Cuckoo hashing  \cite{Cuckoo_table} \cite{Partial_key_Cuckoo}. The Quotient filter (QF) \cite{Quotient_filter} is a hash table of slots to store the fingerprints of elements with quotienting technique \cite{Quotienting}. The fingerprint of an element is divided into two parts, i.e., the $q$ most significant bits as quotient, and the $r$ least significant bits as remainder. A remainder is stored in the slot suggested by the quotient. We vision that in the future, the Cuckoo filters and Quotient filters will be further improved and other alternative data structures will be presented.

\textbf{Extensive applications of BFs}. Hitherto, BFs have been widely employed in various systems. More applications may occur in near future, beyond of communications (e.g., cognitive radio networks, wireless sensor networks, device-to-device communications, smart grid, etc.), networking (e.g., packet forwarding and routing, web caching, gossiping, resource discovering, scheduling, etc.), and database (e.g., information retrieving, recommendation, record linkage, duplication, anonymization, etc.). In Bioinformatics, BFs are employed to represent sequenced genomes, biometric information such as iris, face, handshape, fingerprint, etc. In the coming applications, BFs can be customized for diverse design goals partially with the optimization techniques summarized by us.

\section{Conclusion} \label{sec:Conclusion}
BF has been widely applied in the society of communications and networking. As we stated in this survey, when such a data structure is employed, the users can redesign it to suit their contexts. In this survey, we review the existing BF variants from mainly two dimensions, i.e., performance and generalization. To improve the performance, dozens of variants devote themselves to reducing the false positives and implementation costs. Besides, tens of variants generalize the BF framework in more scenarios by diversifying the input sets and enriching the output functionalities. Specifically, to reduce false positives, the existing BF variants utilizing prior knowledge,  select optimal hash functions, generate multiple BFs and queries, reset the bits in vectors, or represent elements differentially. To further ease the implementation of BF, a dozen variants are proposed to optimize its computation cost, memory access, space efficiency, and energy usage. Besides the representation of general sets, variants are investigated to represent multisets, dynamic sets, weighted sets, key-values,  sequence sets, and spatial sets. Lastly, besides element insertion and query, more BF functionalities are enabled, such as element deletion, element decay,  approximate membership query, and semantic enrichment. Additionally, we also classify the existing optimization techniques from the perspective of the components of BF, and then compare them in terms of both functionality and complexity.  We expect more applications and redesigns of BF in the next generation of networks, communication systems and beyond.

 


\ifCLASSOPTIONcaptionsoff
  \newpage
\fi

%

\begin{IEEEbiographynophoto} {Lailong Luo} received his B.S. and M.S. degree at the school of systems engineering from National University of Defence Technology, Changsha, China, in 2013 and 2015, respectively. He is currently working toward a Ph.D degree in the school of systems engineering, National University of Defense Technology, Changsha, China. His research interests include probabilisitic data structures and data analysis.
\end{IEEEbiographynophoto}

\begin{IEEEbiographynophoto} {Deke Guo}received his B.S. degree in industry engineering from Beijing University of Aeronautic and Astronautic, Beijing, China, in 2001, and the Ph.D. degree in management science and engineering from National University of Defense Technology, Changsha, China, in 2008. He is a Professor with the College of Information System and Management, National University of Defense Technology, Changsha, China. His research interests include probabilisitic data structures, software-defined networking, data center networking, wireless and mobile systems, and interconnection networks. He is a member of ACM and IEEE.
\end{IEEEbiographynophoto}

\begin{IEEEbiographynophoto} {Richard T.B. Ma} received the Ph.D. degree in Electrical Engineering in May 2010 from Columbia University, New York. During his Ph.D. study, he worked as a research intern at IBM T. J. Watson Research Center, Yorktown Heights, NY, USA, and Telefonica Research, Barcelona, Spain. He is currently a Research Scientist in Advanced Digital Science Center, University of Illinois, USA, and an Assistant Professor in School of Computing at National University of Singapore. His research interests include distributed systems and network economics.
\end{IEEEbiographynophoto}

\begin{IEEEbiographynophoto} {Ori Rottenstreich} is an assistant professor at the Department of Computer Science and the Department of Electrical Engineering of the Technion, Haifa, Israel. He is also the chief scientist of Orbs. His main research interest is computer networks and blockchain technologies. In 2015-2017 he was a Postdoctoral Research Fellow at the Department of Computer Science, Princeton University. Earlier, he received the BSc in Computer Engineering (summa cum laude) and PhD degree from the Technion in 2008 and 2014, respectively.
\end{IEEEbiographynophoto}

\begin{IEEEbiographynophoto} {Xueshan Luo} received his B.E. degree in information engineering from Huazhong Institute of Technology, Wuhan, China, in 1985, and his M.S. and Ph.D degrees in system engineering from the National  University of Defense Technology, Changsha, China, in 1988 and 1992, respectively. Currently, he is a professor in the College of Information System and Management, National University of Defense Technology. His research interests are in the general areas of information system and operation research.
\end{IEEEbiographynophoto}



\begin{thebibliography}{99}
\balance

\bibitem{BF}  B. H. Bloom, ``Space/time trade-offs in hash coding with allowable errors,'' \emph{Communications of the ACM}, vol. 13, no. 7, pp. 422-426, 1970.

\bibitem{NDN_1} F. Angius, M. Gerla, G. Pau, ``Bloogo: Bloom filter based gossip algorithm for wireless NDN,'' in \emph{Proc. ACM workshop on Emerging Name-Oriented Mobile Networking Design-Architecture, Algorithms, and Applications,} Hilton Head, SC, USA, 2012.

\bibitem{Multiclass_BF} D. Li, H. Cui, Y. Hu, Y. Xia, X. Wang, ``Scalable data center multicast using multi-class Bloom filter,'' in \emph{ Proc. IEEE ICNP,} Vancouver, BC Canada,  2011.

\bibitem{Multicast_1} X. Tian, Y. Cheng, ``Loop mitigation in Bloom filter based multicast: A destination-oriented approach,'' in \emph{Proc. IEEE INFOCOM,} Orlando, FL, USA, 2012.

\bibitem{Multicast_2} I. Nikolaevskiy, A. Lukyanenko, T. Polishchuk, V. Polishchukc, A. Gurtov, ``isBF: Scalable in-packet Bloom filter based multicast,'' \emph{Computer Communications,} vol. 70, pp. 79-85, 2015.

\bibitem{Routing_1} D. Guo, Y. He, Y. Liu, ``On the feasibility of gradient-based data-centric routing using Bloom filters,'' \emph{IEEE TPDS,} vol. 25, no. 1, pp. 180-190, 2014.

\bibitem{Multicast_ICL} G. Chen, D. Guo, L. Luo, B. Ren, ``Optimization of multicast source-routing based on boom filter,'' \emph{IEEE Communication Letters,} vol. 22, no. 4, pp. 700-703, 2018.

\bibitem{routing_comnet} D. Guo, Y. He, P. Yang, ``Receiver-oriented design of bloom filters for data-centric routing,'' \emph{Computer Networks,} vol. 54, no. 1, pp. 165-174, 2010. 

\bibitem{Paradox_BF} O. Rottenstreich, I. Keslassy, ``The Bloom paradox: When not to use a Bloom filter? '' in \emph{ Proc. IEEE INFOCOM}, Orlando, FL, USA, 2012.

\bibitem{WebCache_1} H. Alexander, I. Khalil, C. Cameron, Z. Tari, A. Zomaya, ``Cooperative web caching using dynamic interest-tagged filtered Bloom filters,''  \emph{IEEE TPDS,} vol. 26, no. 11, pp. 2956-2969, 2015.

\bibitem{Monitoring_1} Y. Li, R. Miao, C. Kim, M. Yu, ``FlowRadar: A better NetFlow for data centers,'' in \emph{Proc. USENIX NSDI,} Santa Clara, CA, USA, 2016. 

\bibitem{Security_1} E.A .Durham, M. Kantarcioglu, Y. Xue, C. Toth, M. Kuzu, B. Malin, ``Composite Bloom filters for secure record linkage,'' \emph{IEEE TKDE,} vol. 26, no. 12, pp. 2956-2968, 2014.

\bibitem{Concatenated_BF} M. Moreira, R. Laufer, P. Velloso, and O. Duarte, ``Capacity and robustness tradeoffs in Bloom filters for distributed applications,''  \emph{IEEE TPDS,} vol. 23, no. 12, pp. 2219-2230, 2012.

\bibitem{ContentDlivery_1} B.M. Maggs, R.K. Sitaraman, ``Algorithmic nuggets in content delivery,'' in \emph{Proc. ACM SIGCOMM,} London, United Kingdom, 2015.

\bibitem{BooleanQueries_1} X. Zhu, R. Hao, S. Jiang, H. Chi, H. Li, ``Verification of boolean queries over outsourced encrypted data based on Counting Bloom filter,'' in \emph{Proc. IEEE GLOBECOM,} San Diego, CA, USA, 2015.  

\bibitem{PublishSubscribe_1} A. Margara, G. Cugola, ``High-performance publish-subscribe matching using parallel hardware,'' \emph{IEEE TPDS,} vol. 25, no. 1, pp. 126-135, 2014.

\bibitem{Spatial_BF} L. Calderoni, P. Palmieri, D. Maio,  ``Location privacy without mutual trust: The spatial Bloom filter,''  \emph{Computer Communications,} vol. 68,  pp. 4-12, 2015.

\bibitem{BloomStore} G. Lu, Y.J. Nam, D.H.C. Du, ``BloomStore: Bloom filter based memory-efficient key-value store for indexing of data de-duplication on flash,'' in \emph{ Proc. IEEE MSST}, Pacific Grove, Canada,  2012.

\bibitem{KBF} S. Xiong, Y. Yao, Q. Cao, T. He, ``kBF: A Bloom filter for key-value storage with an application on approximate state machines,'' in \emph{ Proc. IEEE INFOCOM}, Toronto, Canada,  2014.

\bibitem{ICBF} L. Luo, D. Guo, J. Wu, O. Rottenstreich, Q. He, Y. Qin, X. Luo, ``Efficient multiset synchronization,'' \emph{IEEE/ACM ToN,} vol. 25, no. 2, pp. 1190-1205, 2017.

\bibitem{IBLT} M.T. Goodrich, M. Mitzenmacher, ``Invertible Bloom lookup tables,'' in \emph{Proc. Annual Allerton Conference on Communication, Control, and Computing,} Monticello, Illinois, USA, 2011.

\bibitem{SetRecon} D. Chen, C. Konrad, K. Yi, W. Yu, Q. Zhang, ``Robust set reconciliation,'' in \emph{Proc. ACM SIGMOD,} Snowbird, Utah, USA, 2014.

\bibitem{Set_sync_tkde} L. Luo, D. Guo, X. Zhao, J. Wu, O. Rottenstreich, X. Luo, ``Near-accurate multiset reconciliation,'' \emph{IEEE TKDE,}  DOI10.1109/TKDE.2018.2849997, 2018.

\bibitem{set_sync_CBF} D. Guo, M. Li, ``Set reconciliation via counting bloom filters,'' \emph{IEEE TKDE,} vol. 25,  no. 10, pp. 2367-2380, 2013.

\bibitem{Duplication_1} S. Dutta, A. Narang, S.K. Bera, ``Streaming quotient filter: a near optimal approximate duplicate detection approach for data streams,'' in \emph{Proc. VLDB,} Riva del Garda, Trento, Italy, 2013. 



\bibitem{Biometric_1} J. Hermans, B. Mennink, R. Peeters, ``When a Bloom filter is a doom filter: Security assessment of a novel iris biometric template protection system,'' in \emph{Proc. BIOSIG,} Darmstadt, Germany, 2014.

\bibitem{Biometric_2} M. Gomez-Barrero, C. Rathgeb, J. Galbally, J. Fierrez, C. Busch, ``Protected facial biometric templates based on local gabor patterns and adaptive Bloom filters,'' in \emph{Proc. ICPR,} Stockholm, Sweden, 2014.

\bibitem{Navigation_1} P. Jiang, Y. Ji, X. Wang, J. Zhu, Y. Cheng, ``Design of a multiple Bloom filter for distributed navigation routing,'' \emph{IEEE Transactions on SMC: Systems,} vol. 44, no. 2, pp. 254-260, 2014.

\bibitem{ExistingSurvey_1} S. Tarkoma, C.E. Rothenberg, E. Lagerspetz, ``Theory and practice of Bloom filters for distributed systems,'' \emph{IEEE Communications Surveys and Tutorials,} vol.14, no. 1, pp.131-155, 2012.

\bibitem{ExistingSurvey_2} A. Broder, M. Mitzenmacher, ``Network applications of Bloom filters: A survey,'' \emph{Internet mathematics,} vol.1, no. 4, pp. 485-509, 2004.

\bibitem{ExistingSurvey_3} S. Geravand, M. Ahmadi, ``Bloom filter applications in network security: A state-of-the-art survey,'' \emph{Computer Networks,} vol.57, no. 18, pp. 4047-4064, 2013.

\bibitem{BF_fpr1}  L. L. Gremillion, ``Designing a Bloom filter for differential file access,''  \emph{Communications of the ACM},   vol. 25, no. 7, pp. 600-604, 1982.

\bibitem{BF_fpr2}  J. K. Mullin, ``A second look at Bloom filters,''  \emph{Communications of the ACM},   vol. 26, no. 8, 1983.

\bibitem{BF_fpr3} P. Bose, H. Guo, E. Kranakis, ``On the false-positive rate of Bloom filters,'' \emph{Information Processing Letters},  vol. 108, no. 4, pp. 210-213, 2008.

\bibitem{BF_fpr4} R. L. Graham, D. E. Knuth, O. Patashnik, ``Concrete Mathematics,'' \emph{Addison-Wesley}, 2nd edition, 1994.

\bibitem{Counting_BF} L. Fan, P. Cao, J. Almeida, A.Z. Broder, ``Summary cache: A scalable wide-area web cache sharing protocol,'' \emph{IEEE/ACM ToN,} vol. 8, no. 3, pp. 281-293, 2000.

\bibitem{cache_1} K. Chen, P. Wu, B. Lai, ``Reduce data coherence cost with an area efficient double layer counting bloom filter,''  in \emph{Proc. PPAP,} Taipei, Taiwan, 2012. 

\bibitem{cache_2} E. Papapetrou, E. Pitoura, K. Lillis ``Speeding-up cache lookups in wireless ad-hoc routing using bloom filters,'' in \emph{Proc. IEEE PIMRC,} Berlin, Germany, 2005. 

\bibitem{cache_3} S. Kavitha, R. Thanuja, A. Umamakeswari, ``Updating distributed cache mechanism using bloom filter for asymmetric cryptography in large wireless networks,''  \emph{Indian Journal of Science and Technology,}  vol. 9, no. 48, pp. 1-6, 2016. 

\bibitem{Trie_search} H. Lim, K. Lim, N. Lee, K.H. Park, ``On adding Bloom filters to longest prefix matching algorithms,'' \emph{IEEE TC}, vol. 63, no. 2, pp. 411-423, 2014.

\bibitem{Prefix_1} G. Park, M. Kwon, `` An enhanced bloom filter for longest prefix matching,''   in \emph{Proc. IEEE/ACM IWQoS,}  Montreal, Canada, 2013. 

\bibitem{inpacket_1} C. Rothenberg, C. Macapuna, M. Magalhaes, F. Verdi, A. Wiesmaier, ``In-packet Bloom filters: Design and networking applications,'' \emph{Computer Networks,} vol. 55, no. 6, pp. 1364-1378, 2011. 

 \bibitem{FP_Free_BF} J. Tapolcai, J. Biro, P. Babarczi, A. Gulyas, Z. Heszberger,  D. Trossen, ``Optimal false-positive-free Bloom filter design for scalable multicast forwarding,'' \emph{IEEE/ACM ToN,} vol. 23, no. 6, pp. 1832-1845, 2015.

\bibitem{NDN_2} Y. Wang, T. Pan, Z. Mi, H. Dai, X. Guo, T. Zhang, B. Liu, Q. Dong, ``Namefilter: Achieving fast name lookup with low memory cost via applying two-stage bloom filters,'' in \emph{Proc. IEEE INFOCOM,} Turin, Italy, 2013. 

\bibitem{NDN_3} W. Quan, C. Xu, J. Guan, H. Zhang, L. Grieco, ``Scalable name lookup with adaptive prefix bloom filter for named data networking,'' \emph{IEEE Communications Letters,} vol. 18, no. 1, pp. 102-105, 2014. 

\bibitem{adhoc_1} J. Trindade, T. Vazao, ``HRAN - A scalable routing protocol for multihop wireless networks using bloom filters,''  \emph{Wired/Wireless Internet Communications,}  pp. 434-445, 2011.

\bibitem{adhoc_2} J. Trindade, T. Vazao, ``Routing on large scale mobile ad hoc networks using bloom filters,''  \emph{ Ad Hoc Networks,}  vol. 23, pp. 34-51, 2014. 

\bibitem{adhoc_3} F. Klingler, R. Cohen, C. Sommer, F. Dressler, ``Bloom hopping: Bloom filter based 2-Hop Neighbor Management in VANETs,''  \emph{IEEE TMC,} 10.1109/TMC.2018.2840123, 2018.  

\bibitem{WSN_1} X. Li, J. Wu, J. Xu, ``Hint-based routing in WSNs using scope decay bloom filters,'' in \emph{Proc. IEEE NAS,} Shenyang, China, 2006.  

\bibitem{WSN_2} P. Hebden, A. Pearce, ``Data-centric routing using Bloom filters in wireless sensor networks,''  in \emph{Proc. IEEE ICISIP,} Bangalore, India, 2006.   

\bibitem{WSN_3} A. Reinhardt, O. Morar, S. Santini, S. Zoller, R. Steinmetz, ``CBRF: Bloom filter routing with gradual forgetting for tree-structured wireless sensor networks with mobile nodes,''  in \emph{Proc. IEEE WOWMOM,} San Francisco, USA, 2012.  

 

\bibitem{location_pravicy_1} M. Grissa, A. Yavuz, B. Hamdaoui, ``Cuckoo filter-based location-privacy preservation in database-driven cognitive radio networks,'' in \emph{Proc. IEEE WSCNIS,} Hammamet, Tunisia 2015.   

\bibitem{location_pravicy_2} M. Grissa, A. Yavuz, B. Hamdaoui, ``Location privacy preservation in database-driven wireless cognitive networks through encrypted probabilistic data structures,''  \emph{IEEE TCCN,} vol. 3, no. 2, pp. 255-266, 2017.

\bibitem{Biometric_4} C. Rathgeb, F. Breitinger, C. Busch, H. Baier, ``On application of bloom filters to iris biometrics,'' \emph{IET Biometrics,} vol. 3, no. 4, pp. 207-218, 2014.

\bibitem{Linkage_Privacy} D. Karapiperis, V. Verykios, ``An LSH-based blocking approach with a homomorphic matching technique for privacy-preserving record linkage,''  \emph{IEEE TKDE,} vol. 27, no. 4, pp. 909-921, 2015. 

\bibitem{MSN_pravicy} E. Oriero, K. Rabieh, M. Mahmoud, M. Ismail, E. Serpedin, K. Qaraqe, ``Trust-based and privacy-preserving fine-grained data retrieval scheme for MSNs,'' in \emph{Proc. IEEE WCNC,} Doha, Qatar, 2016. 

\bibitem{Detail_privacy} M. Alaggan, S. Gambs, S. Matwin, M. Tuhin, ``Sanitization of call detail records via differentially-private bloom filters,'' in \emph{Proc. IFIP DBSec,} Fairfax, VA, USA, 2015.  

\bibitem{COMST_info_centric} R. Tourani, S. Misra, T. Mick, and G. Panwar, ``Security, privacy, and access control in information-centric networking: A survey,''  \emph{ IEEE Communications Surveys and Tutorials,} vol. 20, no. 1, pp. 566-600, 2018.   
 
\bibitem{COMST_pitfalls} L. Demir, A. Kumar, M. Cunche,  C. Lauradoux, ``The pitfalls of hashing for privacy,'' \emph{ IEEE Communications Surveys  and Tutorials,} vol. 20, no. 1, pp. 551-565, 2018.
 
\bibitem{COMST_NSF}  M. Ambrosin , A. Compagno, M. Conti,  C. Ghali, G. Tsudik, ``Security and privacy analysis of national science foundation future internet architectures,'' \emph{ IEEE Communications Surveys  and Tutorials,} vol. 20, no. 2, pp. 1418-1442, 2018.
 
\bibitem{COMST_bitcoin_1}  M. Conti, S. Kumar,  C. Lal, S. Ruj, ``A survey on security and privacy issues of bitcoin,''  \emph{IEEE Communications Surveys  and Tutorials,} DOI 10.1109/COMST.2018.2842460, 2018.
 
\bibitem{COMST_bitcoin_2}  M. C. K. Khalilov,  A. Levi, ``A survey on anonymity and privacy in bitcoin-like Digital cash dystems,''  \emph{ IEEE Communications Surveys and Tutorials,} vol. 20, no. 3, pp. 2543-2585, 2018.

\bibitem{security_1} K. Rabieh, M. Mahmoud, K. Akkaya, S. Tonyali, ``Scalable certificate revocation schemes for smart grid AMI networks using bloom filters,'' \emph{IEEE TDSC,} vol. 14, no. 4, pp. 420-432, 2017. 


\bibitem{Dissemination} M. Cisse, N. Usunier, T. Artieres, ``Decentralised Peer-to-Peer data dissemination in wireless sensor networks,'' \emph{Pervasive and Mobile Computing,}  vol. 40,  pp. 242-266, 2017.  

\bibitem{Classification} R. Carbajo, C. Goldrick, ``Robust bloom filters for large multilabel classification tasks,'' \emph{Advances in Neural Information Processing Systems,} pp. 1851-1859, 2013.  

\bibitem{Discovery} K. Choi, D. Wiriaatmadja, ``Discovering mobile applications in cellular device-to-device communications: Hash function and bloom filter-based approach,'' \emph{IEEE TMC,}  vol. 15, vol. 2,  pp. 336-349, 2016.  
 
\bibitem{optihash_BF}L. Carrea, A. Vernitski, M. Reed,  ``Optimized hash for network path encoding with minimized false positives,'' \emph{Computer Networks,} vol. 58, no.11,  pp. 180-191, 2014. 
 
\bibitem{Multistage_BF1} J. Tapolcai, A. Gulyas, Z. Heszbergery, J. Biro, ``Stateless multi-stage dissemination of information: Source routing revisited,''  in \emph{ Proc. IEEE GLOBECOM}, Anaheim, California, USA, 2012.

\bibitem{Multistage_BF2} W. Yang, D. Trossen, J. Tapolcai, ``Scalable forwarding for information-centric networks,'' in \emph{ Proc. IEEE ICC-NGN}, Budapest, Hungary, 2013.

\bibitem{PSIRP} D. Lagutin, K. Visala, S. Tarkoma, ``Publish/subscribe for Internet: PSIRP perspective,''  \emph{Towards the Future Internet-Emerging Trends from European Research,} Amsterdam, The Netherlands: IOS Press, 2010.

\bibitem{Trie_BF} H.M. Ju,  H. Lim, ``On reducing false positives of a Bloom filter in trie-based algorithms,''  in \emph{ Proc. ACM/IEEE ANCS}, Marina del Rey, CA, USA, 2014.

\bibitem{Cross_checking_BF} H. Lim, N. Lee,  J. Lee, C. Yim,  ``Reducing false positives of a Bloom filter using Cross-Checking Bloom filters,'' \emph{Applied Mathematics and Information Sciences,} vol. 8, no. 4, pp. 1865-1877, 2014.

 \bibitem{Complement_BF} H. Lim, J. Lee,  C. Yim,  ``Complement Bloom filter for identifying true positiveness of a Bloom filter,'' \emph{IEEE Communications Letters,} vol. 19, no. 11, pp. 1905-1908, 2015.

 \bibitem{Yes_No_BF} L. Carrea, A. Vernitski, M. Reed, ``Yes-no Bloom filter: A way of representing sets with fewer false positives,'' \emph{arXiv preprint},  arXiv:1603.01060, 2016.

\bibitem{Retouched_BF} B. Chazelle, J. Kilian, R. Rubinfeld, A. Tal, ``Retouched Bloom filters: allowing networked applications to trade off selected false positives against false negatives,'' in \emph{ Proc. ACM CoNEXT}, Lisboa, Portugal,  2006.

\bibitem{Generalized_BF} R.P. Laufer, P.B. Velloso, O.C.MB. Duarte, ``A generalized Bloom filter to secure distributed network applications, '' \emph{Computer Networks Amsterdam,} vol. 55, no.8,  pp. 1804-1819, 2011.

\bibitem{Multi_partitioning_CBF} K. Huang, J. Zhang, D. Zhang, G. Xie, ``A Multi-partitioning approach to building fast and accurate counting Bloom filters,''  in \emph{ Proc. IEEE IPDPS}, Cambridge, MA, USA, 2013.

\bibitem{Power2choice} S. Lumetta, M. Mitzenmacher, ``Using the power of two choices to improve Bloom filters,'' \emph{Internet Mathematics,}  vol.4, no. 1, pp. 17-33, 2007.

\bibitem{Hash_partition} F. Hao, M. Kodialam, T.V. Lakshman, ``Building high accuracy Bloom filters using partitioned hashing,'' in \emph{Proc. ACM SIGMETRICS,} San Diego, CA, USA, 2007.

\bibitem{EGH_filter} S.Z. Kiss, E. Hosszu, J. Tapolcai, L. Rónyai, O. Rottenstreich, ``Bloom filter with a false positive free zone,'' in \emph{Proc. IEEE INFOCOM,}  Honolulu, HI, USA 2018.

\bibitem{ChineseRemainder} D. Pei, S. Arto, C. Ding , ``Chinese remainder theorem: applications in computing, coding, cryptography,'' \emph{World Scientific,}1996.

\bibitem{CGT}  D. Z. Du, F. Hwang, ``Combinatorial group testing and its applications,''  \emph{World Scientific,} 1993.

\bibitem{VIncrement_CBF} O. Rottenstreich, Y. Kanizo, I. Keslassy, ``The variable-increment counting Bloom filter,'' in \emph{ Proc. IEEE INFOCOM}, Orlando, Florida, USA, 2012.

\bibitem{Fingerprints_BF} S. Pontarelli, P. Reviriego, J. A. Maestro, ``Improving counting Bloom filter performance with fingerprints,'' \emph{Information Processing Letters,} vol. 116, no. 4, pp. 304-309, 2016.

\bibitem{LessHash} A. Kirsch, M. Mitzenmacher,``Less hashing, same performance: Building a better Bloom filter,'' \emph{Random Struct. Algorithms,} vol. 33, no. 2, pp. 187-218, 2006.

\bibitem{OneHash_BF} J. Lu, T. Yang, Y. Wang, H. Dai, L. Jin, H. Song, B. Liu, ``One-hashing Bloom filter,'' in \emph{Proc. IEEE/ACM IWQoS}, Portland, OR, USA,  2015. 

\bibitem{LessHash1} A. Kirsch, M. Mitzenmacher,``Building a better Bloom filter,'' in \emph{Proc. Annual European Symposium on Algorithms,} Palma de Mallorca, Spain, 2005.

\bibitem{Space_Code_BF}  A. Kumar, J.J. Xu, L. Li, J. Wang, ``Space-code Bloom filter for efficient traffic flow measurement,'' in \emph{Proc. ACM IMC,}  Miami Beach, FL, USA, 2003.
 
\bibitem{Dynamic_BF} D. Guo, J. Wu, H. Chen, Y. Yuan, X. Luo, ``The Dynamic Bloom filters,''  \emph{IEEE TKDE,} vol. 22, no. 1, pp. 120-133, 2010.

\bibitem{DBF_Array} J. Wei, H. Jiang, K. Zhou, D. Feng,  ``DBA: A dynamic Bloom filter array for scalable membership representation of variable large data sets,'' in \emph{ Proc. IEEE MASCOTS,} Raffles Hotel, Singapore, 2011.

\bibitem{Par_BF} Y. Liu, X. Ge, D.H.C. Du, X. Huang, ``Par-BF: A parallel partitioned Bloom filter for dynamic data sets,''  \emph{The International Journal of High Performance Computing Applications,} vol. 30, no. 3, pp. 259-275, 2016.

\bibitem{Bloom_1} Y. Qiao, T. Li, S. Chen, ``Fast Bloom filters and their generalization,'' \emph{IEEE TPDS,} vol. 25, no. 1, pp. 93-103, 2014.

\bibitem{OMASS} M. Mitzenmacher, P. Reviriego, S. Pontarelli, ``OMASS: One memory access set separation,''  \emph{IEEE TKDE,} vol. 28, no. 7, pp. 1940-1943, 2016.

\bibitem{Parallel_BF} B. Xiao, Y. Hua, ``Using Parallel Bloom filters for multiattribute representation on network services,''  \emph{IEEE TPDS,} vol. 12, no. 1, pp. 20-32, 2010.

\bibitem{Bloomier_BF} Y. Lu, B. Prabhakar, F. Bonomi, ``The Bloomier filter: An efficient data structure for static support lookup tables,'' in \emph{ Proc. ACM-SIAM,} New Orleans, Louisiana, USA, 2004.

\bibitem{LoadBance_BF} H. Song, F. Hao, M. Kodialam, T. V. Lakshman,  ``IPv6 lookups using distributed and Load Balanced Bloom filters for 100Gbps core router line cards,'' in \emph{Proc. IEEE INFOCOM,} Rio de Janeiro, Brazil, 2009.

\bibitem{Combinatorial_BF} F. Hao, M. Kodialam, T. V. Lakshman, H. Song, ``Fast dynamic multiset membership testing using Combinatorial Bloom filters,'' in \emph{Proc. IEEE INFOCOM,} Rio de Janeiro, Brazil, 2009.

\bibitem{UltraFast_BF} J. Lu, Y. Wan, Y. Li, C. Zhang, H, Dai, Y. Wang, G.Zhang, B. Liu, ``Ultra-fast Bloom filters using SIMD techniques,'' in \emph{Proc. IEEE/ACM IWQoS}, Vilanova i la Geltrú, Spain, 2017. 

\bibitem{Compressed_BF} M. Mitzenmacher, ``Compressed Bloom filters,'' \emph{IEEE/ACM ToN,} vol. 10, no. 5, pp. 604--612, 2002. 

\bibitem{Compacted_BF} N. Mosharraf, A.P. Jayasumana, I. Ray, ``Compacted Bloom filter,'' in \emph{Proc. IEEE CIC,} Pittsburgh, PA, USA, 2016.

\bibitem{d_left_hash} F. Bonomi, M. Mitzenmacher, R. Panigrahy, S Singh, G. Varghese, ``An improved construction for counting Bloom filters,''  in \emph{ Proc. European Symposium on Algorithms,} Zurich, Switzerland, 2006.

\bibitem{additional_hash_BF}   M. Ahmadi, S. Wong, "A memory-optimized Bloom filter using an additional hashing function," in \emph{ Proc. IEEE GLOBECOM,} New Orleans, LA, USA, 2008.

\bibitem{Matrix_BF} S. Geravand, M. Ahmadi, ``A novel adjustable matrix Bloom filter-based copy detection system for digital libraries,'' in \emph{ Proc. IEEE CIT,} Paphos, Cyprus, 2011.

\bibitem{almost_perfect} A. Broder, M. Mitzenmacher, ``Using multiple hash functions to improve IP lookups'' in \emph{Proc. IEEE INFOCOM,} Alaska, USA,  2001.

\bibitem{almost_perfect1} Y. Lu, B. Prabhakar, F. Bonomi, ``Perfect hashing for network applications,'' in \emph{Proc. IEEE ISIT,}  Seattle, WA, USA, 2006.
 
\bibitem{Forest_structured_BF} G. Lu, B. Debnath, D.H.C. Du, ``A Forest-structured Bloom filter with flash memory,'' in \emph{ Proc. IEEE MSST,} Denver, Colorado, 2011.

\bibitem{Pipelined_BF1} I. Kaya, T. Kocak, ``Energy-efficient pipelined Bloom filters for network intrusion detection,'' in \emph{IEEE ICC,} Istanbul, Turkey, 2006.

\bibitem{Pipelined_BF2} T. Kocak, I. Kaya, ``Low-power Bloom filter architecture for deep packet inspection,'' \emph{IEEE Communications Letters,} vol. 10, no. 3, pp. 210-212, 2006.

\bibitem{FullyPipelined_BF}  M. Paynter, T. Kocak, ``Fully pipelined Bloom filter architecture,'' \emph{IEEE Communications Letters,} vol. 12, no. 11, pp. 855-857, 2008.

\bibitem{Energy_Efficient_BF} Y. Zhou, T. Song, X. Wang, ``EABF: Energy efficient self-adaptive Bloom filter for network packet processing,''  in \emph{ Proc. IEEE ICC,} Ottawa, Canada, 2012.

\bibitem{L_CBF} E. Safi, A. Moshovos, A. Veneris, ``L-CBF: A low-power, fast counting Bloom filter architecture,'' \emph{IEEE TVLSI,} vol. 16, no. 6, pp. 628-638, 2008.


\bibitem{Spectral_BF} S. Cohen, Y. Matias,  ``Spectral Bloom filters,'' in \emph{ Proc. ACM SIGMOD,} Madison, Wisconsin, USA, 2003.

\bibitem{Loglog_BF} Y. Yao, S. Xiong, H. Qi, Y. Liu, L.M. Tolbert, Q. Cao, ``Efficient histogram estimation for smart grid data processing with the loglog-Bloom-filter,'' \emph{IEEE Transactions on Smart Grid,} vol. 6, no. 1, pp. 199-208, 2015.

\bibitem{Adaptive_BF} Y. Matsumoto, H. Hazeyama, Y. Kadobayashi, ``Adaptive Bloom filter: A space-efficient counting algorithm for unpredictable network traffic,'' \emph{ IEICE Transactions on Information and Systems,}  vol. 91, no. 5, pp. 1292-1299, 2008. 

\bibitem{Shifting_BF} T. Yang, A. Liu, M. Shahzad, Y. Zhong, Q. Fu, Z. Li, ``A shifting Bloom filter framework for set queries,'' in \emph{Proc. IEEE VLDB,} New Delhi, India, 2016.

\bibitem{coungting_estimate}  M Durand, P Flajolet, ``Loglog counting of large cardinalities,'' \emph{European Symposium on Algorithms,} Berlin, Heidelberg, 2003.

 \bibitem{Variable_length_BF} Y. Lu, B. Prabhakar, F. Bonomi, ``Bloom filters: Design innovations and novel applications,'' in \emph{Proc. The Annual Allerton Conference on Communication, Control and Computing,} Monticello, Illinois, USA, 2005.

\bibitem{Scalable_BF} P. S. Almeida, C. Baquero, N. Preguica, D. Hutchison, ``Scalable Bloom filters,'' \emph{Information Processing Letters}, vol. 101, no. 6, pp. 255-261, 2007.

\bibitem{Weighted_BF} J. Bruck, J. Gao, A. Jiang,  ``Weighted Bloom filter,'' in \emph{ Proc. IEEE ISIT,} Seattle, WA, USA, 2006.

\bibitem{Popularity_Conscious_BF} M. Zhong, P. Lu, K. Shen, J. Seiferas, ``Optimizing data popularity conscious Bloom filters, '' in \emph{ Proc. ACM PODC,} Toronto, Ontario, Canada, 2008.

\bibitem{Dynamo} G. Decandia, D. Hastorun, M. Jampani, G. Kakulapati, A. Lakshman, A. Pilchin, S. Sivasubramanian, P. Vosshall, W. Vogels,``Dynamo: Amazon's highly available key-value store,''  in \emph{Proc. ACM SOSP,} Stevenson, Washington, USA,  2007.

\bibitem{Memcached} Memcached Website. Available: https://memcached.org, Oct. 2017.

\bibitem{Cassandra} Apache Foundation. Available: Cassandra Website, http:// cassandra.apache.org, Oct. 2017.

\bibitem{Redis} Redis Website. Available: http://redis.io/, Oct. 2017. 

\bibitem{BigTable} F. Chang, J. Dean, S. Ghemawat, W. C. Hsieh, D. A. Wallach, M. Burrows, T. Chandra, A. Fikes, R. E. Gruber, ``Bigtable: A distributed storage system for structured data,'' in \emph{Proc. USENIX OSDI}, Seattle, Washington, USA, 2006.

\bibitem{Btree} C. H. Wu, T. W. Kuo,  L. P. Chang, ``An efficient b-tree layer implementation for flash-memory storage systems,'' \emph{ACM TECS}, vol. 6, no. 3, 2007.

\bibitem{Bptree} S. Nath, A. Kansal, ``FlashDB: Dynamic self-tuning database for NAND flash,'' in \emph{Proc. ACM/IEEE IPSN}, Cambridge, Massachusetts, USA, 2007.

\bibitem{HashTable} A. Anand, C. Muthukrishnan, S. Kappes, A. Akella, S. Nath, ``Cheap and large cams for high performance data-intensive networked systems,'' in  \emph{ Proc. USENIX NSDI,}  San Jose, California, USA, 2010.

\bibitem{Kraken} D.E. Wood, S.L. Salzberg, ``Kraken: Ultrafast metagenomic sequence classification using exact alignments,'' \emph{Genome Biology}, vol. 15, no. 3, 2014.

\bibitem{Velvet}  D.R. Zerbino,  E. Birney ``Velvet: Algorithms for de novo short read assembly using de Bruijn graphs,'' \emph{Genome Research}, vol. 18, no. 5, pp. 821-829, 2008.

\bibitem{k-mer_BF} D. Pellow, D. Filippova, C. Kingsford, ``Improving Bloom filter performance on sequence data using  k-mer Bloom filters,''  \emph{Journal of Computational Biology,} vol. 24, no. 6, pp. 547-557, 2017.

\bibitem{Trie_BF0} G. Holley, R. Wittler, J. Stoye, ``Bloom filter Trie: An alignment-free and reference-free data structure for pan-genome storage,''  \emph{Algorithms for Molecular Biology,}  vol. 11, no. 3,  2016.

\bibitem{Deletable_BF} C. Rothenberg, C. Macapuna, F. Verdit,  M. Magalhaes, ``The deletable Bloom filter: A new member of the Bloom family,'' \emph{IEEE Communication Letters,} vol. 14, no. 6, pp. 557-559, 2010.

\bibitem{Ternary_BF} H. Lim, J. Lee, H. Byun, C. Yim, ``Ternary Bloom filter replacing counting Bloom filter,'' \emph{IEEE Communications Letters,} vol. 21, no. 2, pp. 278-281, 2017.

\bibitem{Stable_BF} F. Deng, D. Rafiei, ``Approximately detecting duplicates for streaming data using stable Bloom filters,'' in \emph{Proc. ACM SIGMOD,} Chicago, Illinois, USA, 2006.

\bibitem{TCBF} Y. Zhao, J. Wu, ``The design and evaluation of an information sharing system for human networks,'' \emph{IEEE TPDS,} vol. 25, no. 3, pp. 796-805, 2014.  

\bibitem{Double_Buffering}  F. Chang, C. Wu, and K. Li, ``Approximate caches for packet classification,'' in \emph{Proc. IEEE INFOCOM,} Honkong, China, 2004.

\bibitem{Two_Active}  M.K. Yoon, ``Aging Bloom filter with two active buffers for dynamic sets,'' \emph{IEEE TKDE,} vol. 22, no. 1, pp. 134-138, 2010.

\bibitem{Forgetful_BF} R. Subramanyam, I. Gupt, L.M. Leslie, W. Wang, ``Idempotent distributed counters using a forgetful Bloom filter,'' \emph{Cluster Computing,} vol. 19, no. 2, pp. 879-892, 2016.

\bibitem{locality_hashing} M. Datar, N. Immorlica, P. Indyk, V.S. Mirrokni, ``Locality-sensitive hashing scheme based on p-stable distributions,'' in \emph{Proc. ACM SoCG,} Brooklyn, New York, USA, 2004.

\bibitem{locality_hashing1} P. Indyk, R. Motwani, ``Approximate nearest neighbors: towards removing the curse of dimensionality,'' in \emph{Proc.  ACM STOC,}  Dallas, TX, USA, 1998.

\bibitem{Multigranularity_LBF} J. Qian, Q. Zhu, H. Chen, ``Multi-granularity locality-sensitive Bloom filter,'' \emph{IEEE TC,}  vol. 64, no. 12, pp. 3500-3514, 2015.

\bibitem{DS_BF} A. Kirsch, M. Mitzenmacher, ``Distance-sensitive Bloom filters,'' in \emph{Proc. The Workshop on Algorithm Engineering and Experiments,} Miami, Florida, USA, 2006.

\bibitem{LSBF}  Y. Hua, B. Xiao, B. Veeravalli, D. Feng, ``Locality-sensitive Bloom filter for approximate membership query,''  \emph{IEEE TC,} vol. 61, no. 6, pp. 817-830, 2012.

\bibitem{IBF}  D. Eppstein, M.T. Goodrich, F. Uyeda, G. Varghese, ``What's the difference?: Efficient set reconciliation without prior context,'' in \emph{Proc. ACM SIGCOMM,} Toronto, Ontario, Canada, 2011.

\bibitem{protect_1} P. Reviriego, S. Pontarelli, J.A. Maestro, M. Ottavi, ``A method to protect Bloom filters from soft errors'', in \emph{IEEE DFT,}  Amherst, MA, USA,  2015.

\bibitem{protect_2}  A. Sánchez-Macián, P. Reviriego, J.A. Maestro, S. Liu, ``Single event transient tolerant Bloom filter implementations'', \emph{IEEE TC,} vol. 66, no. 10, pp. 1831-1836, 2017.


\bibitem{learned_BF_1} T. Kraska, A. Beutel, E. Chi, J. Dean, N. Polyzotis, ``The case for learned index structures,'' in \emph{Proc. ACM SIGMOD,} Houston, TX, USA, 2018.   

\bibitem{learned_BF_2} M. Mitzenmacher, ``A model for learned Bloom filters and related structures,''  \emph{arXiv preprint,} arXiv:1802.00884, 2018.



\bibitem{Cuckoo_filter} B. Fan, D. Andersen, M. Kaminsky, M. Mitzenmacher, ``Cuckoo filter: practically better than Bloom,'' in \emph{Proc. ACM CoNEXT,} Sydney, Australia, 2014. 

\bibitem{Quotient_filter} M. Bender, M. Farach-Colton,  B. Kuszmaul, B. Kuszmaul, D. Medjedovic, P. Montes, P. Shetty, R. Spillane, E. Zadok, ``Don't thrash: How to cache your hash on flash,'' in \emph{Proc. USENIX HotStorage,}  Portland, OR, USA, 2011.

\bibitem{Simplification_Cuckoo} D. Eppstein, ``Cuckoo filter: Simplification and analysis,'' \emph{arXiv preprint,} arXiv:1604.06067, 2016.

\bibitem{Adaptive_Cuckoo} M. Mitzenmacher, S. Pontarelli, P. Reviriego, ``Adaptive Cuckoo filters,'' in \emph{Proc. SIAM ALENEX,} New Orleans, Louisiana, USA, 2018.

\bibitem{Dynamic_Cuckoo} H. Chen, L. Liao, H. Jin, J. Wu, ``The dynamic Cuckoo filter,'' in \emph{IEEE ICNP,} Toronto, Canada, 2017.


\bibitem{Consistent_CF} L. Luo, D. Guo, O. Rottenstreich, X. Luo, R. T.B. Ma, B. Ren, ``The consistent Cuckoo filter,'' in \emph{IEEE INFOCOM,} Paris, France, 2019.

\bibitem{Counting_Quntient} P. Pandey, M. Bender, Rob. Johnson, R. Patro, ``A general-purpose counting filter: Making every bit count,'' in \emph{Proc. ACM SIGMOD,} Chicago, Illinois, USA, 2017

\bibitem{Cuckoo_table} R. Pagh, F. Rodler, ``Cuckoo hashing,'' \emph{Journal of Algorithms,} vol. 51, no. 2, pp. 122-144, 2004.

\bibitem{Partial_key_Cuckoo} B. Fan, D. G. Andersen, M. Kaminsky, ``MemC3: Compact and concurrent memcache with dumber caching and smarter hashing,'' in \emph{ Proc. USENIX NSDI,} Lombard, IL, USA, 2013.

\bibitem{Quotienting} D. E. Knuth, ``The art of computer programming: Sorting and searching,'' vol. 3, Addison Wesley, 1973. 

\end{thebibliography}
\end{document}